\documentclass[12pt,preprint]{aastex}

\newcommand{\text}[1]{\rm #1}

\slugcomment{}

\shorttitle{Effects of Rotation on Stochasticity of GWs in Supernovae}
\shortauthors{Kotake et al.}

\begin{document}

\title{Effects of Rotation on Stochasticity of Gravitational Waves 
in Nonlinear Phase of Core-Collapse Supernovae}

\author{Kei Kotake\altaffilmark{1}, Wakana Iwakami Nakano\altaffilmark{2}, 
 and Naofumi Ohnishi\altaffilmark{2}}
\affil{$^1$Division of Theoretical Astronomy, National Astronomical Observatory of Japan, 2-21-1, Osawa, Mitaka, Tokyo, 181-8588, Japan}
\email{kkotake@th.nao.ac.jp}
\affil{$^2$Department of Aerospace Engineering, Tohoku University,
6-6-01 Aramaki-Aza-Aoba, Aoba-ku, Sendai, 980-8579, Japan}
\begin{abstract}
By performing three-dimensional (3D) simulations that demonstrate the 
neutrino-driven core-collapse supernovae aided by the standing accretion shock 
instability (SASI),  we study how the spiral modes of the SASI can have impacts on the properties of 
 the gravitational-wave (GW) emission. To see the effects of rotation in 
 the non-linear postbounce phase, 
we give a uniform rotation on the flow advecting from the outer boundary 
of the iron core, whose specific angular momentum is 
assumed to agree with recent stellar evolution models. 
We compute fifteen 3D models in which the initial angular momentum as well as the 
input neutrino luminosities from the 
protoneutron star are changed in a systematic manner. By performing 
a ray-tracing analysis, we accurately estimate 
 the GW amplitudes generated by anisotropic neutrino emission. Our results show that
 the gravitational waveforms from neutrinos in models that include rotation 
 exhibit a common feature otherwise they 
vary much more stochastically in the absence of rotation. The breaking of the 
 stochasticity stems from the excess of the neutrino 
emission parallel to the spin axis. This is because the compression of matter 
is more enhanced in the vicinity of the equatorial plane due to the growth of the 
spiral SASI modes, leading to the formation of the spiral flows circulating around the 
spin axis with higher temperatures. We point out that a recently 
proposed future space interferometers like Fabry-Perot type DECIGO would permit the 
detection of these signals for a Galactic supernova. 
\end{abstract}

\clearpage

\section{Introduction}
 The successful detection of neutrinos from SN1987A paved the way for 
{\it Neutrino Astronomy} \citep{hirata}, an alternative to 
conventional astronomy by electromagnetic waves.
 Core-collapse supernovae are now 
expected to be opening yet another astronomy, {\it Gravitational-Wave Astronomy}.
 Currently long-baseline laser interferometers such as
LIGO \citep{firstligonew},
VIRGO$^1$\footnotetext[1]{http://www.ego-gw.it/},
GEO600$^2$\footnotetext[2]{http://geo600.aei.mpg.de/},
and TAMA300 \citep{tamanew} are operational 
(see, e.g., \citet{hough} for a recent review).  
For these detectors, core-collapse supernovae have been proposed as one of 
the most plausible sources of gravitational waves (GWs) 
(see, for example, \citet{kota06,ott_rev} for recent reviews).

Although the mechanism of explosion is not completely understood yet,
  current multi-dimensional simulations based on refined numerical models
 show several promising scenarios. Among the candidates 
is the neutrino heating mechanism aided by convection and standing accretion shock
 instability (SASI) (e.g., \citet{marek,bruenn,suwa}), 
the acoustic mechanism \citep{burr06}, or the magnetohydrodynamic (MHD) mechanism
 (e.g., \citet{taki04,taki09,ober06b,burr07}, and \citet{kota06} for 
collective references therein). 
For the former two to be the case, 
the explosion geometry is expected to be unipolar and bipolar, and for 
 the MHD mechanism to be bipolar.
  Since the GW signatures imprint a live information
 of the asphericity at the moment of explosion, they could provide us invaluable
 clues to understand the supernova mechanism.

 Traditionally, most of the theoretical predictions of GWs
   have focused on the bounce signals
(e.g., \citet{mm,zwer97,kotakegw,kota04a,shibaseki,ott,ott_prl,ott_2007,dimm02,dimmelprl,dimm08,simon1}) and references therein). However recent stellar evolution 
calculations suggest that rapid rotation assumed in most of the previous studies 
is not canonical for progenitors with neutron star formations
\citep{hege05,ott_birth}.
Besides the rapid rotation of the core, 
convective matter motions and anisotropic neutrino emission in the much 
later postbounce phase are expected to be the primary GW sources 
with comparable amplitudes to the bounce signals. 
Thus far, various physical ingredients for producing asphericities
 and the resulting GWs in the postbounce phase have been studied, such as the roles 
of pre-collapse density inhomogeneities \citep{burohey,muyan97,fryersingle,fryer04a},
 moderate rotation of the iron core \citep{mueller04}, nonaxisymmetric rotational 
instabilities \citep{rampp,ott_3D}, g-modes \citep{ott_new} and r-modes pulsations 
\citep{nils} of protoneutron stars (PNSs), and SASI 
(\citet{kotake07,kotake_ray,kotake09,marek_gw,murphy}).

Here SASI, becoming 
very popular in current supernova researches, 
is a uni- and bipolar sloshing of the stalled supernova shock 
with pulsational strong expansion and contraction (e.g., \citet{blon03,sche04,Ohni05,fogli07,iwakami1}). Based on two-dimensional (2D) simulations that parametrize the 
 neutrino heating and cooling by a light-bulb scheme 
 to obtain explosions (which we shortly call as 
  {\it parametric SASI simulations}), we pointed out 
that the GW amplitudes from anisotropic neutrino emission increase almost monotonically 
with time, and that such signals
 may be visible to next-generation detectors for a Galactic source 
\citep{kotake07,kotake_ray}. By performing such a parametric simulation but without 
 the excision inside the PNS, \citet{murphy} showed 
that the GW signals from matter motions can be a good indicator to get the
information of the explosion geometry.
These features qualitatively agree with the ones obtained by
\citet{yukunin} who reported 
  exploding 2D simulations in which a multi-group flux limited diffusion 
transport is solved with the hydrodynamics.
 \citet{marek_gw} analyzed the GW emission based on their long-term 2D (ray-by-ray) 
Boltzmann simulations, which seem very close to produce 
 explosions \citep{marek}. They also confirmed that the GWs from neutrinos 
with continuously growing amplitudes (but with the different sign of the 
amplitudes in \citet{kotake07,kotake_ray,yukunin}), 
are dominant over the ones from matter motions. 
They proposed that the third-generation class detectors 
such as the Einstein Telescope are required for detecting the GW signals 
with a good signal-to-noise ratio.

Thanks to a growing computational power, three-dimensional (3D)
 simulations, though mostly limited to the parametric 
simulations at present, are now becoming practicable. 
In 3D, the modes of SASI are divided into {\it sloshing} modes 
and {\it spiral} modes (e.g., \citet{iwakami1}).
 Asymmetric $m = 0$ modes so far studied in 2D models 
and axisymmetric $m\neq 0$ modes are classified into the {\it sloshing} modes,
where $m$ stands for the azimuthal index of the spherical harmonics $Y_{l}^{m}$.
In the latter situation, the $\pm m$ modes degenerate
so that the $+m$ modes has the same amplitudes as the $-m$ modes.
If random perturbations or uniformly rotating flow are imposed on 
these axisymmetric flows in the postbounce phase, 
the degeneracy is broken and the rotational modes emerge
 \citep{blo07,iwakami2}.
In this situation, the $+m$ modes has the different amplitudes from $-m$ modes.
Such rotating non-axisymmetric $m\neq0$ modes are called as {\it spiral} modes.
 These non-axisymmetric modes are expected to bring about a 
breakthrough in our supernova theory, because they can help to produce explosions 
 more easily compared to 2D due to its extra degree of freedom 
\citep{nordhaus}, and also because they may have 
a potential to generate pulsar spins (\citet{blo07}, see also \citet{yamasaki,annop}).

 In a series of our previous papers, we have studied
  the effects of the sloshing SASI modes on the GW signatures based on the 
 2D parametric SASI simulations \citep{kotake07,kotake_ray}. 
In the latter study, we proposed a ray-tracing method to accurately 
 estimate GWs generated by anisotropic neutrino emission. Then moving on to the
 3D non-rotating parametric simulations, we pointed out that the gravitational 
waveforms in 3D vary much more stochastically than for the corresponding 2D models 
because the explosion anisotropies depend sensitively on the growth of
the SASI which develops chaotically in all directions \citep{kotake09}.
As a sequel of these studies, we hope to study the role
of the spiral SASI modes on the GW emission in this work.
 We include the effects of rotation
 as in \citet{iwakami2}. That is, when the SASI gradually transits from the 
 linear to nonlinear phase,  
 we give a uniform rotation to the flow advecting from the outer boundary of the 
iron core, whose specific angular momentum is assumed to agree with recent stellar evolution models \citep{hege05}.  
After the initial rotational flows advect to the PNS surface, 
rotation begins to have influence over the GW signals 
 due to the non-axisymmetric flow motions and anisotropic neutrino emission outside the 
 PNS. Such an approach (excision inside the PNS), could be justified by some striking
  evidences that support the slow rotation of iron cores \citep{hege05},
   and also that the postbounce density structure of the PNS for such a progenitor
  (model m15b6 in \citet{hege05}) has a very similar structure to the ones in the
non-rotating models \citep{ott_birth}.
 To see clearly the effects of rotation, we compute fifteen 3D models in which 
 the initial angular momentum as well as the input neutrino luminosities from the PNS 
 are changed in a systematic manner. To estimate the 
GWs from anisotropic neutrino radiation,  we perform the ray-tracing analysis.
 With these computations, we hope to clarify 
how rotation that gives a special direction to the system (i.e., the 
 spin axis), could affect the stochastic GW features obtained in 
  our previous study (e.g., \citet{kotake09}).

The plan of this paper is as follows. In section \ref{sec2}, 
 we shortly summarize the information how to construct our 3D models and also how to 
  compute the gravitational waveforms.
 The main results are shown in Section \ref{sec3}.  We summarize our results 
and discuss their implications in Section \ref{sec4}.  

\section{Numerical Methods and Models}\label{sec2}
\subsection{Extraction of Gravitational Waveforms}

The numerical methods to extract the gravitational waveforms have been already described
 in \citet{kotake_ray}. We extract the GW amplitudes from mass motions
 using the standard quadrupole formula (e.g., \citet{zhuge,rampp}), and their spectra by 
\citet{flanagan} with the FFT techniques. For the GWs generated by anisotropic neutrino 
emission \citep{epstein,muyan97}, the two modes of the GWs 
  can be derived as follows,
\begin{eqnarray}
    h_{+} &=& 
C \int_{0}^{t} \int_{4 \pi} d\Omega' 
(1 + s(\theta') c (\phi') s(\xi) + c(\theta') c(\xi)) \times \nonumber \\ 
& & \frac{( s(\theta')c(\phi') c(\xi) - c(\theta')s(\xi))^2 - s^2(\theta') s^2(\phi')}
{[s(\theta')c(\phi')c(\xi) - c(\theta')s(\xi)]^2 + s^2(\theta')s^2(\phi')} 
\frac{dl_{\nu}(\Omega',t')}{d\Omega'},
\label{t+}
\end{eqnarray}
and 
\begin{eqnarray}
     h_{\times} &=& 2C  \int_{0}^{t}
\int_{4\pi} d\Omega' (1 + s(\theta') c (\phi') s(\xi) + c(\theta') c(\xi)) \times \nonumber \\ 
& &  \frac{s(\theta') s(\phi')(s(\theta') c(\phi') c(\xi) - c(\theta')s(\xi))}{[s(\theta')c(\phi')c(\xi) - c(\theta')s(\xi)]^2 + s^2(\theta')s^2(\phi')}
 \frac{dl_{\nu}(\Omega',t')}{d\Omega'},
\label{tcc}
\end{eqnarray}
where $s(A)\equiv \sin(A), c(B)\equiv \cos B$,
 $C \equiv 2G/(c^4 R)$ with $G,c$ and $R$, being the gravitational constant, 
 the speed of light, the distance of the source to the observer respectively,
 $dl_{\nu}/d\Omega$
represents the direction-dependent neutrino luminosity emitted per unit
of solid angle into direction of $\Omega$, and  $\xi$ is the viewing angle 
(e.g., \citet{kotake_ray}).
For simplicity, we consider here two cases, in which the observer is 
 assumed to be situated along 
'polar' ($\xi=0$) or 'equatorial'($\xi=\pi/2$) direction. It is important to note 
that in the case of 
2D axisymmetric case, the only non-vanishing component is the plus mode for 
 the equatorial observer, 
\begin{eqnarray}
h^{\rm e}_{+} &=&  2C \int_{0}^{t} dt^{'}
\int_{0}^{\pi}~d\theta'~\Phi(\theta')~\frac{dl_{\nu}(\theta',t')}
{d\Omega'},
\label{tt}
\end{eqnarray}
 where the function of $\Phi(\theta^{'})$ has positive values in the north polar cap 
for $0 \leq \theta' \leq 60^{\circ}$ and in 
the south polar cap for $120^{\circ} \leq \theta' \leq 180^{\circ}$, 
but becomes negative values between $60^{\circ} < \theta' < 120^{\circ}$
 (see Figure 1 of \citet{kotake07}). 

To determine $dl_{\nu}/d\Omega$ in equations (\ref{t+}, \ref{tcc}), 
we perform a ray-tracing calculation which we shortly summarize in the following
 (see \citet{kotake_ray} for more detail). 
In the ray-tracing approach, we consider transfer along the ray specified by 
a constant impact parameter $p$. The coordinate along $p$ is called $s$, satisfying
\begin{equation}
r = (p^2 + s^2)^{1/2},
\end{equation}
where $r$ is the radial coordinate. In order to get the numerical 
convergence of $dl_{\nu}({\bf\Omega})/d\Omega$, 
 we need to set 45,000 rays for each direction, which consists of 
$500 \times 90$ rays, where the former is 
for the impact parameters covering from the inner- ($p_{\rm in}$ = 50 km) to the outer- 
boundary ($p_{\rm out} $ = 2000 km)
 of the computational domain and the latter is for covering 
the circumference (e.g., $2\pi$) of the concentric circles on the plane perpendicular to the rays.

The transfer equation of the neutrino occupation probability 
$f_{\nu}(\epsilon_{\nu},p,s)$ for a given neutrino energy 
 $\epsilon_{\nu}$ along each ray is given by,
\begin{equation}
\frac{d f_{\nu}(\epsilon_{\nu},p,s)}{d s} = j(\epsilon_{\nu},p,s)(1 - f_{\nu}(\epsilon_{\nu},p,s)) - \frac{f_{\nu}(\epsilon_{\nu},p,s)}{\lambda},
\label{transfer}
\end{equation}
 where $j$ and $\lambda$ is the 
 emissivity and absorptivity via neutrino absorptions and emission by free nucleons 
($\nu_{\text{e}} + \text{n} \rightleftarrows \text{e}^{-} + \text{p}$) 
 (\citet{bruenn85,Ohni05}), which 
 are dominant processes outside the PNSs. The optical depth for those reactions
 are estimated by $\tau_{\nu} = \int_{r}^{\infty} 1/ \lambda$.
 For the sake of simplicity, 
the neutrino scattering and the velocity-dependent terms in the transport 
equation are neglected. 
Along each ray, $f_{\nu}$ is transferred by the line integral.
 When the line integral starts from the surface on the PNS,
 we set the initial value of
\begin{equation}
 f(\epsilon_{\nu})
  = \frac{1}{1+\exp(\epsilon_{\nu}/k_{\text{B}}T_{\nu})} \cdot
\frac{1}{4\pi},
\end{equation}
 assuming that the neutrino distribution function at the surface 
is approximated by the Fermi-Dirac distribution with a vanishing chemical potential. 
 Here the neutrino temperature is set to be constant near 
$T_{\nu_{\text{e}}} = 4$ MeV,
 whose values change slightly depending on the input neutrino luminosity.
 Note that these values are constant in time for each model. This is necessary
to realize the steady unperturbed initial states (e.g., \citet{Ohni05}).  
 For the rays that do not hit the PNS, we start the line integral 
from the outer most boundary antipodal to the line of sight, where $f_{\nu}$ is 
essentially zero.

By a post process, we perform the line integral up to the outer-most boundary for 
each hydro-timestep. With $f(\epsilon_{\nu},p,s_{\rm out})$, which is obtained by the 
line integral 
 up to the outer-most boundary, the neutrino energy fluxes 
along a specified direction of ${\bf\Omega}$ can be estimated,
 \begin{equation}
\frac{dl_{\nu}({\bf\Omega},p)}{d\Omega~dS}  = \int f(\epsilon_{\nu},p,s_{\rm out})\cdot(c\epsilon_{\nu}) 
\cdot \frac{\epsilon_{\nu}^2 d\epsilon_{\nu}}{(2 \pi \hbar c)^3}.
\label{flux}
\end{equation}
 By summing up the energy fluxes with the weight of the area 
in the plane perpendicular to the rays, 
we can find $dl_{\nu}/d\Omega$ along a specified direction ${\bf\Omega}$,
\begin{equation}
\frac{dl_{\nu}({\bf\Omega})}{d\Omega} = \int \frac{dl_{\nu}({\bf\Omega},p)}{d\Omega dS}~dS = 
\int_{p_{\rm in}}^{p_{\rm out}} dp~2 \pi p~ \frac{dl_{\nu}({\bf\Omega},p)}{d\Omega dS}.
\label{final}
\end{equation}
Repeating the above procedures, $dl_{\nu}({\bf\Omega})/d\Omega$ 
 can be estimated for all the directions, by which we can find the amplitudes of 
the neutrino GWs through equations (\ref{t+},\ref{tcc}). In the following 
computations, we assume that the distance to the GW source is comparable to 
 our galactic center ($R = 10~\rm{kpc}$) unless stated otherwise. 

\subsection{Construction of 3D Models with Rotation}

The employed numerical methods and model concepts are essentially the same as 
those in our previous paper \citep{iwakami2}. 
Using the ZEUS-MP code \citep{hayes} as a hydro-solver, we solve the 
dynamics of the standing accretion shock flows of matter 
attracted by the protoneutron star and irradiated by neutrinos 
emitted from the PNS.
We employ the so-called light-bulb approximation (see, e.g.,
 \citet{jankamueller96,Ohni05}) and adjust the neutrino luminosities from the
PNSs to trigger explosions.
This method makes it possible for us to study the properties of 
GWs in the postbounce phase, namely from the
shock-stall, through the growth of the SASI, to 3D explosions (see also
 \citet{nordhaus,annop}), and is found to work well in 2D, capturing 
the essential features obtained by more realistic simulations 
(see references in \cite{sche04,scheck06}).

The computational grid is comprised of 300 logarithmically spaced,
radial zones to cover from the absorbing inner boundary of 
$\sim 50 ~{\rm km}$ to the outer boundary of $2000 ~{\rm
km}$, and 30 polar ($\theta$) and 60
azimuthal ($\phi$) uniform mesh points, which are used to cover the whole
solid angle (see for the resolution tests in \citet{iwakami1}).
The initial conditions are provided in the same manner of 
\cite{Ohni05}, which describes the spherically symmetric steady accretion
flow through a standing shock wave \citep{yamasaki}.
In constructing the initial conditions, we assume
 a fixed density $\rho_{\rm in} = 10^{11}$~g~cm$^{-3}$ at the inner boundary.
And the initial mass accretion rates and the initial 
mass of the central object are set 
to be $\dot{M} = 1~M_{\odot}$~s$^{-1}$ and $M_{\rm in} = 1.4~M_{\odot}$,
respectively.
To include rotation, we impose a rigid rotation on the outer boundary of the 
computational domain as follows,
\begin{equation}
v_\phi^{2D}(r,\theta) = \beta_\phi v_r^{1D} (r) \sin \theta,
\label{eq:rot}
\end{equation}
where $v_\phi^{2D}$ denotes the unperturbed $\phi$ component of velocity, $v_r^{1D}(r) = - \dot{M}/4\pi r^2\rho(r)$ is the unperturbed radial velocity,  and $\beta_\phi$ 
 denotes the rotation parameter (see \citet{iwakami2} for more detail). 
We examine the flow characteristics in the following three ways 
$\beta_\phi=0, 0.01$, and 0.015 which corresponds to the specific angular momentum 
${\it L}\sim (4-6) \times 10^{15}$~cm$^2$~s$^{-1}$ on the equatorial plane. This 
 choice is close to the recent rotating presupernova models that include the 
 effects of magnetic breaking during stellar evolution \citep{hege05}. 
 It is computationally prohibitive at present to follow a long-term postbounce 
 evolution in 3D simulations that implement the spherical coordinates due to its
 severe Courant-Friedlichs-Lax condition around the pole. As a prelude to 
the full 3D simulations starting from gravitational collapse to 
explosions without excision inside 
 the PNS, we choose to take into account the effect of rotation by the exploratory 
 method as mentioned above.
 To induce non-spherical instability we add random velocity perturbations to be
 less than 1 $\%$ of the unperturbed radial velocity.
By changing also the neutrino temperature at the PNS surface (e.g., equation 
(6)), we compute fifteen 3D models whose input neutrino luminosities vary in the range of 
$L_{\nu_{e}} = 6.0 - 6.8 \times 10^{52}$~erg ~s$^{-1}$ (see Table \ref{table1}). 

  In Table \ref{table1}, $\Delta t$ represents the simulation time when the average 
shock radius that continuously increases with the growth of SASI, reaches the outer 
boundary of the computational domain with a typical explosion energy of 
$\sim 10^{51}$ erg. On the other hand, for models denoted by
 "$-$", we terminated the simulation at about 1000 ms, not seeing the 
increase of the shock radius (represented by "No" in the table). It can be seen that 
 models with higher neutrino luminosities such as models of series D and E produce 
 explosions without rotation (models D0 and E0), while models with the intermediate 
luminosities such as models B2, C1 and C2 produce explosions only when they possess 
rotation (compare models B0 and C0). This is mainly because the centrifugal
 force induced by rotation makes not only the shock radius but also the 
 gain region larger, which works to assist the onset of explosion.
Regardless of rotation,  no explosions are obtained 
 in the lowest luminosity models (model of series A)
 for the rotation parameters investigated here. 

To analyze clearly the role of rotation on the GW emission,
 models of series A that do not produce explosions, are most favorable.
 As will be explained more in detail later, this is because 
  in the higher luminosity models, the 
violent fluid motions as well as vigorous convective overturns, which works to 
smear out the effects of rotation, make the analysis more complicated. 
 For example, $\Delta t$ becomes 
shorter with larger angular momentum for the intermediate luminosity 
(models of series B and C in Table \ref{table1}). 
However this trend is no more true for models of series D (e.g., model D2). 
 This non-monotonic effect can be understood as a consequence of stochasticity 
 in a regime of high neutrino luminosity.
 
$\alpha_{\rm sk}$ in Table 1 
represents the angle between spin and kick directions that the central PNS may receive 
at the final simulation time, which we estimate according to
 \citet{annop}. The obtained spin-kick angle ranges 
from 25$^\circ$ (model B2) in which the spin direction is marginally aligned 
 with the kick direction, 88$^\circ$ (model E0) 
in which the two directions are closely perpendicular, to 170$^\circ$ (model C1)
 in which the two directions are closely opposite. 
 When a rapid rotation is imposed in our models (e.g., for our models of series 2), 
the final spin direction is closely aligned with the direction of the induced rotation
 (namely, perpendicular to the equatorial plane), while even in this case, 
stochasticity of the kick direction seems rather unaffected.
 To clarify the correlation clearly, a more elaborate study is needed 
in which the initial neutrino luminosity, perturbations, and rotation, are varied 
 much more systematically, which we plan to study as a sequel of this study (Iwakami et al. in 
preparation).

 As already mentioned, we can adjust the epoch by hand when the rotational flow
  advects to the PNS surface
 (see Figure 1 in \citet{iwakami2}). 
 The deformation of the standing 
shock becomes remarkable typically at about $t = 100$ ms, marking the epoch when 
the SASI transits from the linear to non-linear regime.
For the non-exploding models of A1 and A2, we choose to inject the rotational flow 
 so that it advects to the PNS surface at around $t =400$ ms when 
the SASI has been long developed into the non-linear phase. 
 For the other luminosity models, we take the time 
at around $t = 100$ ms (therefore soon after the SASI enters the 
non-linear phase). This is because some models with higher neutrino luminosities 
show a trend of explosions before the rotational flows advect to the central regions 
if we delay the epoch of the injection and also because we want to fix the
 epoch common to these models.

 \begin{deluxetable}{ccccccccccc}
\tabletypesize{\scriptsize}
\tablecaption{Model Summary \label{table1}}
\tablewidth{0pt}
\tablehead{
\colhead{Model}
 & $L_{{\nu_{e}}}$ 
 & $\beta_{\phi}$
 & \colhead{$\Delta t$}
 & Explosion?
 & \colhead{$|h^{\rm pol}_{\rm max}|$}
 & \colhead{$|h^{\rm equ}_{\rm max}|$}
 & \colhead{$|h^{\rm pol}_{\nu, \rm max}|$}
 & \colhead{$|h^{\rm equ}_{\nu, \rm max}|$}
 & \colhead{$E_{{\rm GW}}$ }
 & $\alpha_{\rm sk}$
 \\
 & ($10^{52}$ erg/s)
 & 
 & (ms)
 & 
 & ($10^{-22}$)
 & ($10^{-22}$)
 & ($10^{-22}$)
 & ($10^{-22}$)
 & \colhead{($10^{-11}M_{\odot}c^2$)}
 & ($^\circ$)}
\startdata
A0& 6.0 &0     &-- & No  & 0.98(+) & 1.59($\times$)  & 0.72(+) & 1.13($\times$) & 2.54 
& 100 \\
A1& 6.0  &0.01  &-- & No & 1.98(+) & 7.62(+)         & 1.18(+) & 6.74(+)  & 6.76
& 104\\
A2& 6.0  &0.015 &-- & No  & 2.58(+) & 16.7(+)         & 0.91($\times$)  & 15.1(+)& 13.4 
& 63\\
\hline
B0& 6.2  &0    &-- & No &  2.93(+)  &  3.19(+) & 2.45(+)  & 2.38(+) &5.65 & 68\\ 
B1& 6.2  &0.01 &-- &No &  2.88($\times$)  &11.7(+)& 1.37($\times$) & 10.3(+)  &15.1 & 55 \\ 
B2& 6.2 &0.015&721  &Yes & 5.03($\times$)  &  11.2(+) & 2.86($\times$) & 5.90(+) &16.1
 & 25\\ 
\hline
C0& 6.4  &0    &-- & No & 4.11(+)  & 5.00(+)  & 3.46(+) & 4.28(+) & 10.7 & 41\\
C1& 6.4 &0.01 &650   & Yes & 4.62(+)  & 6.46(+)  & 2.05($\times$) & 4.97(+) & 12.1 & 170\\
C2& 6.4  &0.015&515 & Yes & 3.43(+)  & 9.38(+)  & 1.11($\times$)  &6.09(+) & 8.96 &141\\
\hline
D0& 6.6 &0    &550 &Yes & 4.31(+)  & 2.93(+)  & 2.25(+) & 2.14($\times$) & 8.18 &113\\
D1& 6.6&0.01 &460  &Yes & 4.73($\times$)  & 4.70($\times$) & 3.72($\times$) &3.91($\times$) & 6.17 & 45\\
D2& 6.6 &0.015&525 &Yes  & 2.31($\times$)  & 4.20(+)& 1.07($\times$) & 3.50(+) & 6.87
 & 82 \\
\hline
E0& 6.8 &0    &510 &Yes &3.44($\times$) &2.56($\times$) &2.03($\times$) &2.06($\times$) & 6.60 & 88\\
E1& 6.8&0.01 &425  &Yes  &2.94($\times$)  &2.33($\times$)  &0.94(+) &0.96($\times$) & 6.67 & 150\\
E2& 6.8&0.015&410  &Yes  &2.34($\times$) &3.49($\times$)  &9.69($\times$)& 1.69(+) & 5.44
 & 168\\
\enddata
\tablecomments{
$L_{{\nu}_e}$ denotes the input luminosity.
$\Delta t$ represents the simulation time. $\beta_{\phi}$ is a rotational parameter
 introduced in equation (9).
$h^{\rm pol}_{\rm max}$, $h^{\rm equ}_{\rm max}$ represents 
the maximum amplitudes (neutrino + matter) during the simulation time seen from 
 the pole or the equator, respectively, while $h^{\rm pol}_{\nu, \rm max}$, 
$h^{\rm equ}_{\nu, \rm max}$ are the ones including only the neutrino contribution.
 The polarization of the GWs at the maximum is indicated by $+$ and
$\times$.
$E_{{\rm GW,}}$ is the radiated energy in the form of the 
 neutrino GWs in unit of $M_{\odot} c^2$.  $\alpha_{\rm sk}$ is the spin-kick angle (see text for more detail). The distance to the source is assumed to be 10 kpc. }
\end{deluxetable}

\clearpage 
\section{Results\label{sec3}}

First of all, we summarize how the inclusion of rotation could have impacts on
 the hydrodynamics as well as the overall trends in the GW emission in 
section \ref{sec3.1}. Then in section \ref{sec3.2}, we move on to analyze the 
gravitational waveforms more in detail.


\subsection{Stochastic Hydrodynamics and GWs \label{sec3.1}}

\begin{figure}[hbt]
\begin{center}
\epsscale{.74}
\plottwo{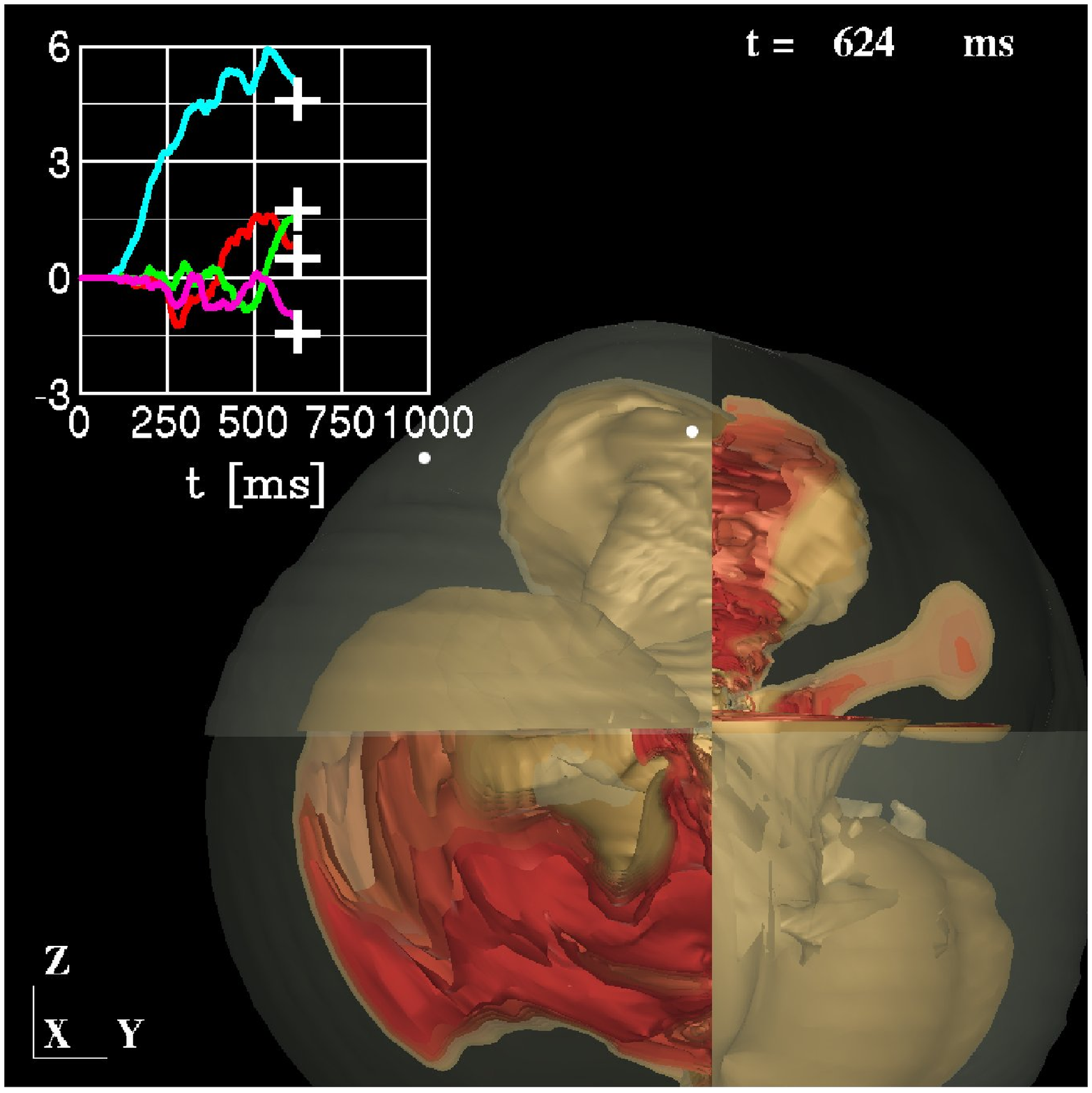}{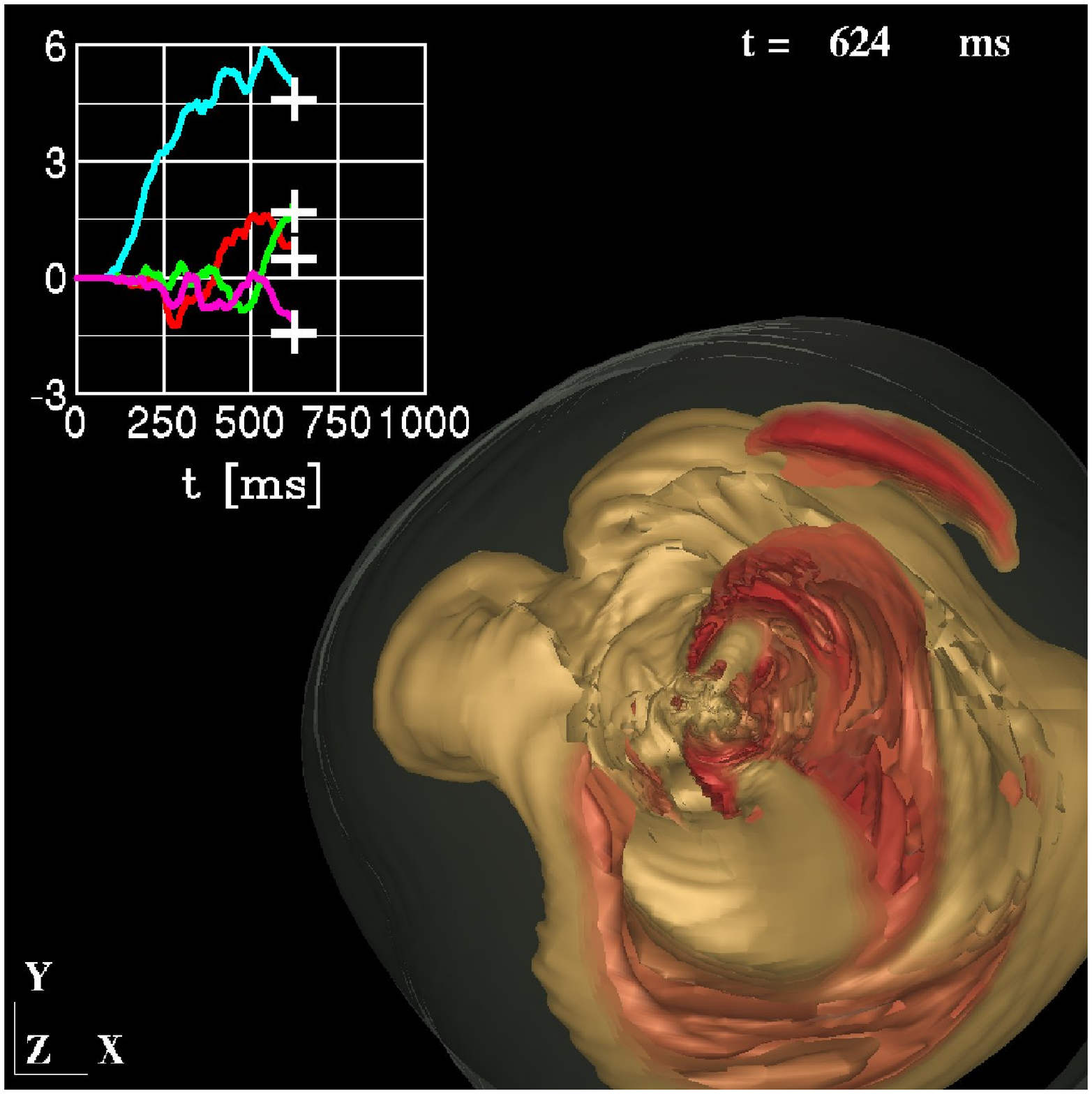}
\vspace{0.1cm}
\plottwo{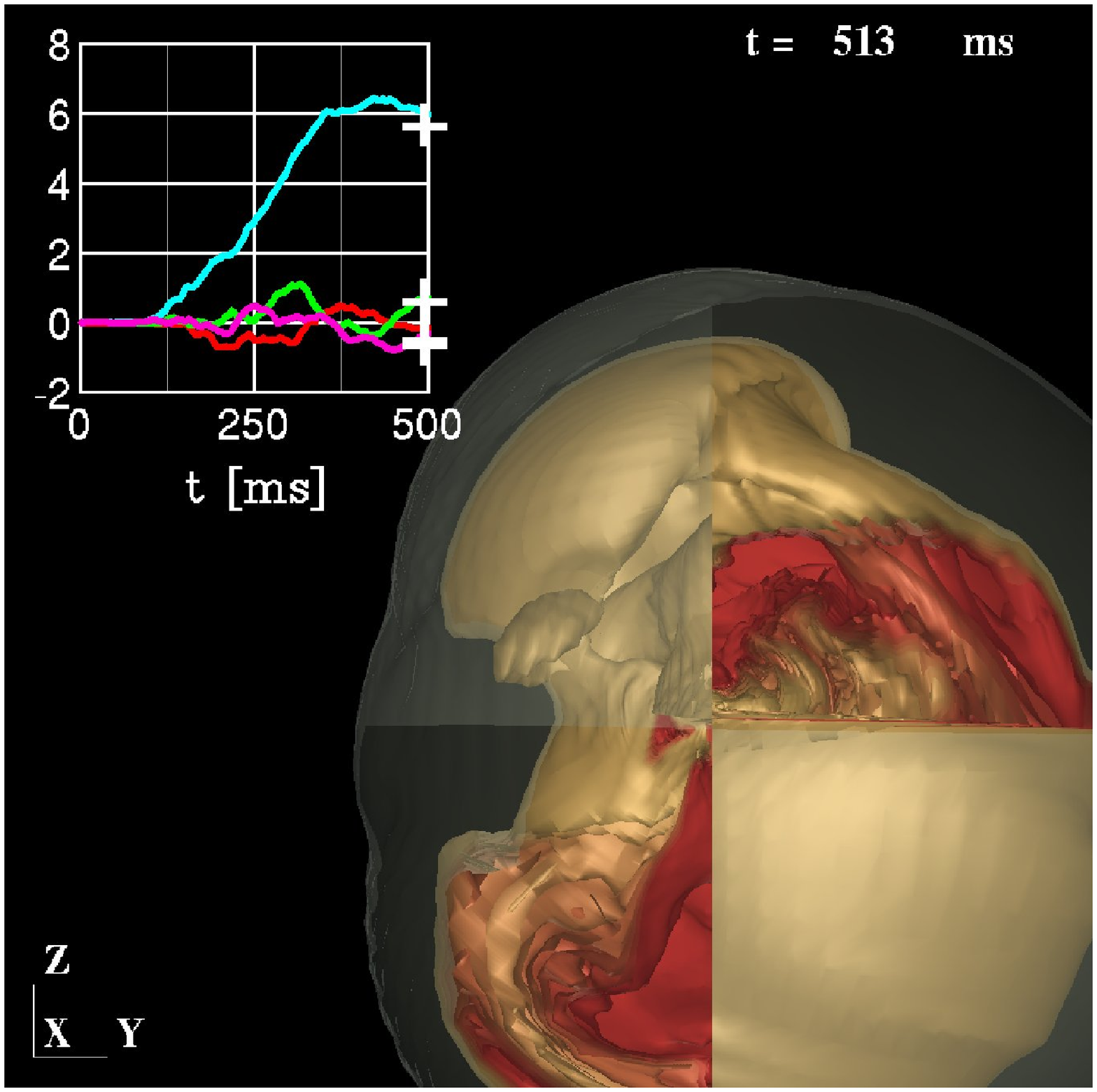}{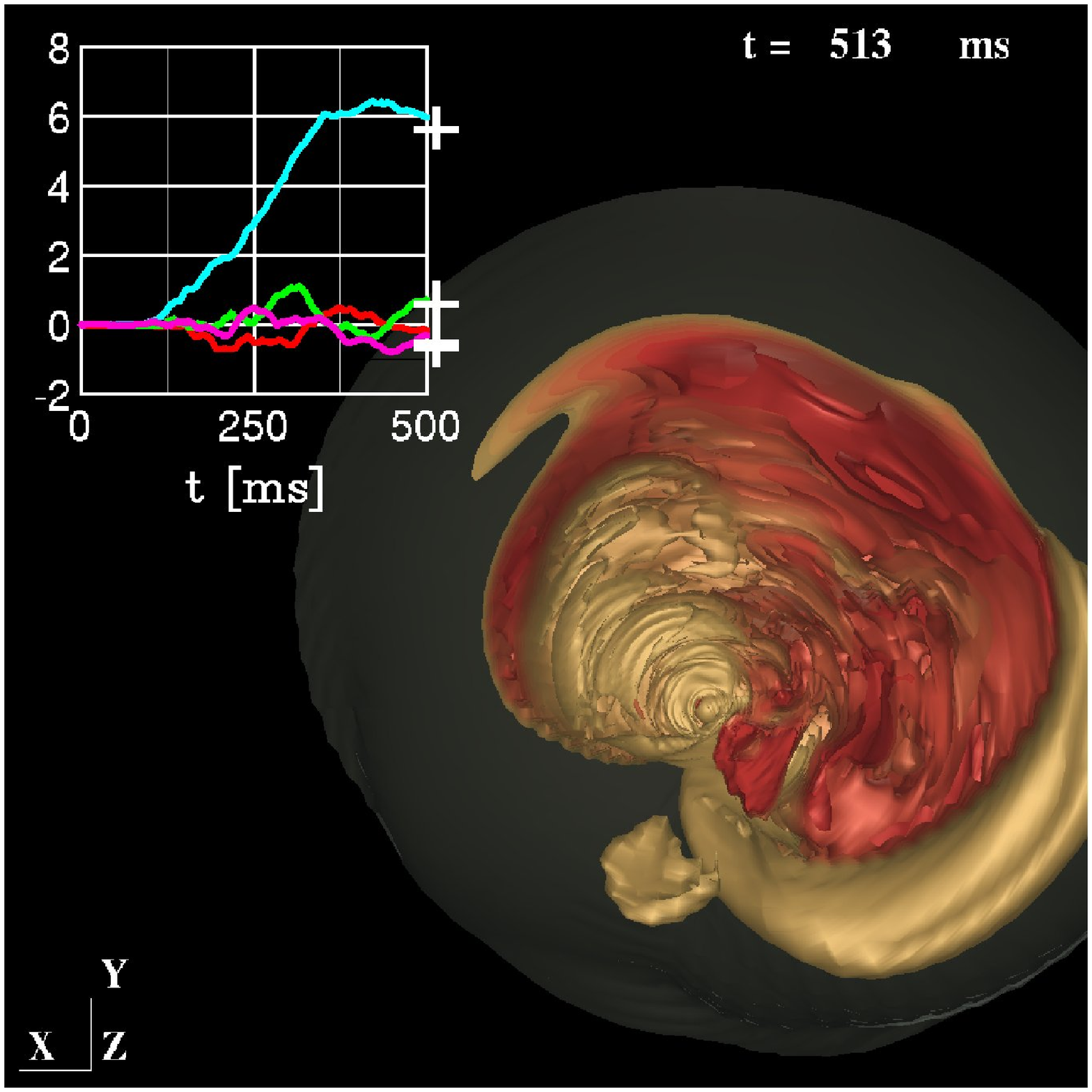}
\end{center}
\caption{Snapshots of the entropy distributions for models B2 (top panels, $t = 624$ 
ms) and C2 (bottom panels, $t=513$ ms) seen from the equator (left panels, $\phi =0$) or 
 the pole (right panels, $\theta =0$), respectively.
The second and fourth quadrant of each panel shows 
the surface of the standing shock wave.
In the first and third
quadrant, the profiles of the 
high entropy bubbles (colored by red) inside the section cut by the 
$ZY$ (left) or $YX$ (right) plane are shown. The side length of each plot is 1000km.
The insets show the gravitational waveforms with '$+$' on each curves 
representing the time of the snapshot. Note that the colors of the curves are
taken to be the same as the top panel of Figure \ref{fig3}.}
\label{fig1}
\end{figure}

\begin{figure}[hbt]
\begin{center}
\epsscale{.74}
\plottwo{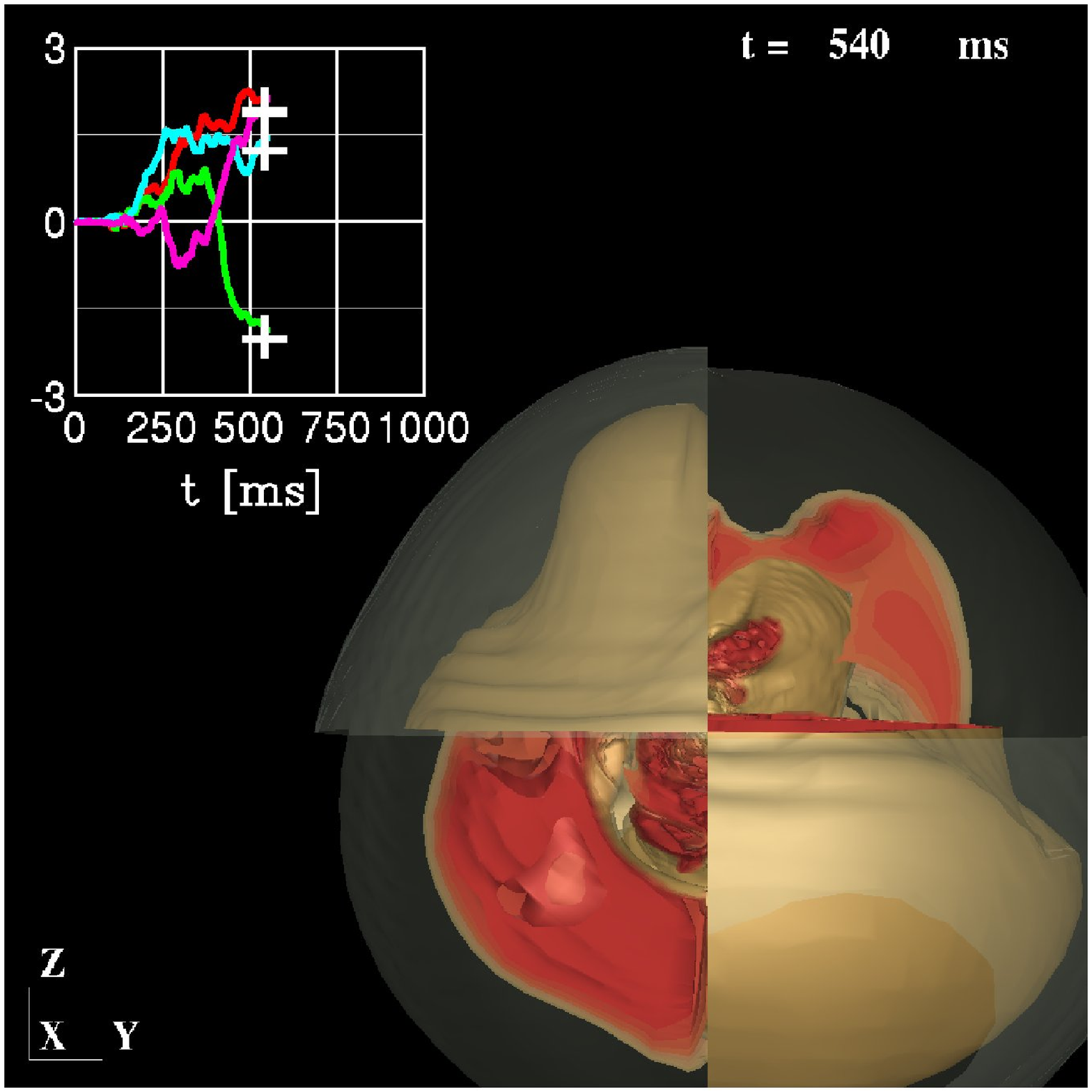}{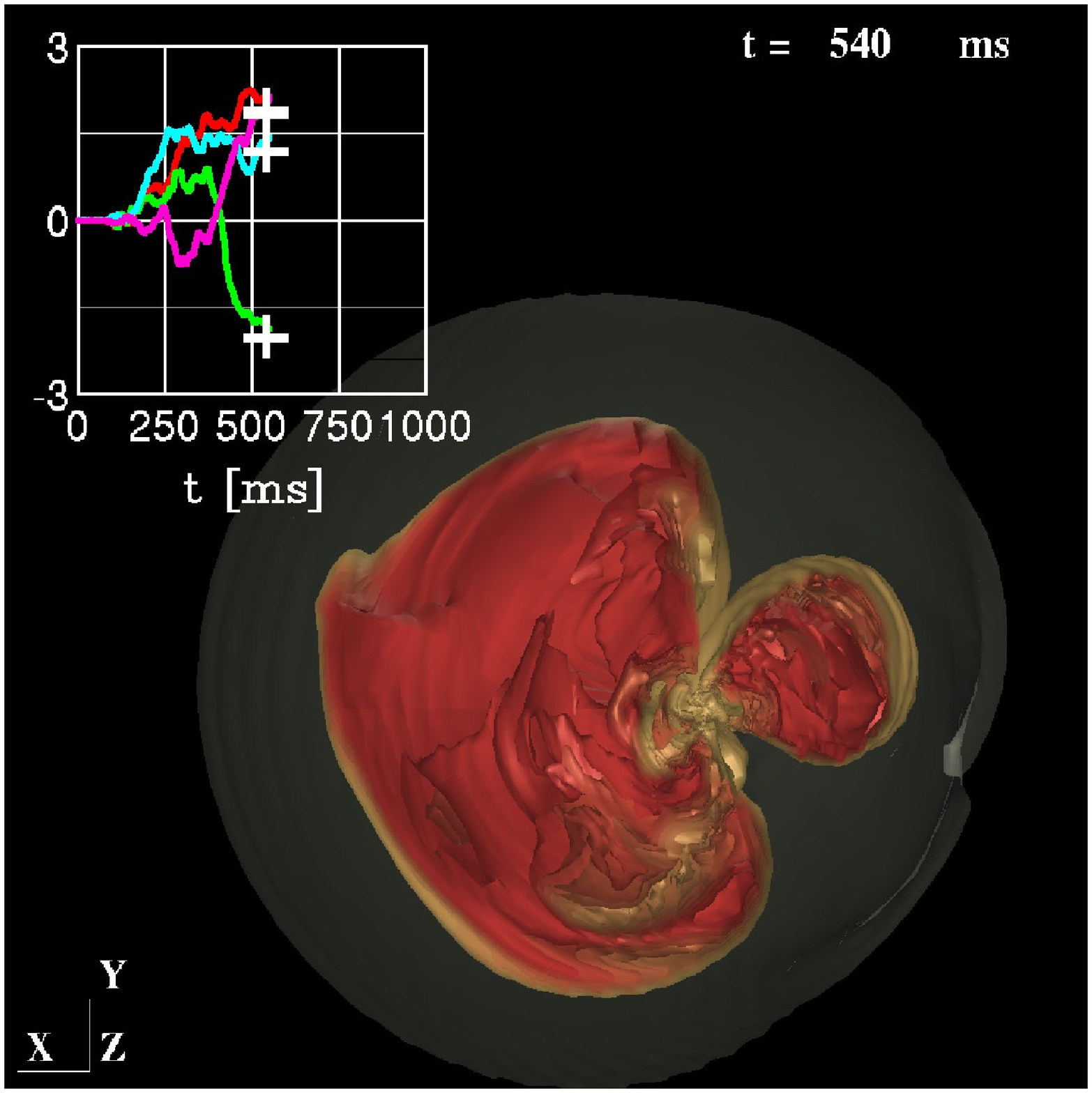}
\vspace{0.1cm}
\plottwo{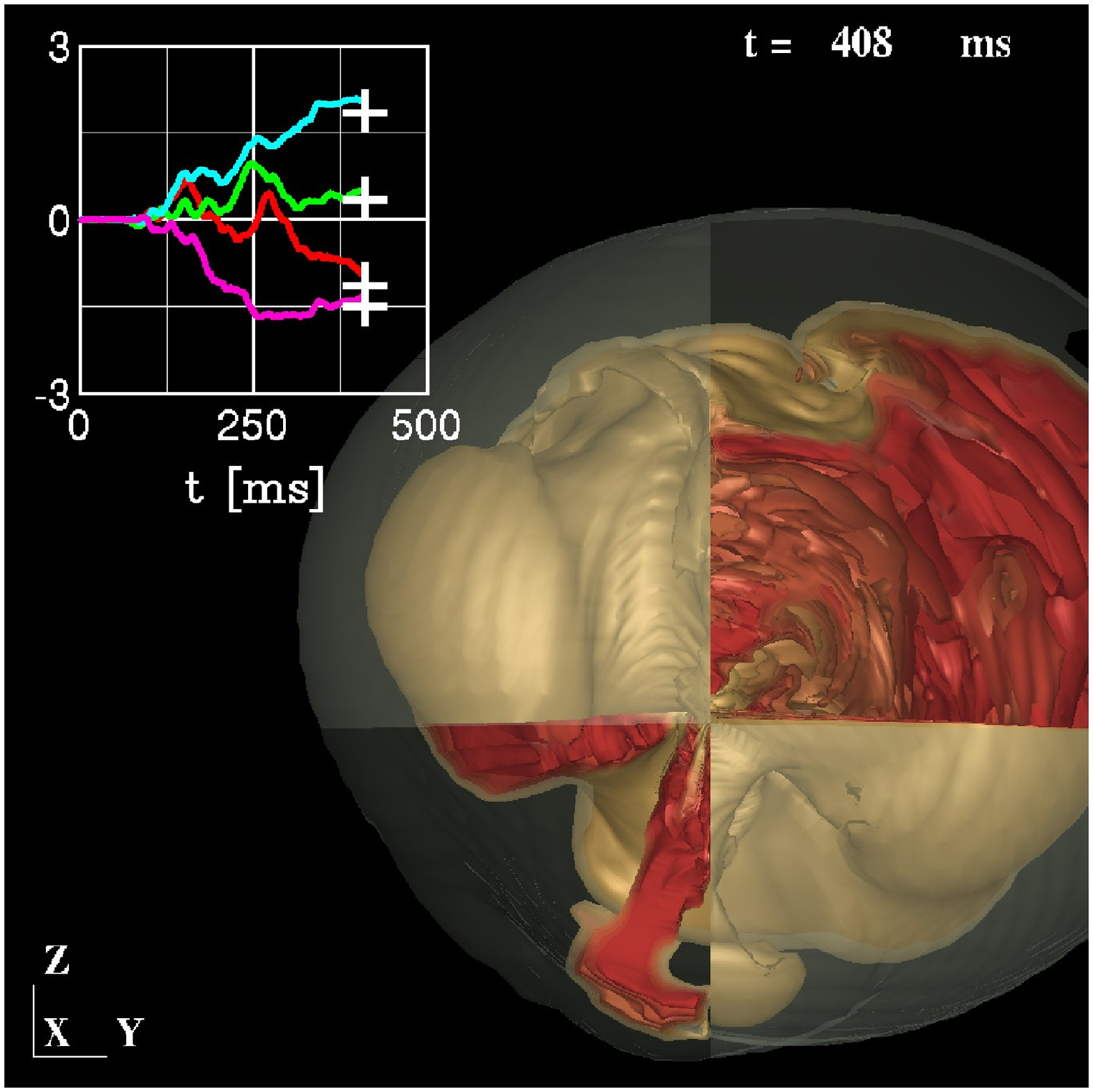}{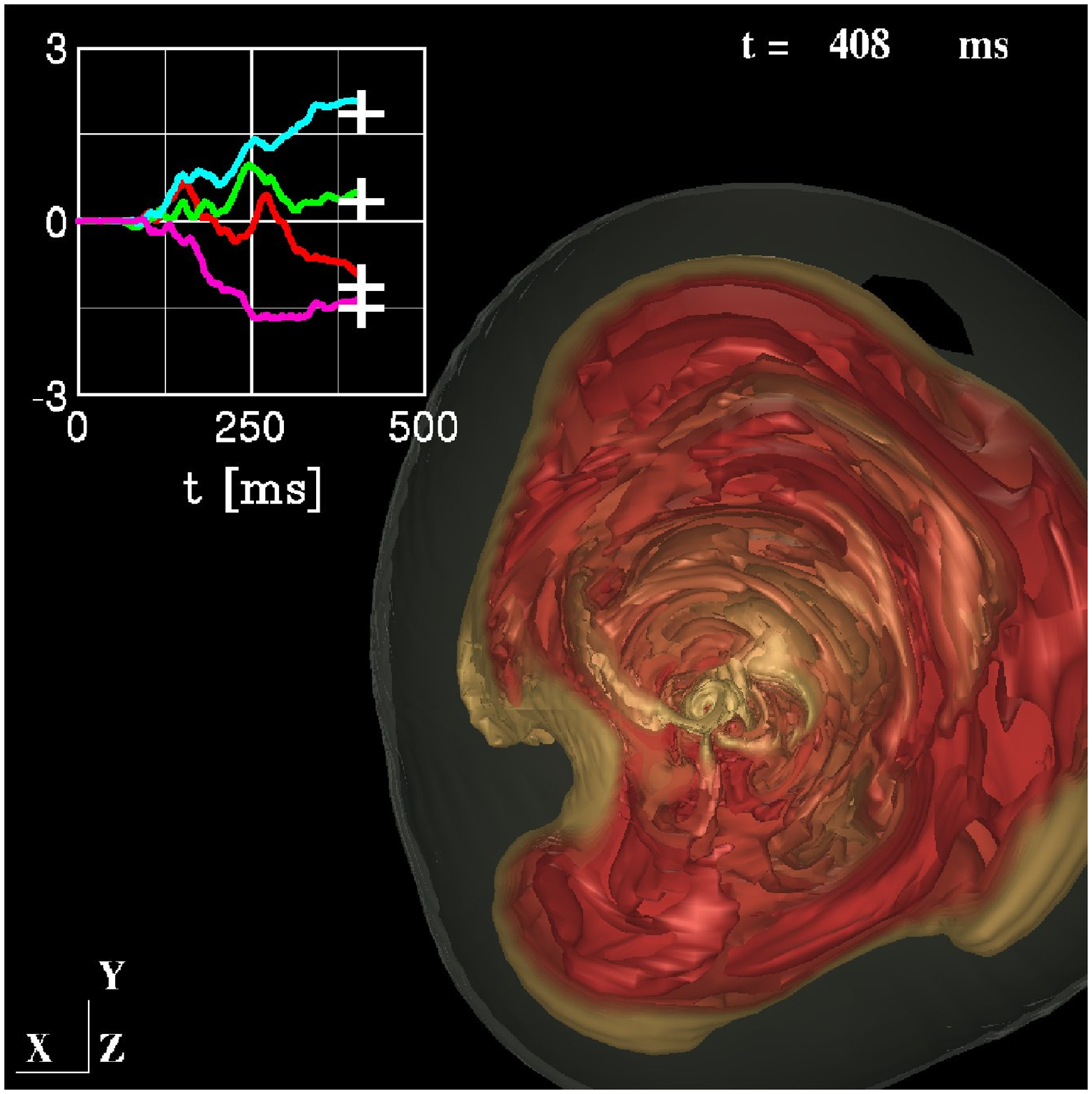}
\end{center}
\caption{Same as Figure \ref{fig1} but for models D0 (top, $t=540$ ms) and E2 
(bottom, $t=408$ ms).}
\label{fig2}
\end{figure}

Figures \ref{fig1},\ref{fig2} show the 3D hydrodynamic features of the SASI 
near at the shock breakout from the outer boundary of the computational domain.
 Comparing the top left (model B2) to the bottom left panel (model C2) 
in Figure \ref{fig1}, it can be inferred that the direction of explosion varies models 
to models. Given the same initial angular momentum, the major axis of the growth of 
SASI happens to be toward the rotational axis for model B2 (top left panel showing 
 the propagation of the high entropy bubbles (colored by red) 
  closely parallel to the rotational axis (:$z$-axis)),
 while it is shown to be toward the equatorial direction (:$y$-axis) 
for model C2 (bottom left).
 A common feature seen for the pair models 
is the spiral flows inside the shock that rotates globally around the $z$-axis, 
which is seen like arcs (right panels).
 This 
is as a result of introducing the rotation whose axis corresponded to the $z$-axis. 
As a faster rotation was added to the flows, spiral flows run with higher velocity
 from the triple points that are formed inside the standing shock, 
to near the PNS, and then flows rotate with higher velocity inside the 
high-entropy blobs. 

Figure \ref{fig2} shows the hydrodynamic features
 for models with higher neutrino luminosities (models D0 (top) and E2 (bottom)). 
 Due to the higher neutrino luminosity,
 the volumes of high entropy blobs (colored by red)
 behind the shock become much larger than the ones in Figure \ref{fig1} (right panels).
  Some spiral structures are depicted in the bottom right panel of Figure \ref{fig2}, 
 however, it is more difficult to see the structures compared to Figure 1. 
Observing from the horizontal direction, model D0 explodes in a rather spherical manner 
(top left), 
while the major axis of SASI for model E2 is tilted about $45$ degrees from the 
rotational axis (bottom left). Our 3D results presented here imply that the direction of 
explosion as well as the major axis of the growth of SASI cannot be predicted a priori,
 reflecting the chaotic fluid motions driven by the interplay of the neutrino heating,
 SASI, and convection. This feature is qualitatively 
same as the previous 3D 
results but without rotation \citep{iwakami1,nordhaus,annop}.

As the explosion dynamics becomes chaotic, there is 
no systematic dependence of the maximum GW amplitudes and the radiated GW energies 
on the input neutrino luminosities (see Table \ref{table1}). 
In fact, the largest emitted GW energy is obtained for model B2 with the 
intermediate neutrino luminosity (see, $E_{\rm GW}$ Table \ref{table1}). 
Regarding the polarization ($+$ or $\times$) of the maximum GW amplitudes, it may be 
 hard to decipher any characteristics in Table \ref{table1} due to its messiness.
 But looking very carefully, one may be able to find that the polarization of the 
maximum GWs from neutrinos seen from the equator ($h^{\rm equ}_{\nu, \rm max}$) 
is always plus ($+$).

\begin{figure}
  \begin{center}
    \begin{tabular}{cc}
      \resizebox{80mm}{!}{\includegraphics{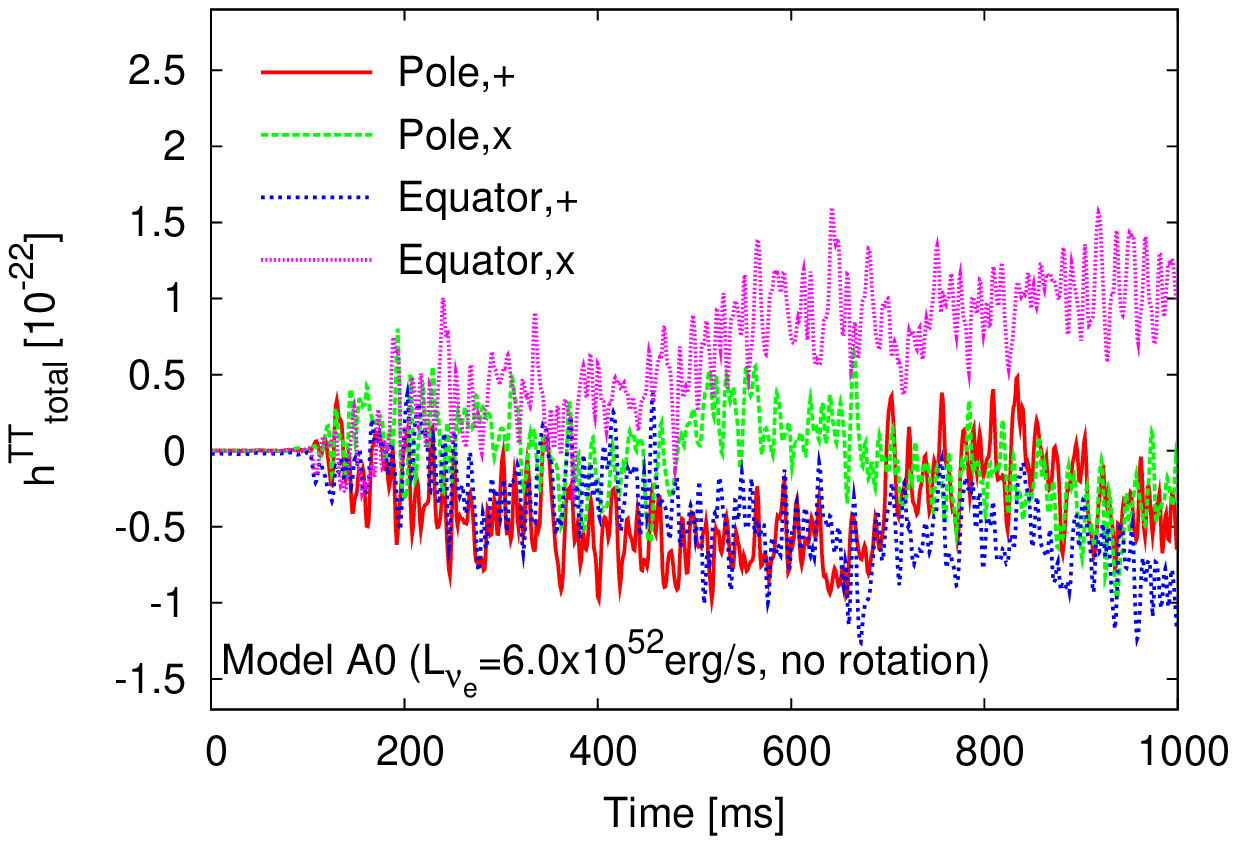}} &
      \resizebox{80mm}{!}{\includegraphics{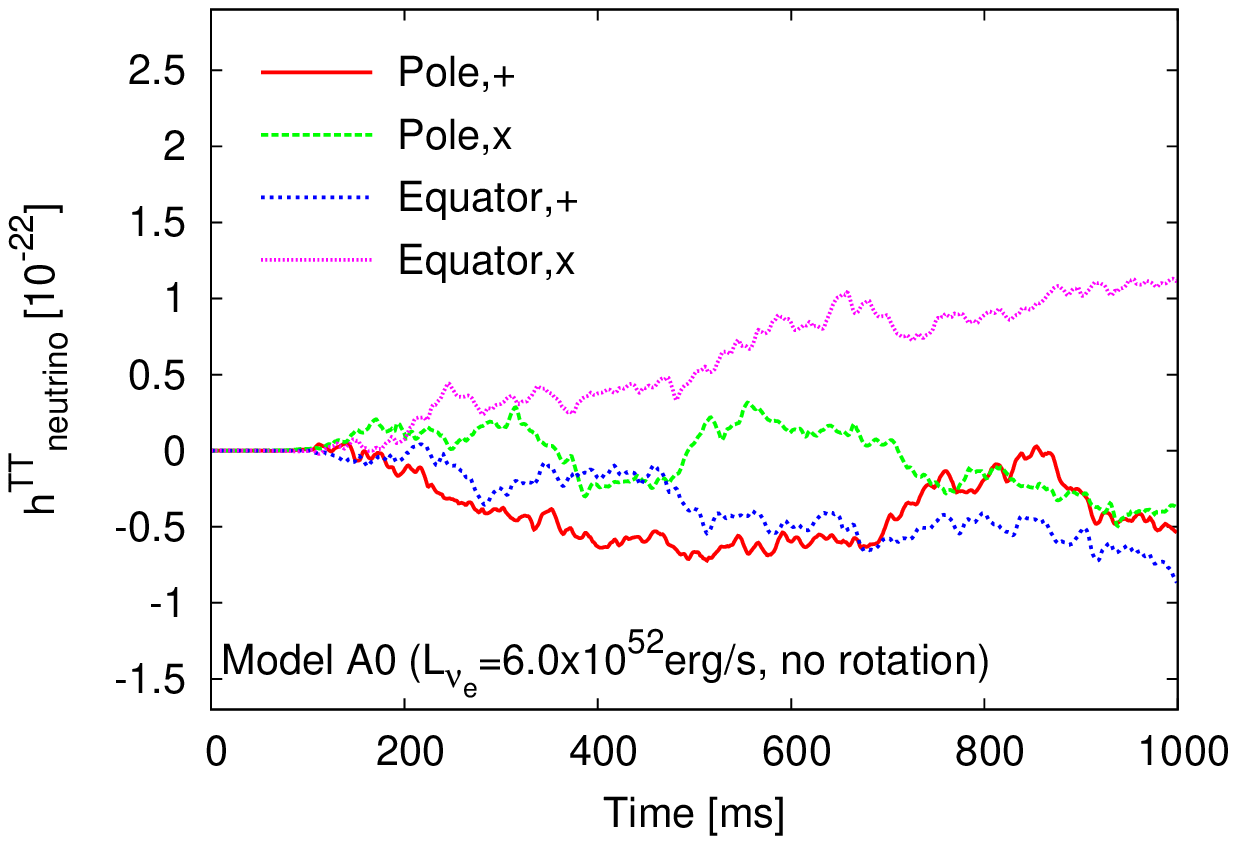}} \\
      \resizebox{80mm}{!}{\includegraphics{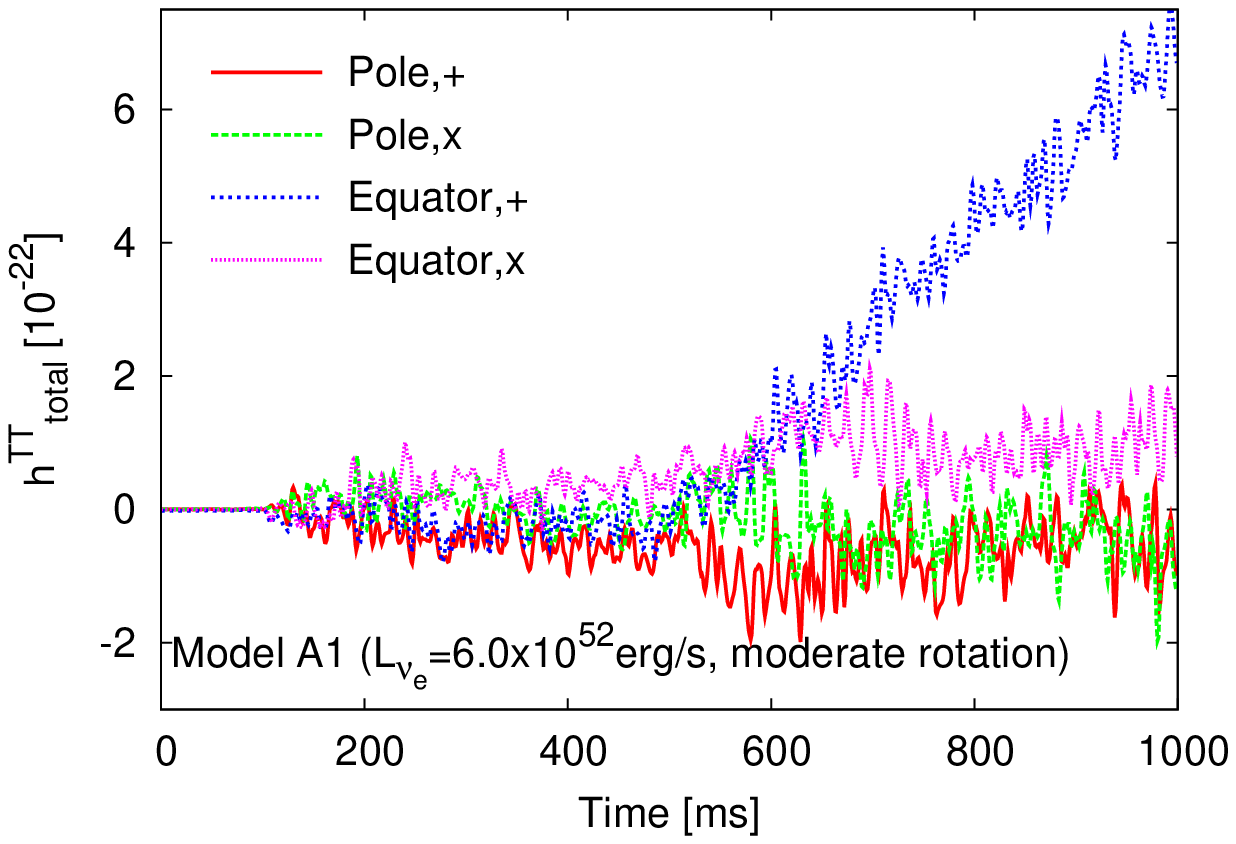}} &
      \resizebox{80mm}{!}{\includegraphics{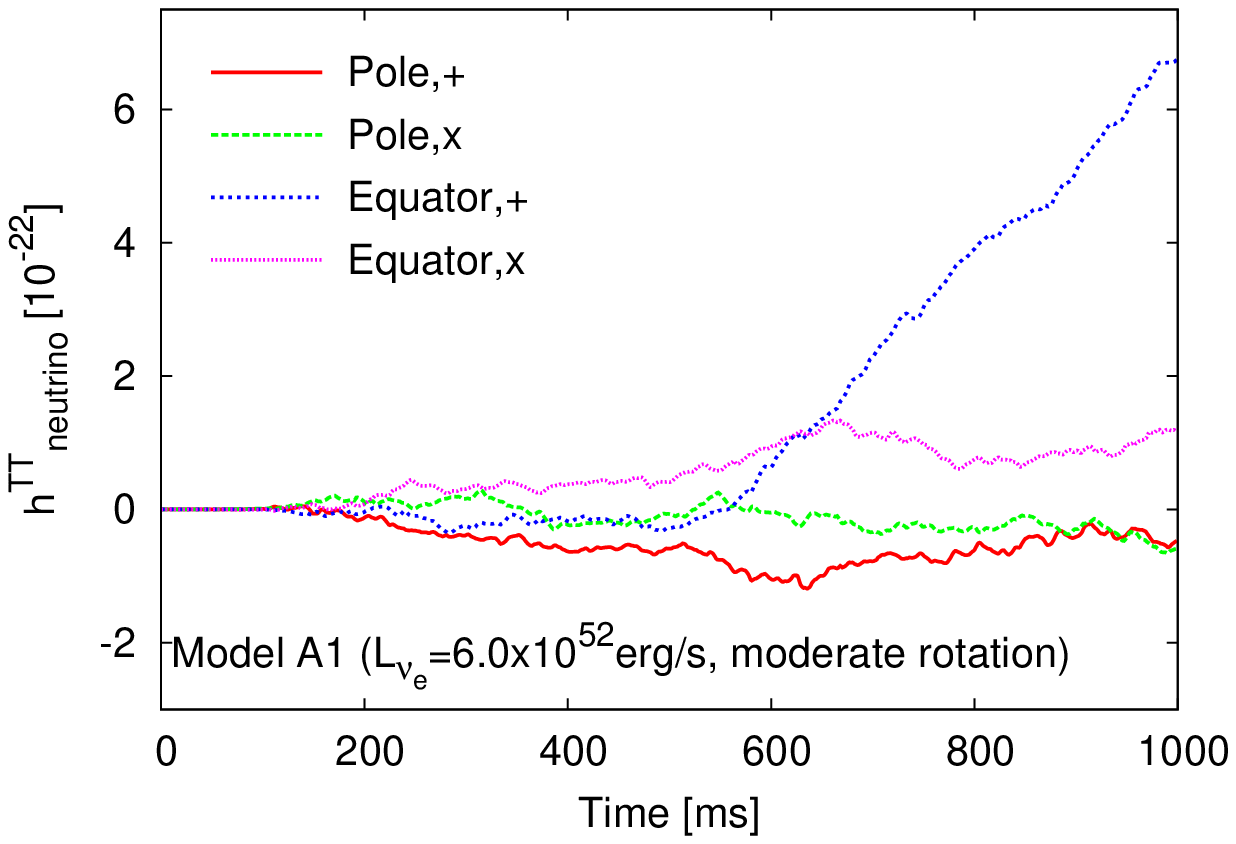}} \\
\resizebox{80mm}{!}{\includegraphics{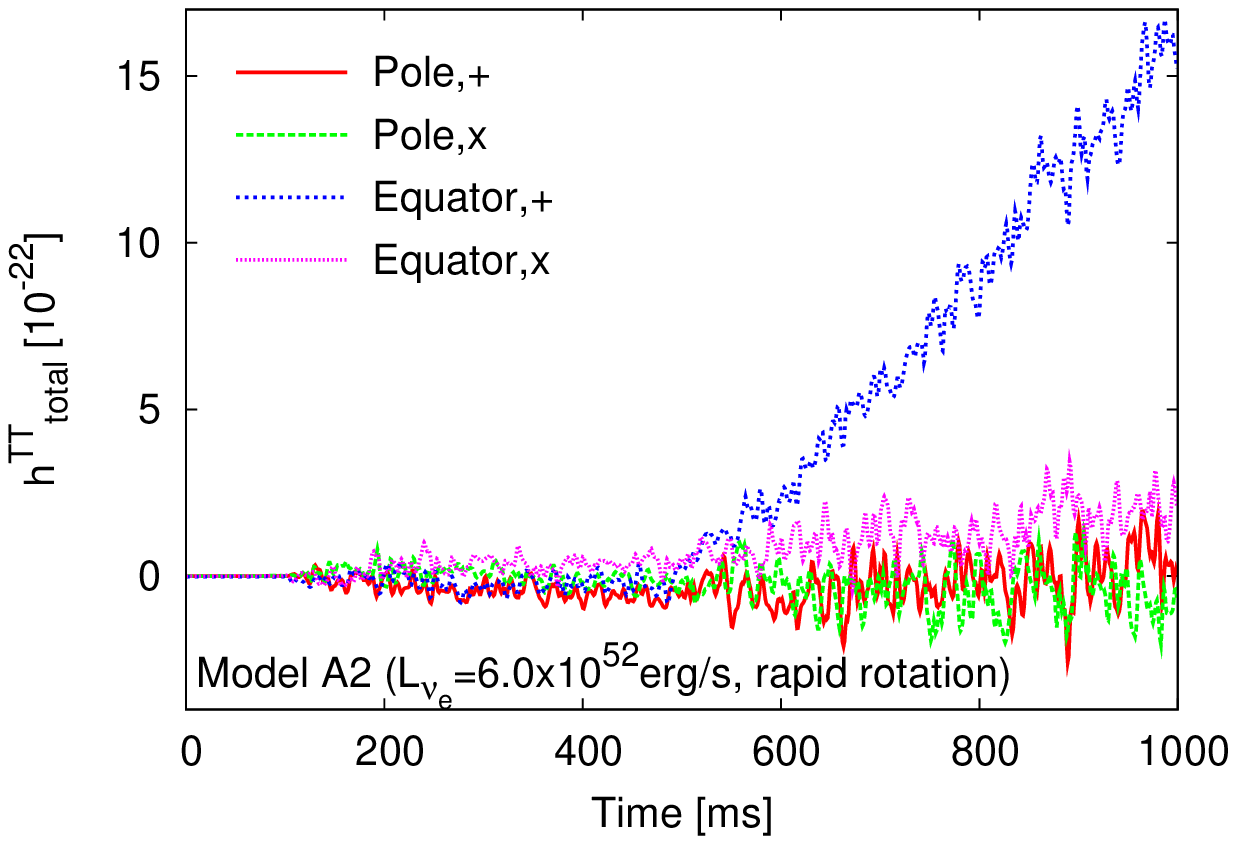}} &
\resizebox{80mm}{!}{\includegraphics{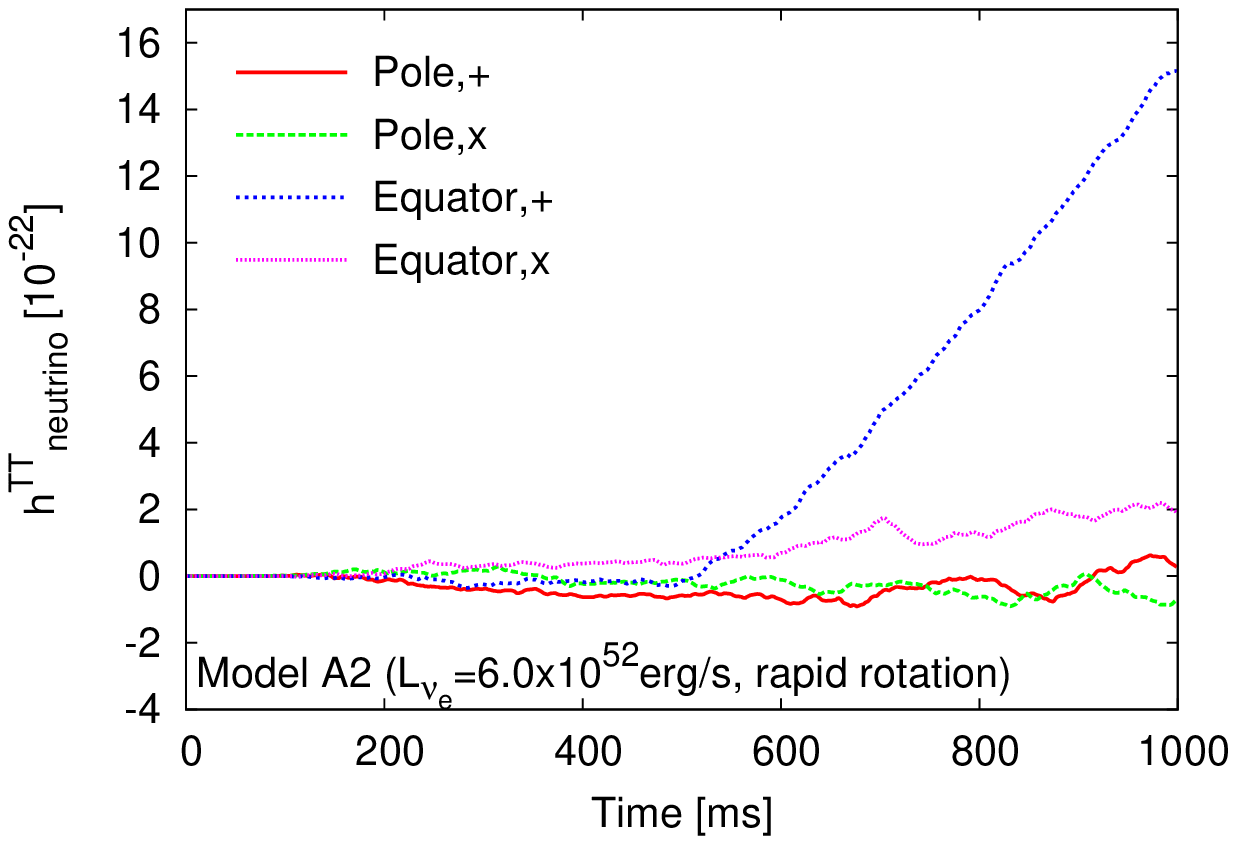}} \\
    \end{tabular}
    \caption{Gravitational waveforms from the sum of neutrinos and matter motions 
(left) and only from neutrinos (right) for models A0 (top),
 A1 (middle), and A2 (bottom). The time is measured from 
the epoch when the neutrino luminosity is injected from the surface of the 
neutrino sphere. In all the computed models, SASI gradually transits from 
 the linear to non-linear regime 
at about $100$ ms, simultaneously making the amplitudes deviate from zero. For models
 of series A shown here, the rotational flow is adjusted to advect to the PNS surface 
at around 
$t =400$ ms (see text for more detail). The supernova is assumed to be located at the distance
 of 10 kpc.} 
 \label{fig3}
  \end{center}
\end{figure}

\begin{figure}
  \begin{center}
    \begin{tabular}{cc}
      \resizebox{80mm}{!}{\includegraphics{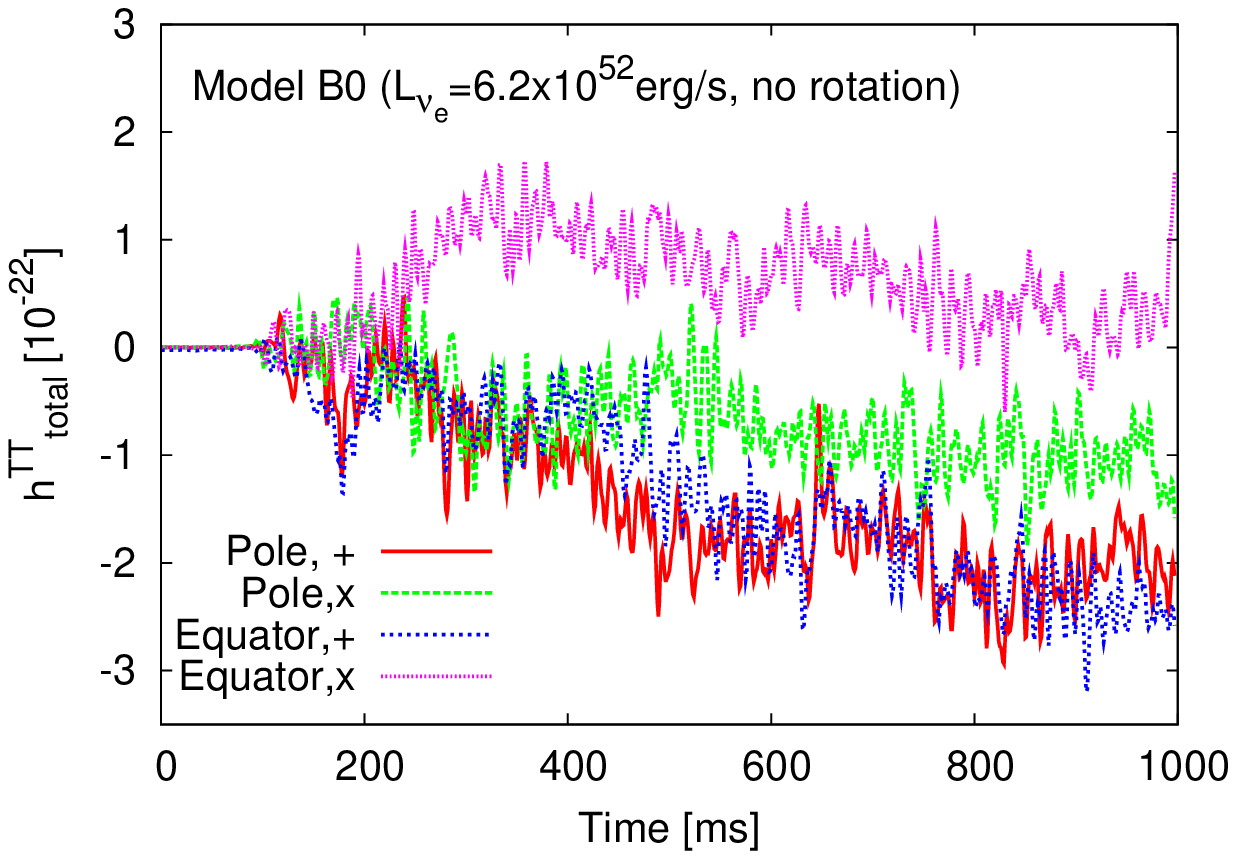}} &
      \resizebox{80mm}{!}{\includegraphics{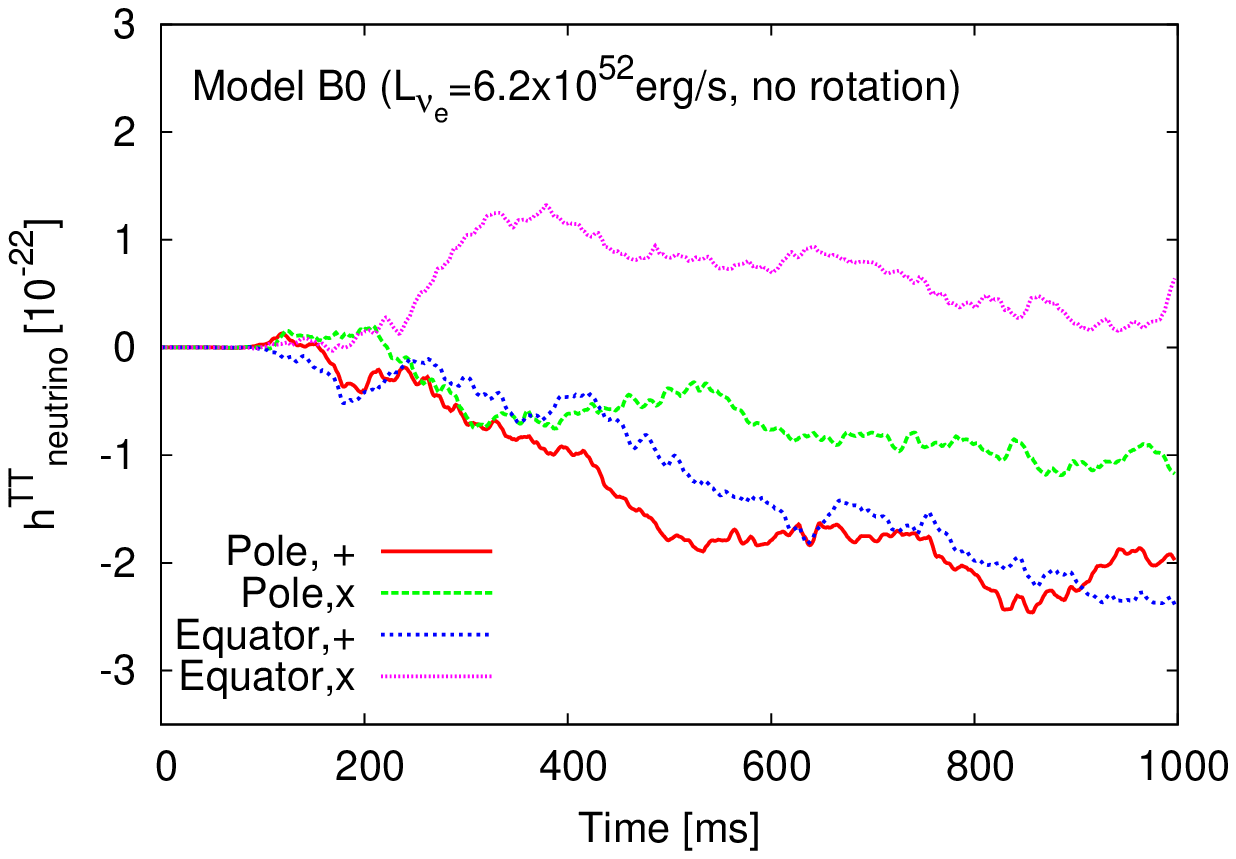}} \\
      \resizebox{80mm}{!}{\includegraphics{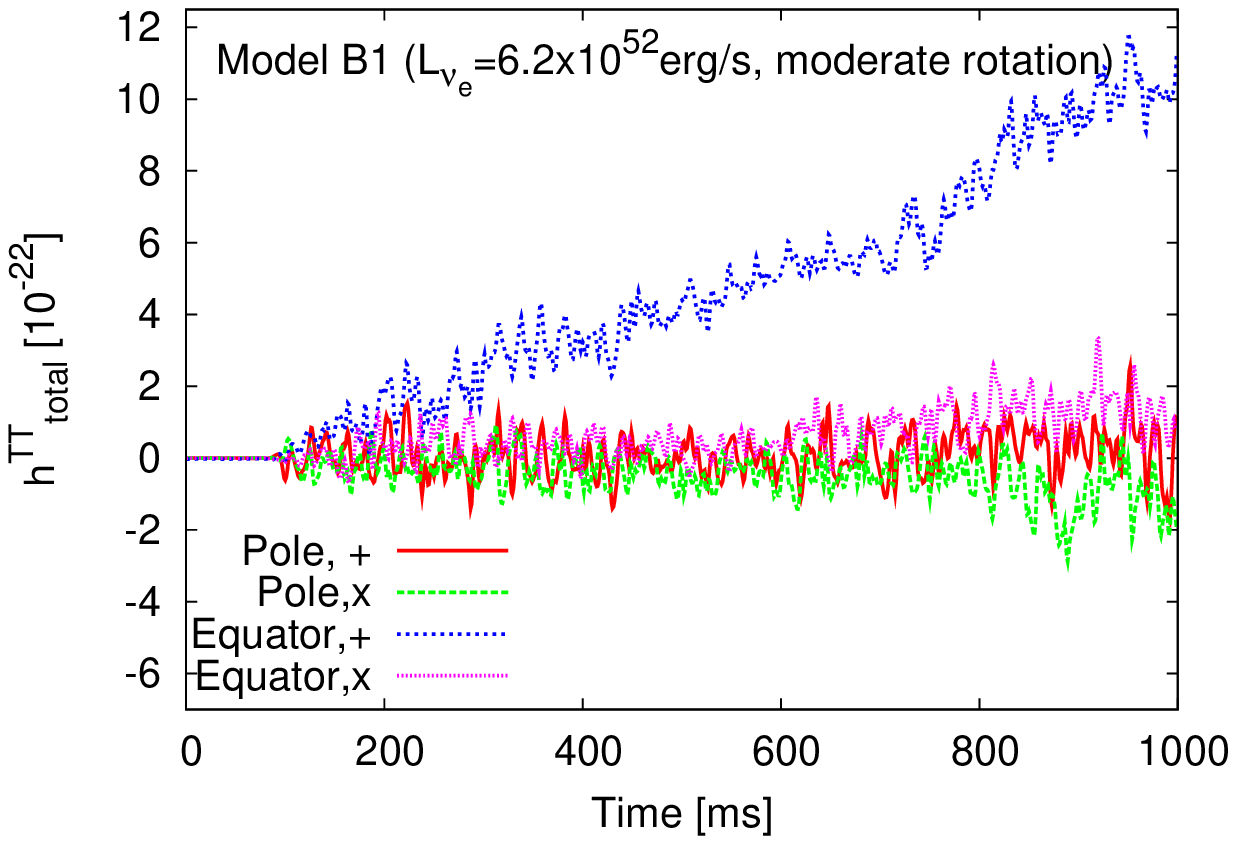}} &
      \resizebox{80mm}{!}{\includegraphics{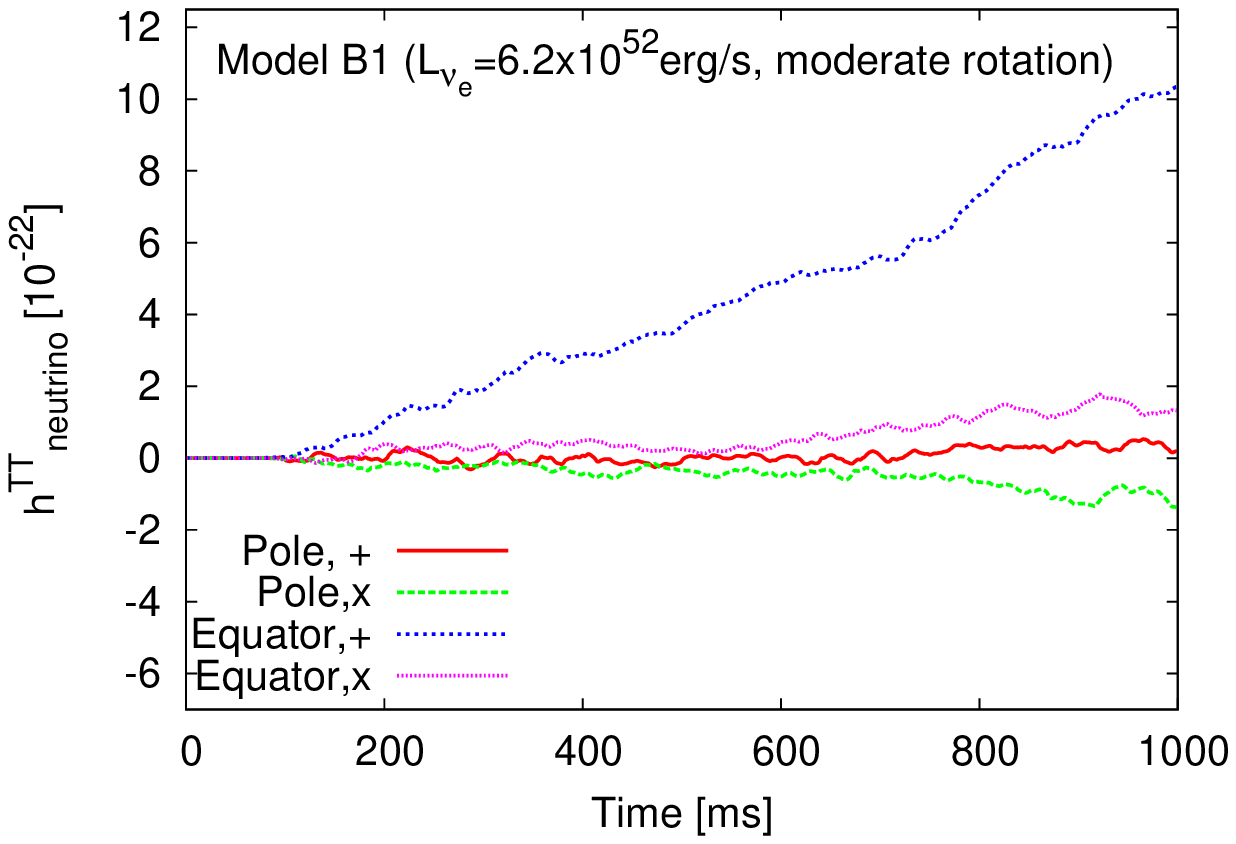}} \\
\resizebox{80mm}{!}{\includegraphics{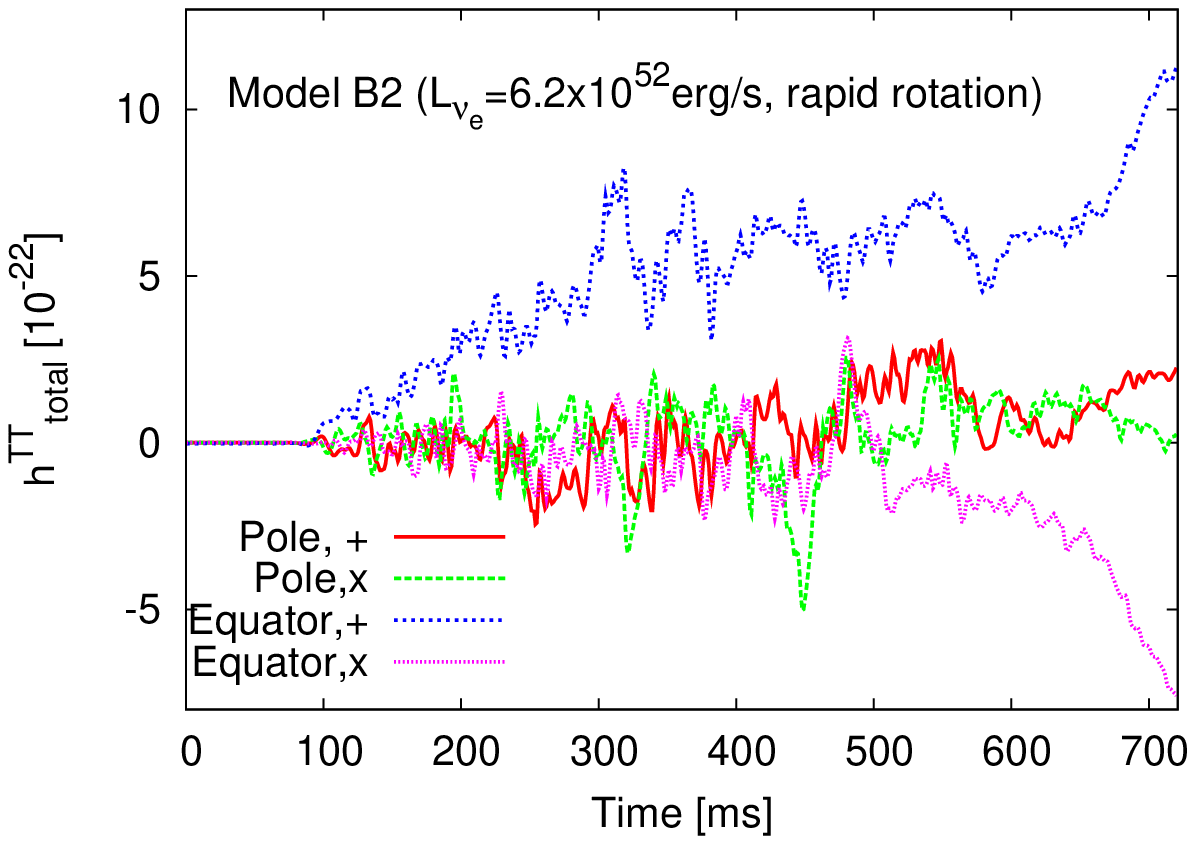}} &
\resizebox{80mm}{!}{\includegraphics{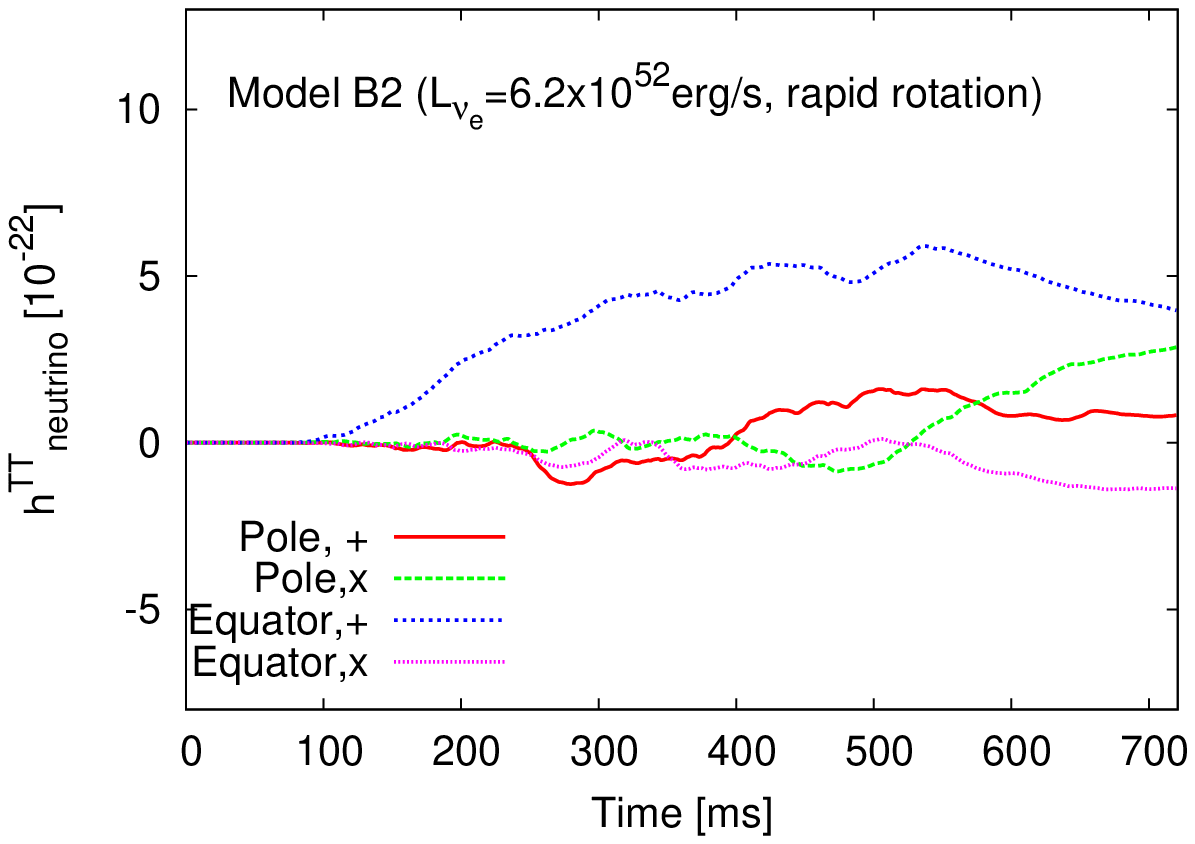}} \\
    \end{tabular}
    \caption{Same as Figure \ref{fig3} but for models of series B
. Note that except for models of series A, the rotational flows are adjusted to advect to the 
PNS surface at around $t = 100$ ms.}
    \label{fig4}
  \end{center}
\end{figure}

\begin{figure}
  \begin{center}
    \begin{tabular}{cc}
      \resizebox{80mm}{!}{\includegraphics{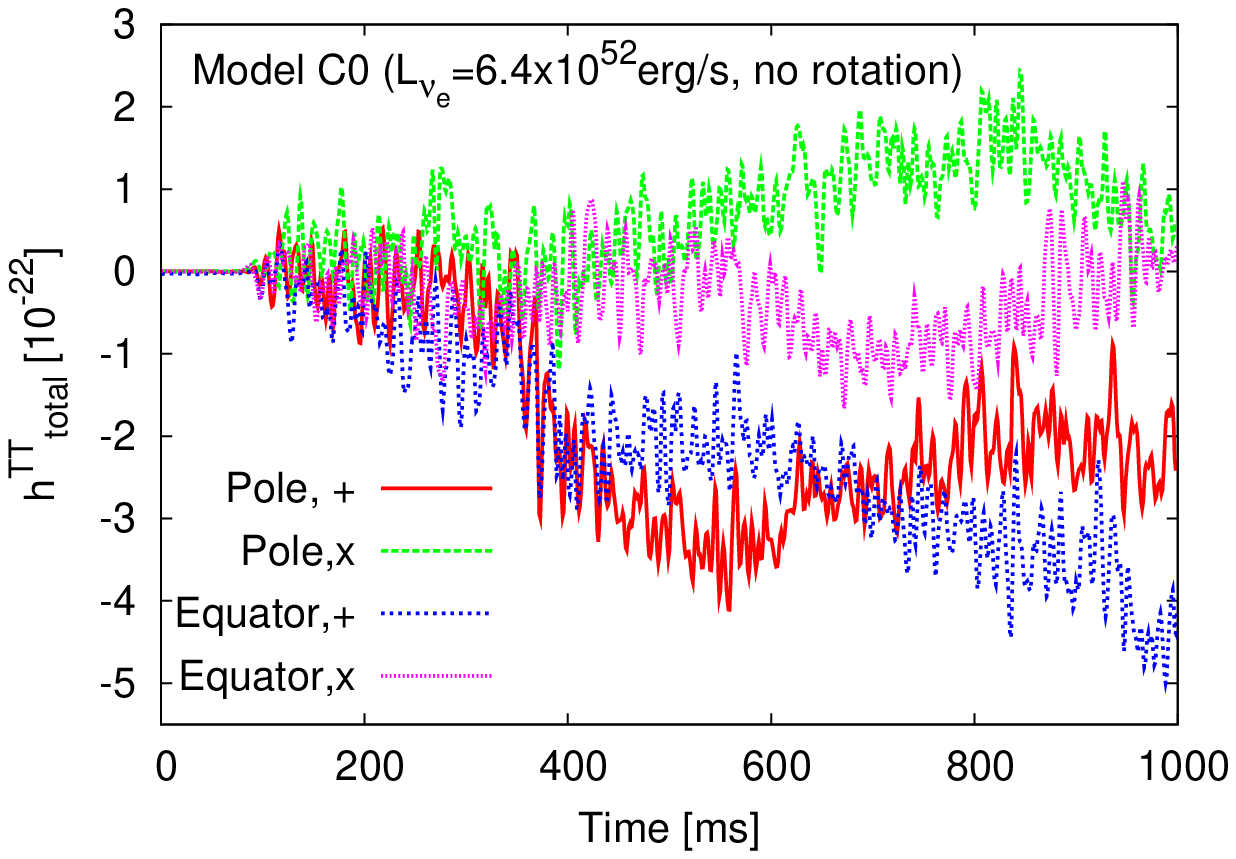}} &
      \resizebox{80mm}{!}{\includegraphics{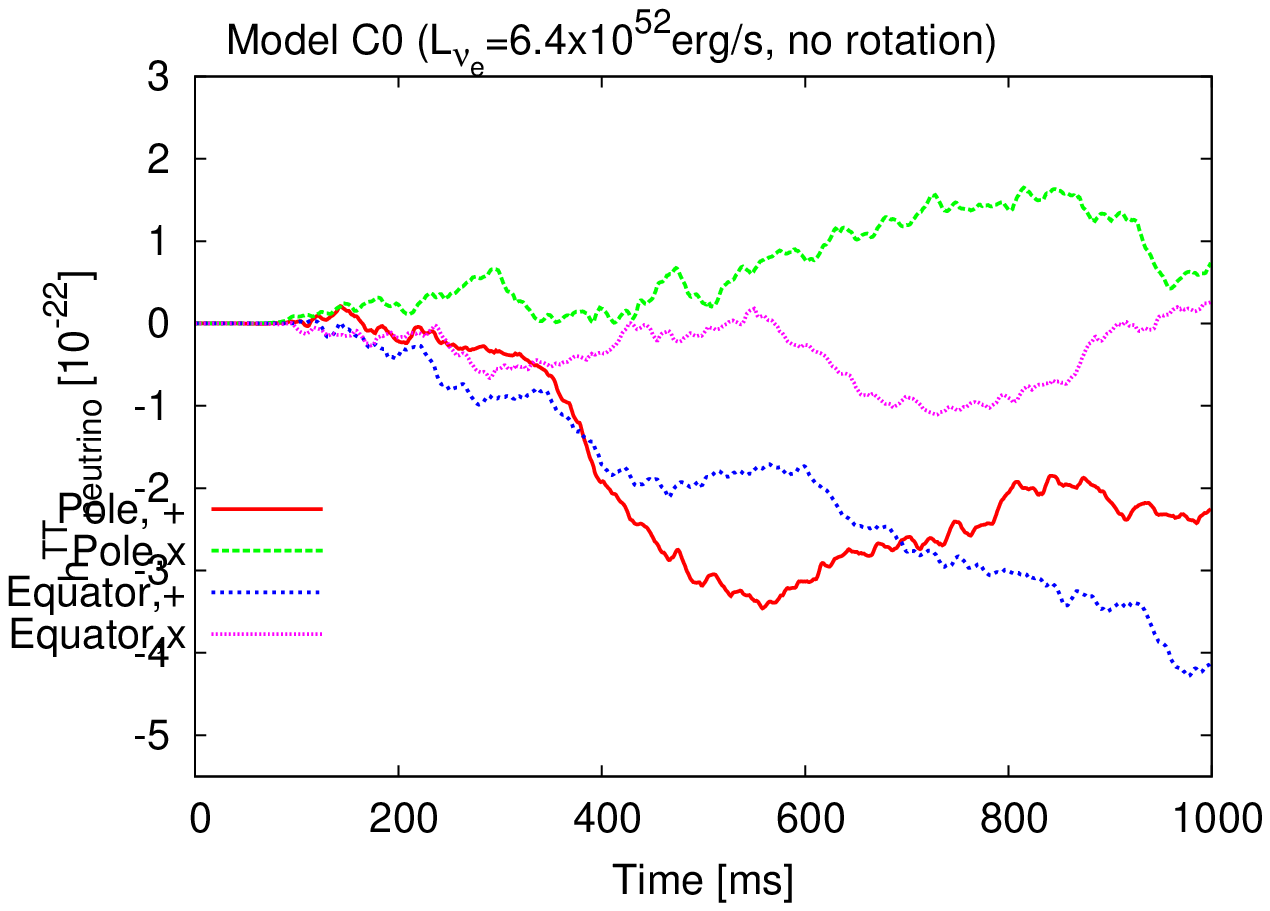}} \\
      \resizebox{80mm}{!}{\includegraphics{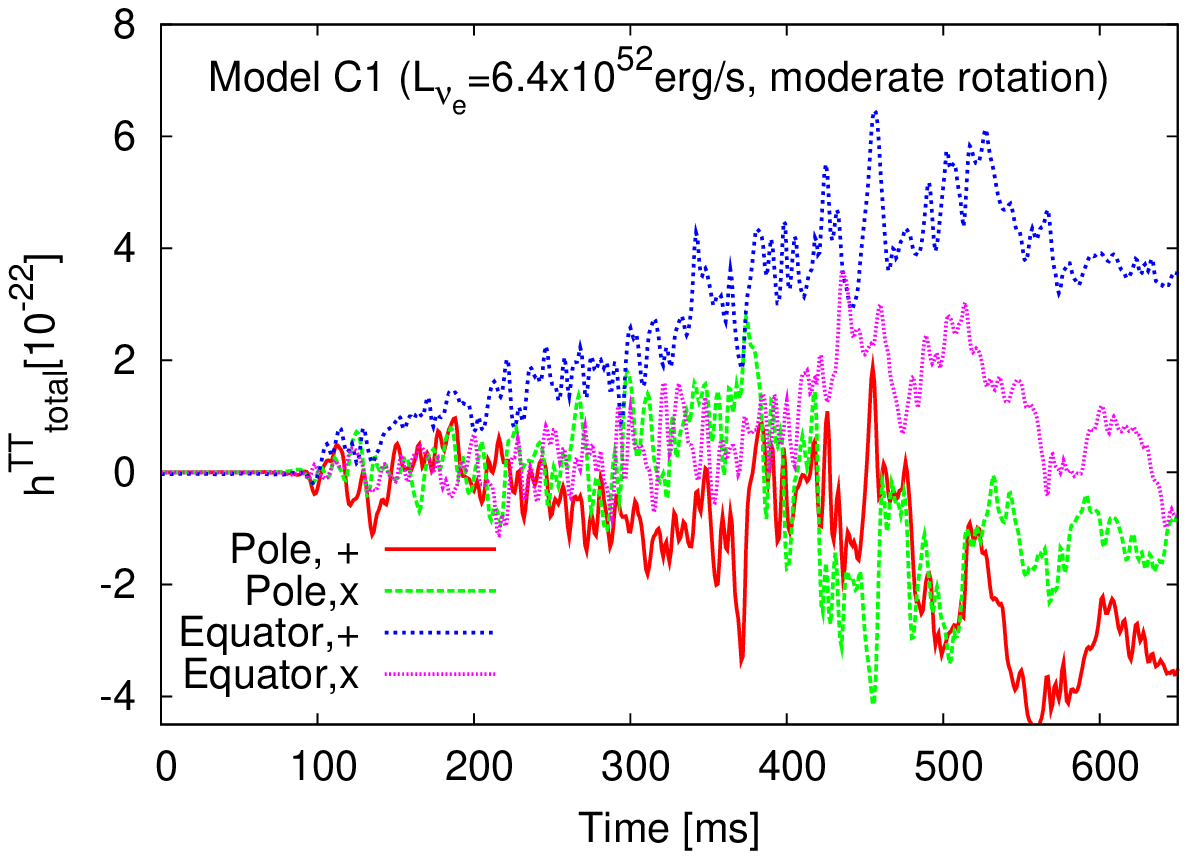}} &
      \resizebox{80mm}{!}{\includegraphics{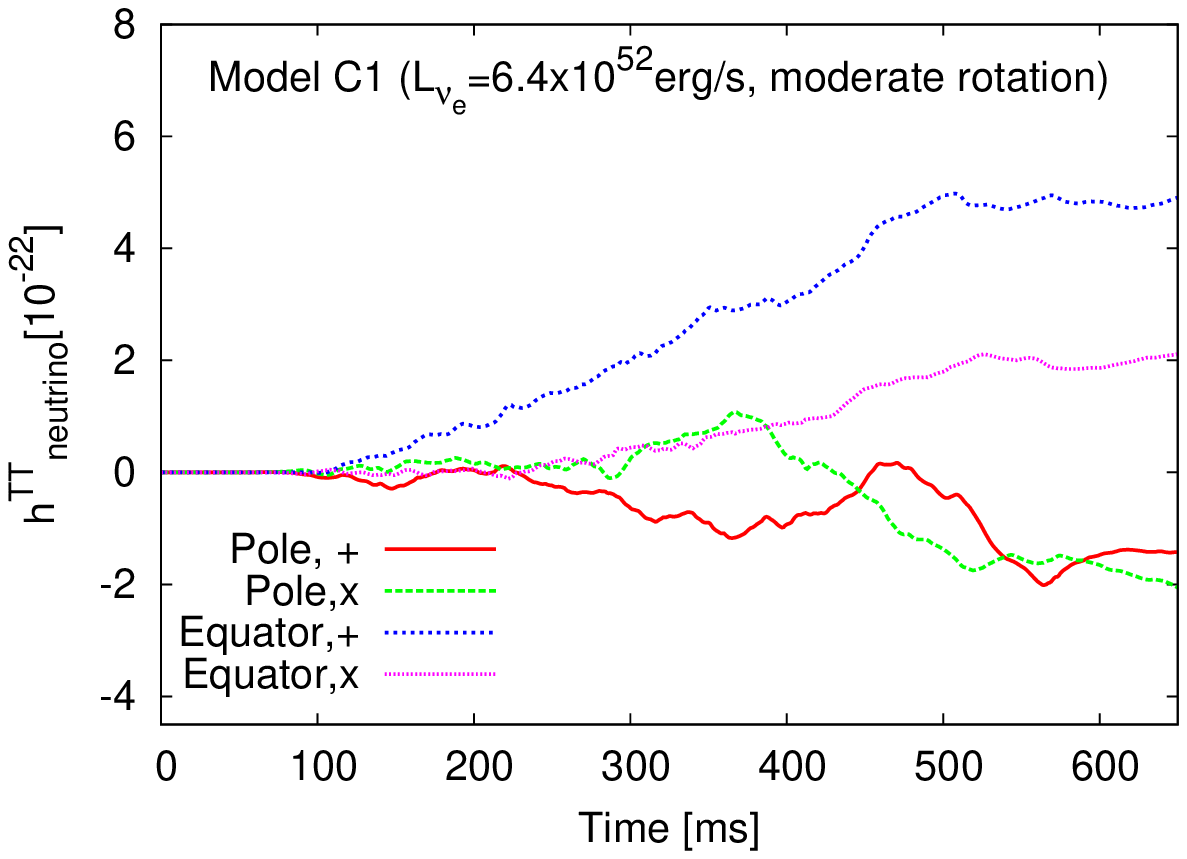}} \\
\resizebox{80mm}{!}{\includegraphics{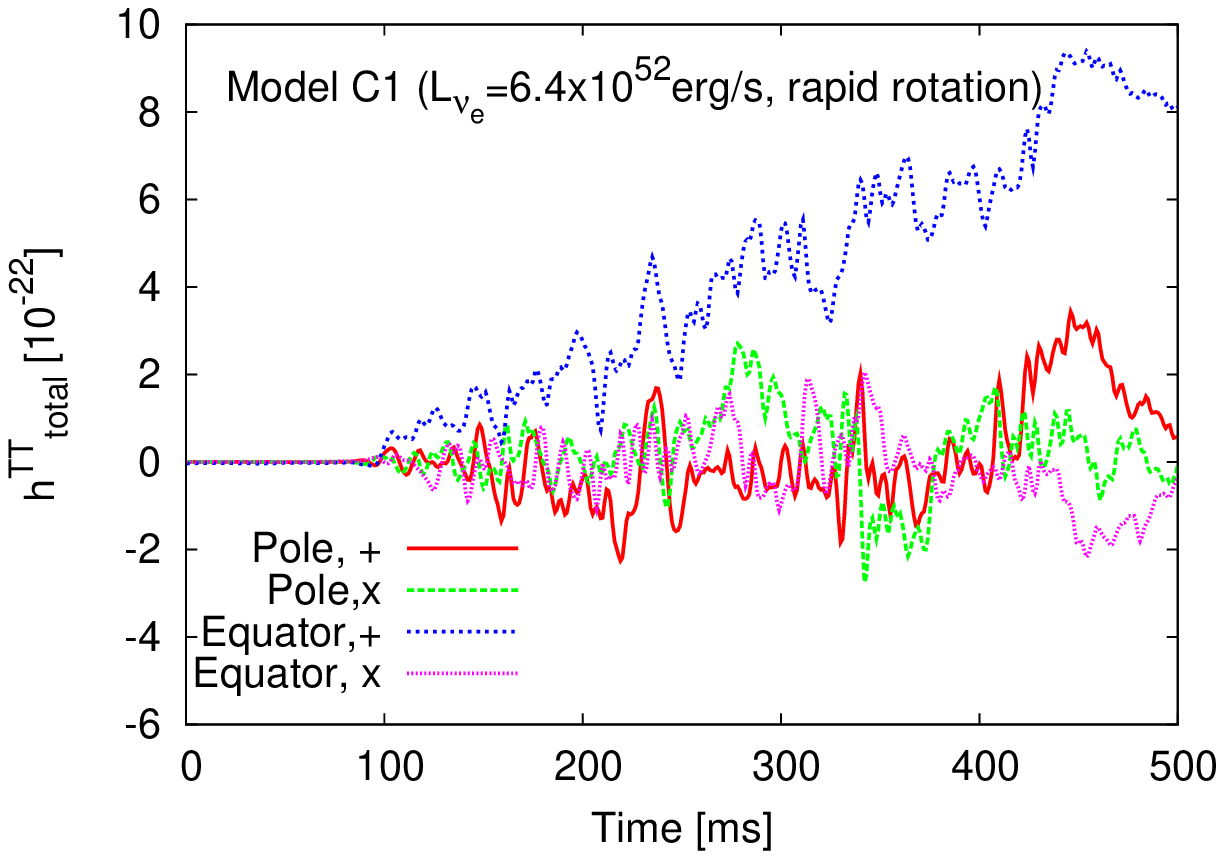}} &
\resizebox{80mm}{!}{\includegraphics{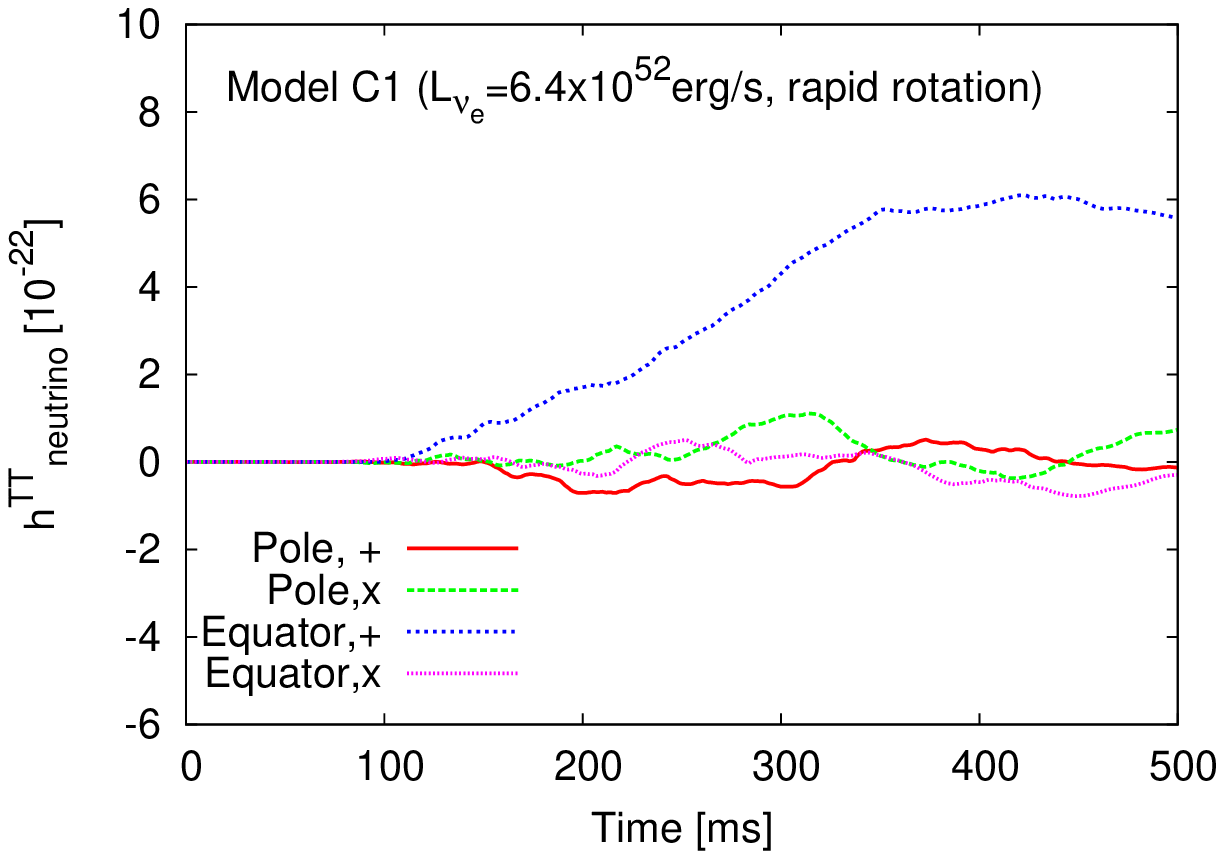}} \\
    \end{tabular}
    \caption{Same as Figure \ref{fig4} but for models of series C.}
    \label{fig5}
  \end{center}
\end{figure}

\begin{figure}
  \begin{center}
    \begin{tabular}{cc}
      \resizebox{80mm}{!}{\includegraphics{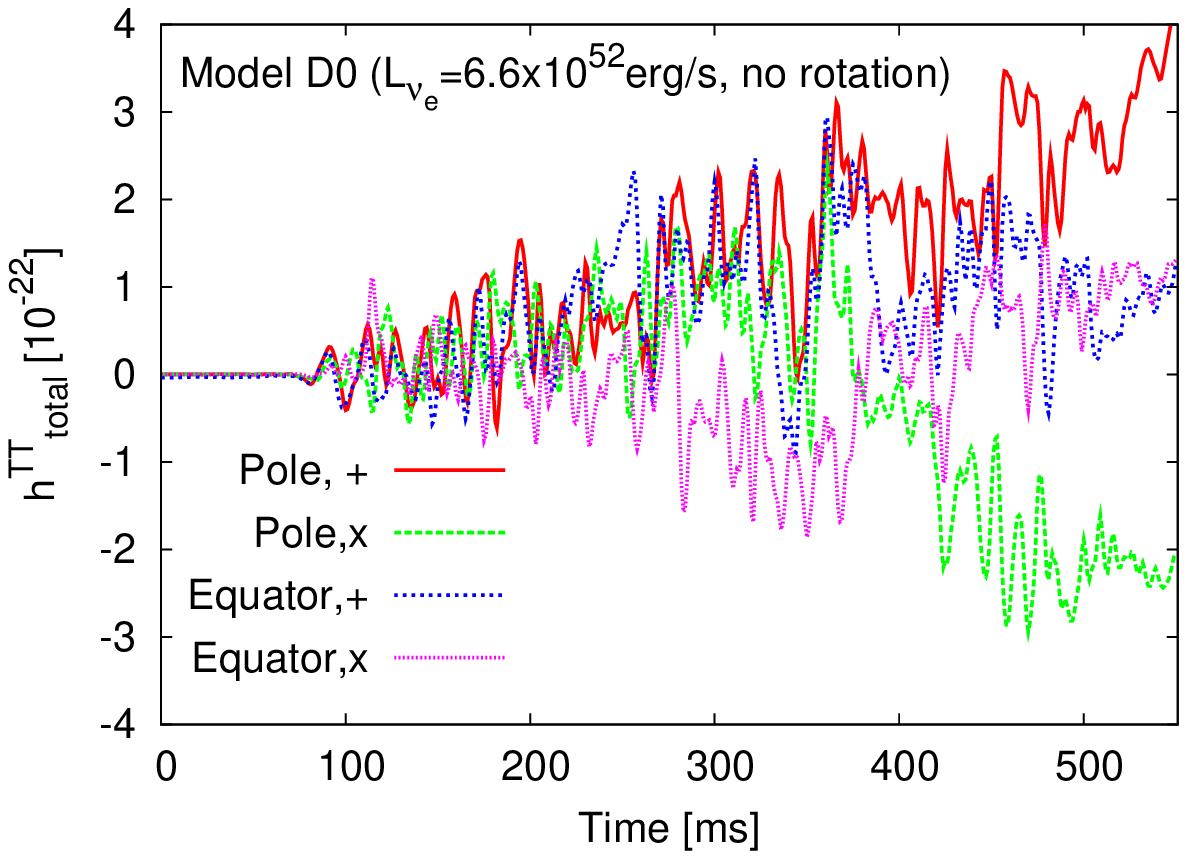}} &
      \resizebox{80mm}{!}{\includegraphics{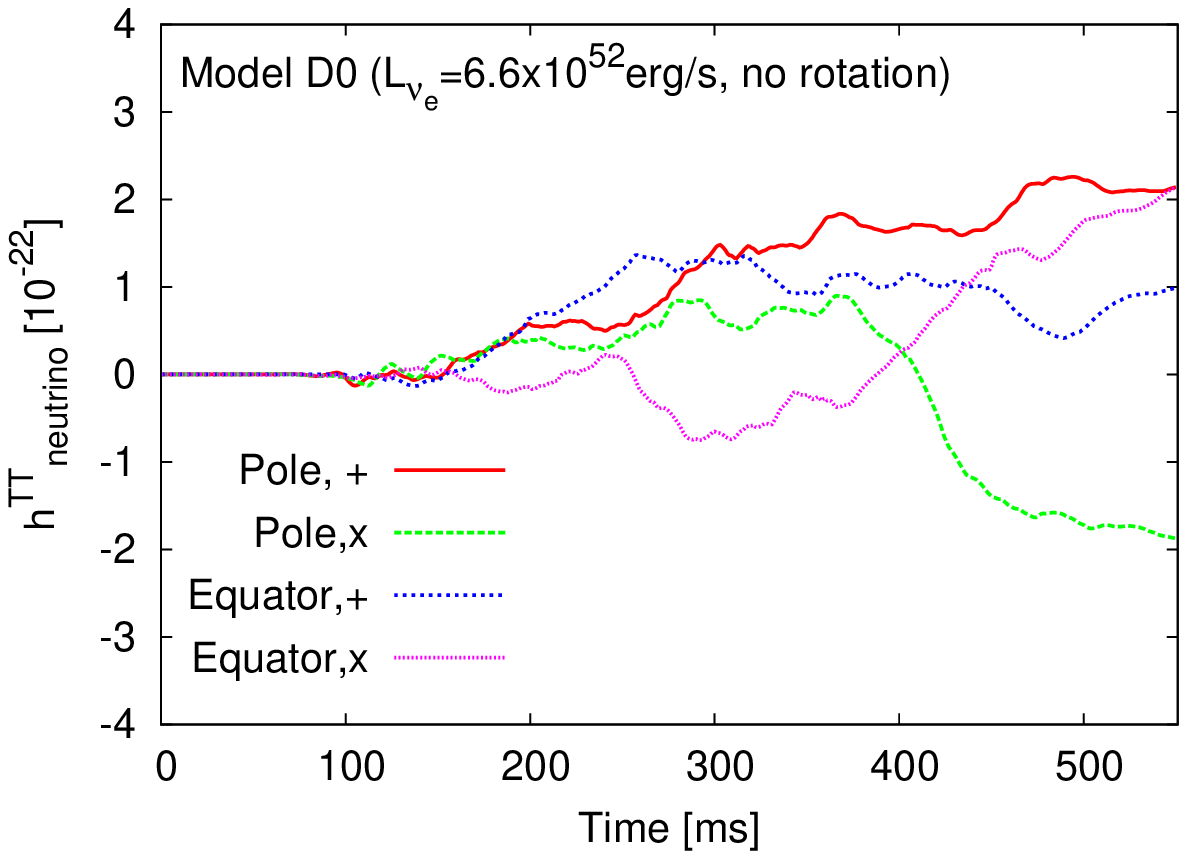}} \\
      \resizebox{80mm}{!}{\includegraphics{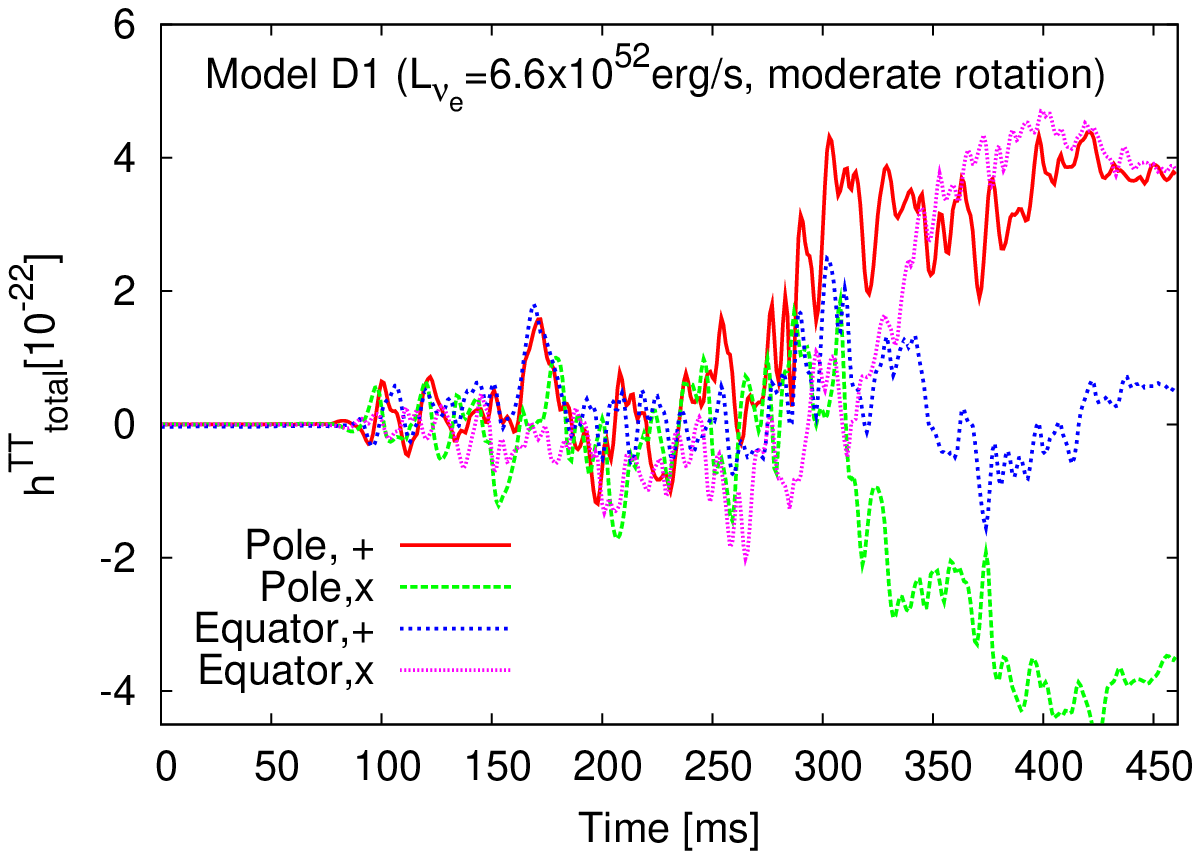}} &
      \resizebox{80mm}{!}{\includegraphics{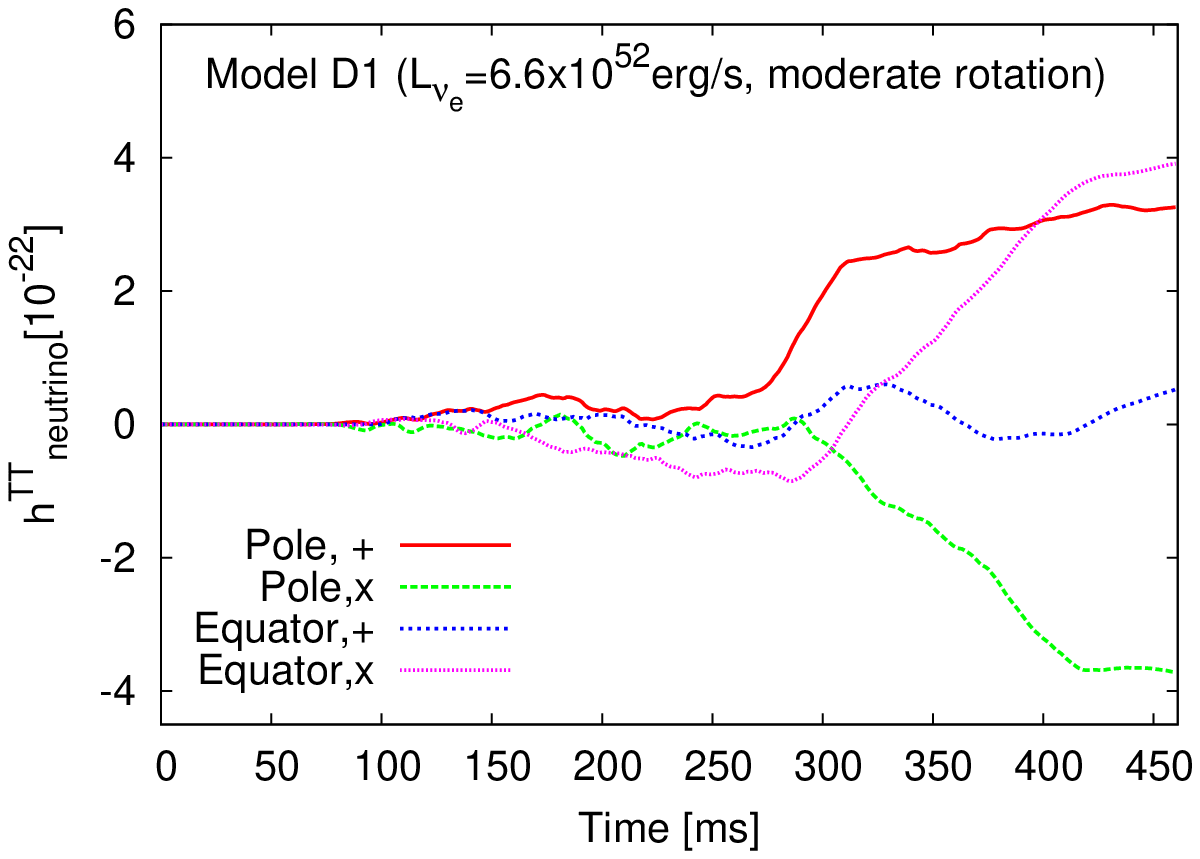}} \\
\resizebox{80mm}{!}{\includegraphics{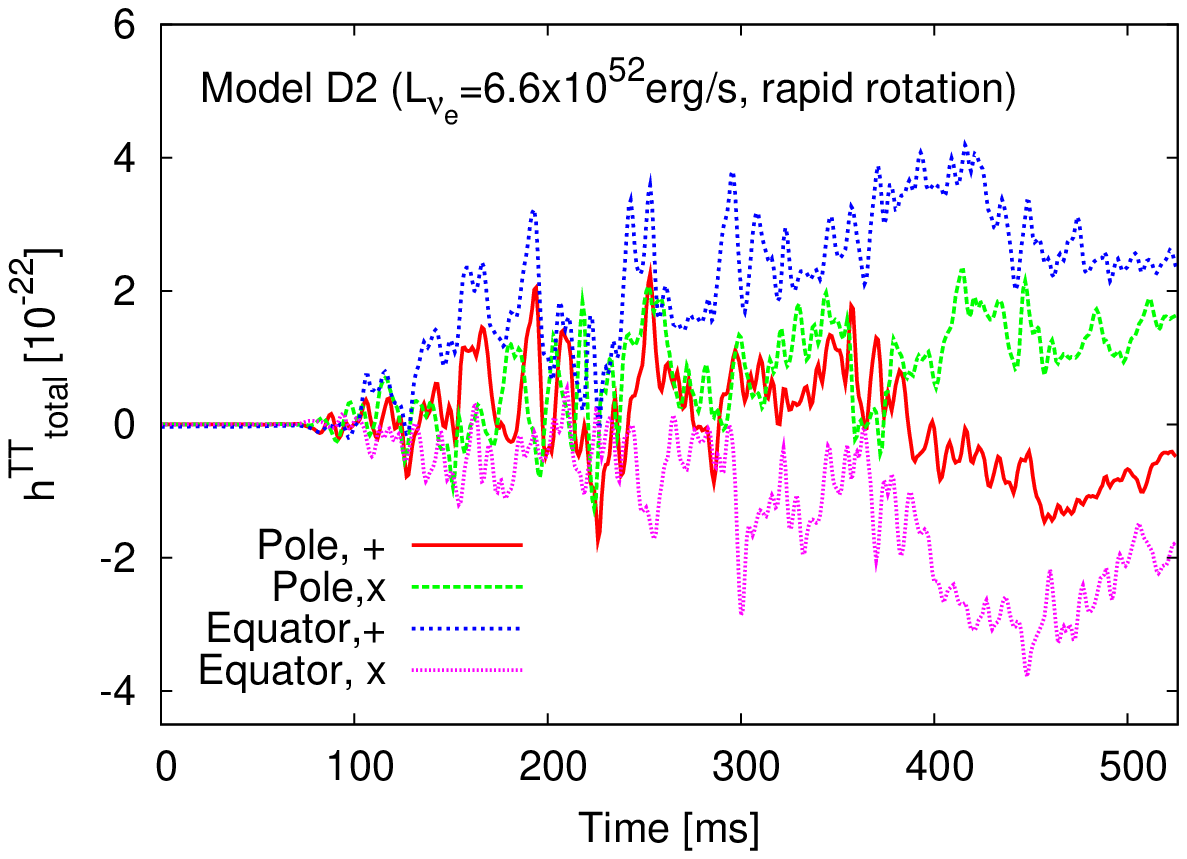}} &
\resizebox{80mm}{!}{\includegraphics{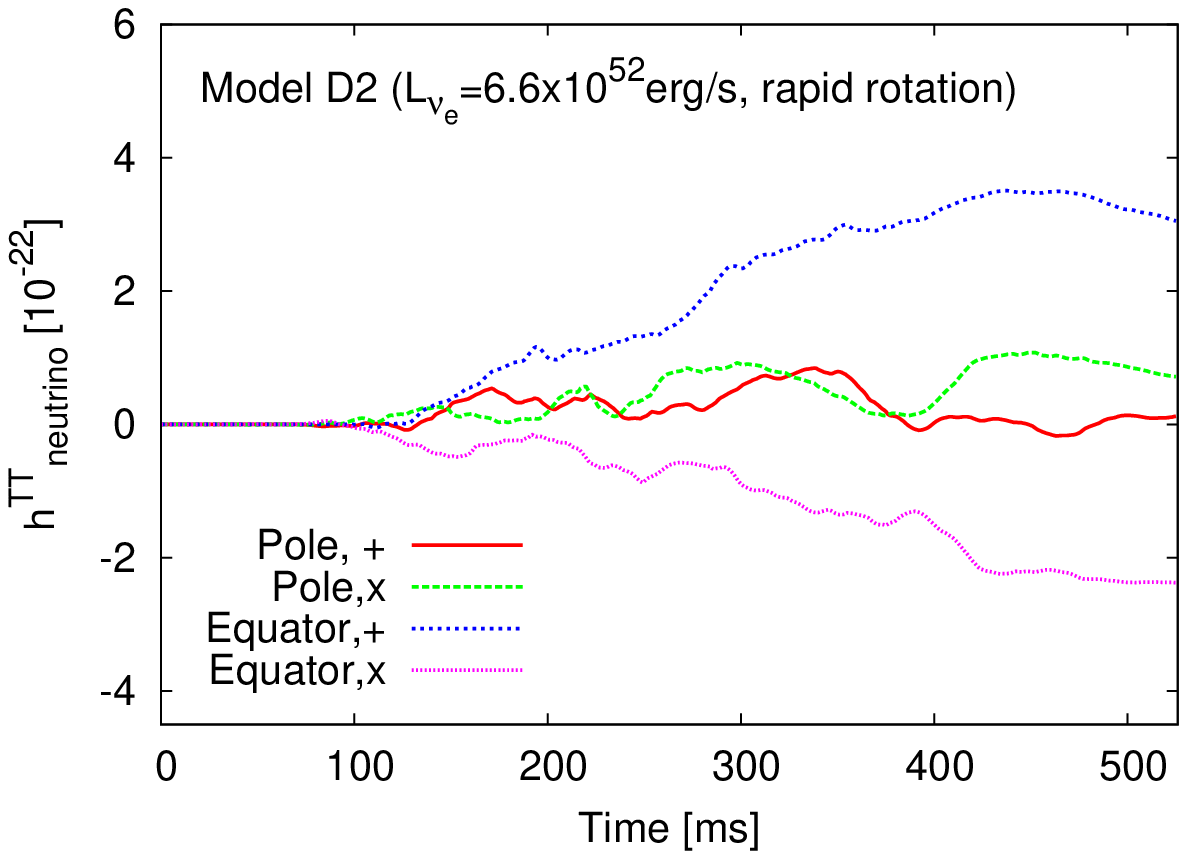}} \\
    \end{tabular}
    \caption{Same as Figure \ref{fig4} but for models of series D.}
    \label{fig6}
  \end{center}
\end{figure}

\begin{figure}
  \begin{center}
    \begin{tabular}{cc}
      \resizebox{80mm}{!}{\includegraphics{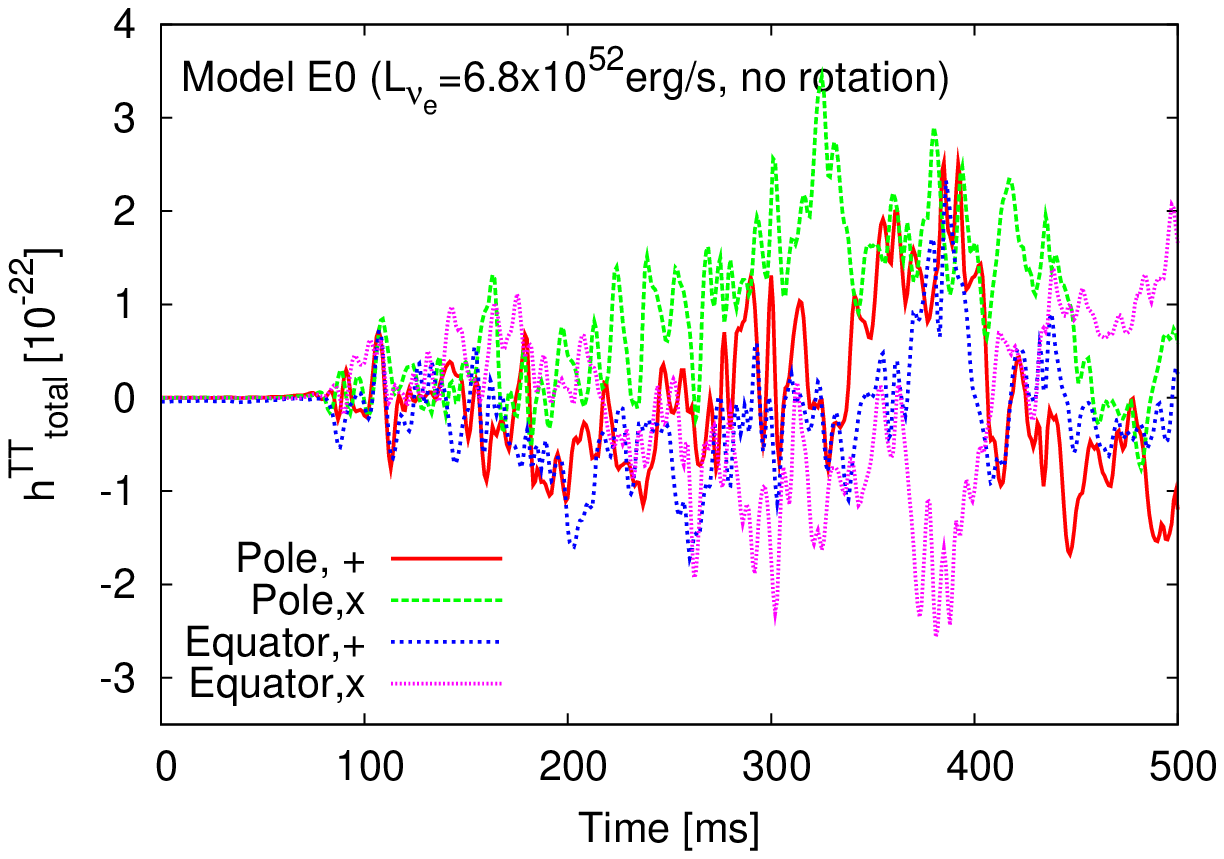}} &
      \resizebox{80mm}{!}{\includegraphics{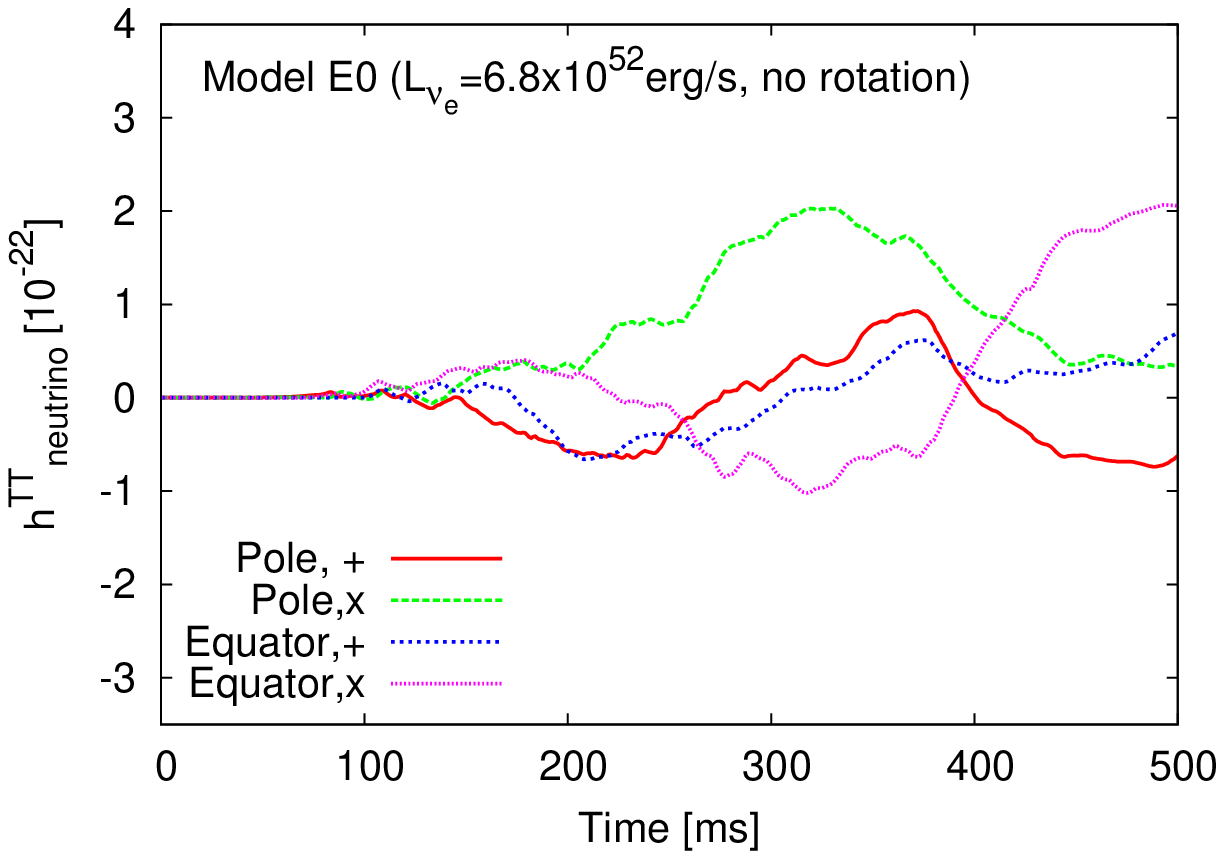}} \\
      \resizebox{80mm}{!}{\includegraphics{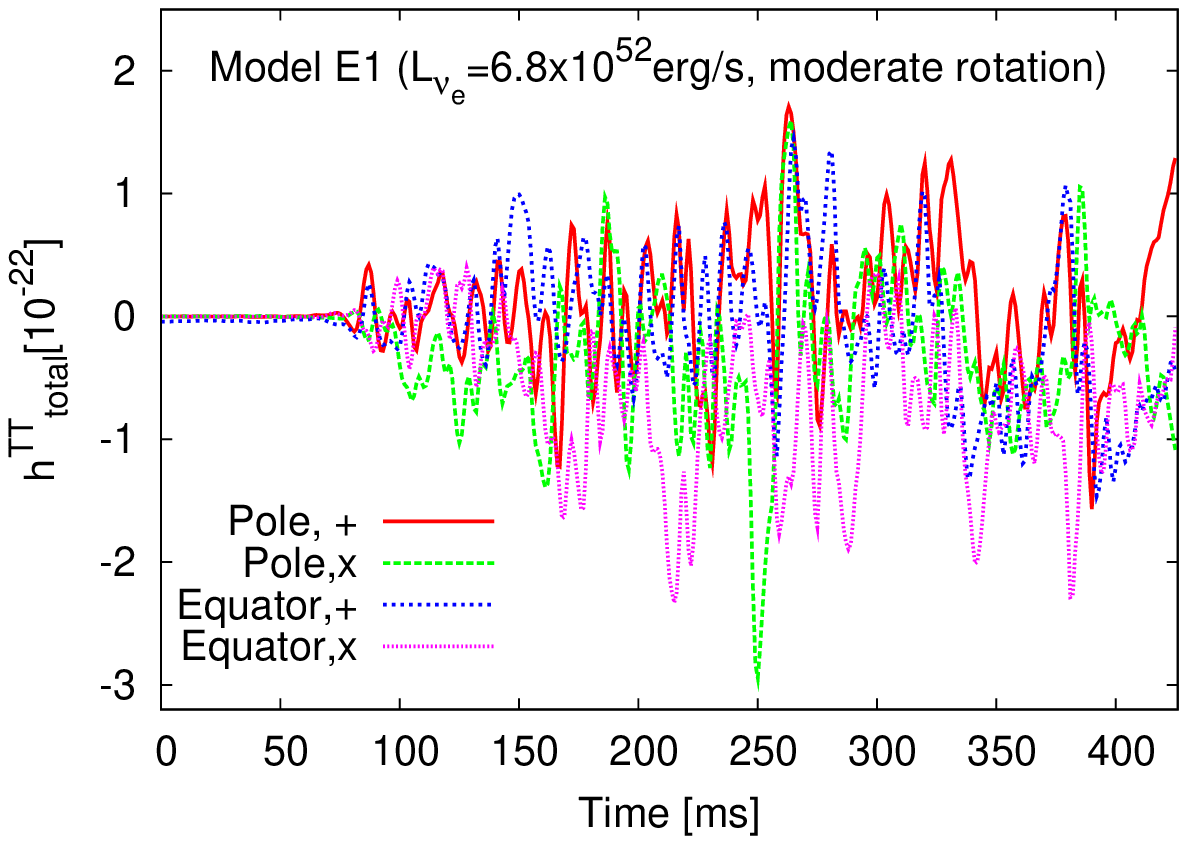}} &
      \resizebox{80mm}{!}{\includegraphics{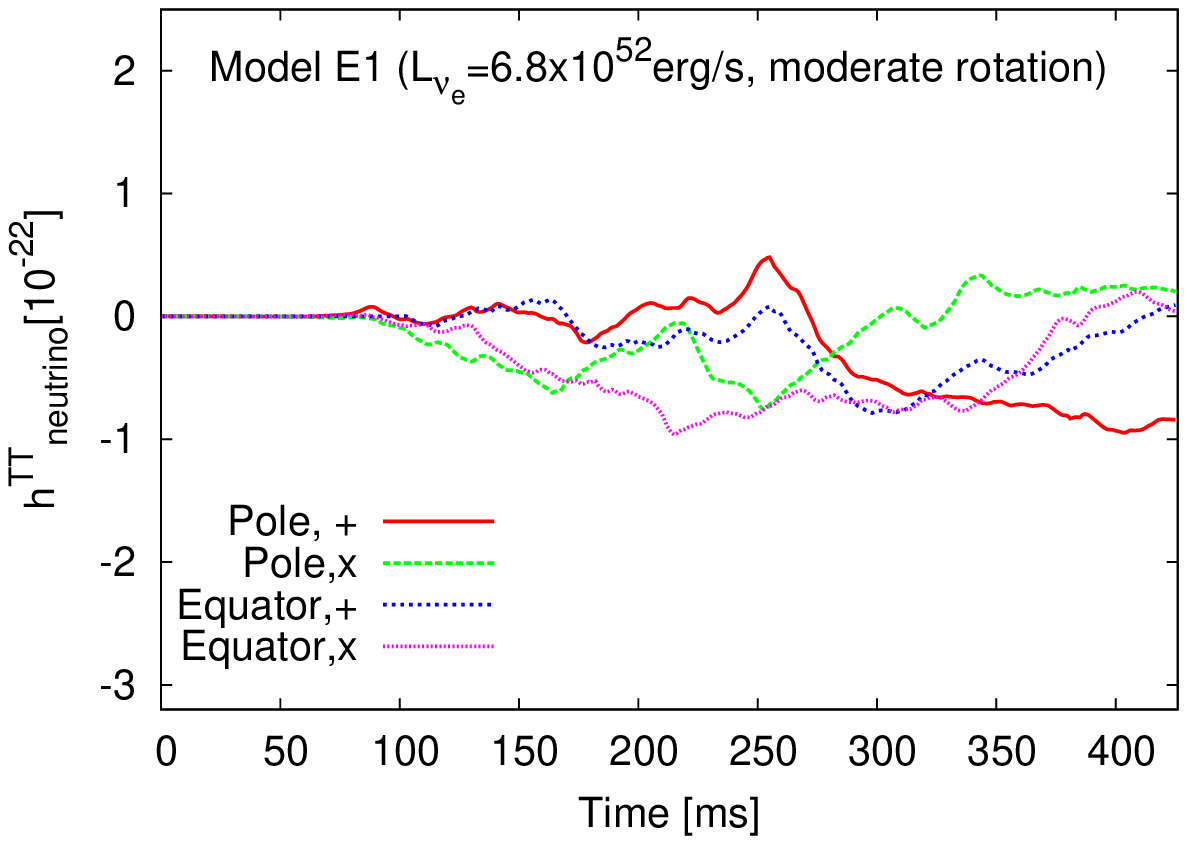}} \\
\resizebox{80mm}{!}{\includegraphics{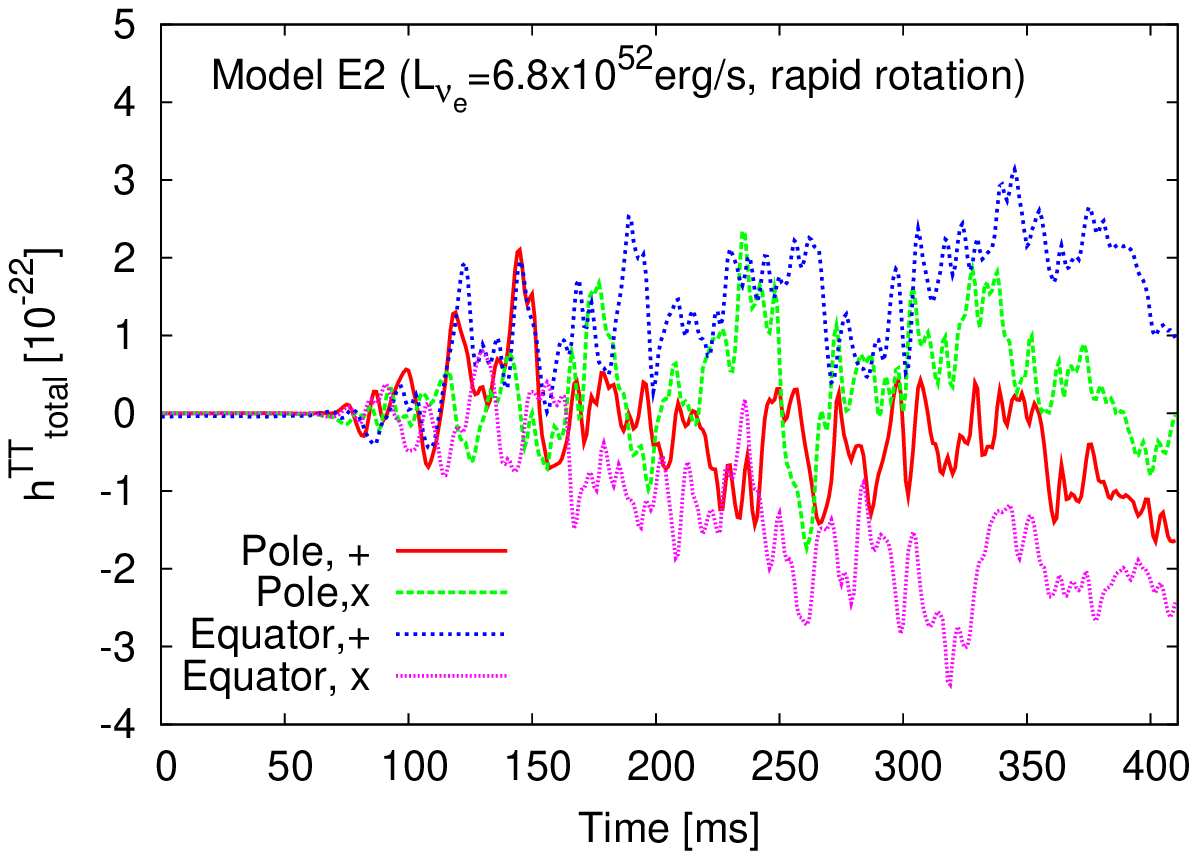}} &
\resizebox{80mm}{!}{\includegraphics{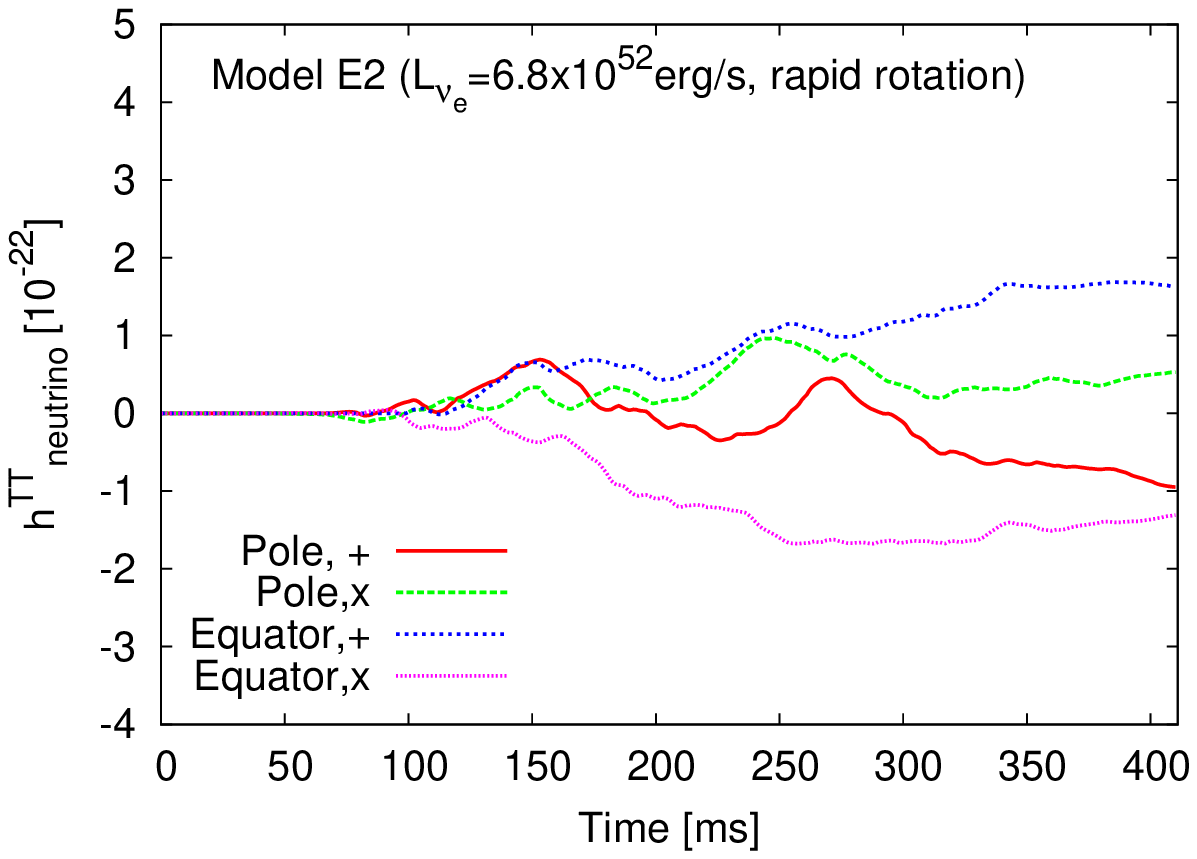}} \\
    \end{tabular}
    \caption{Same as Figure \ref{fig4} but for models of series E.}
    \label{fig7}
  \end{center}
\end{figure}

This feature can be more clearly seen in the waveform catalogues presented from 
Figure \ref{fig3} to \ref{fig6}. 
From Figure 3, it can be seen that $h^{\rm equ}_{\nu,+}$ (blue line)
becomes prominent among others after around $t \sim 500-600$ ms for models
 with rotation (middle and bottom panels).
 This trend seems quite general as seen in the other luminosity models (see middle and 
 bottom panels in Figure \ref{fig4} to \ref{fig6}). Note again that the time 
 when the rotational flow is adjusted to touch the PNS surface is 
 $t=400$ ms for models of series A (and $t=100$  ms for the other), 
which accounts for the 
time difference in the rise of $h^{\rm equ}_{\nu,+}$ (compare Figure 3 to Figure 4 to 7). In the absence of rotation,
 the waveforms do not exhibit such 
a feature and vary much more stochastically (e.g., top panels in Figure 
\ref{fig3} to \ref{fig7}). Comparing the bottom panels
 in Figures \ref{fig3} to \ref{fig7}, it can be seen that the increasing trend becomes 
more fainter as the input luminosity becomes larger. This is mainly because 
vigorous matter motions irradiated by intense neutrino heating smear out the 
 growth of the spiral SASI modes as well as the structures of the spiral flows. To 
extract purely the rotational effects on the GWs, it is most 
convenient to analyze models of series A
 that fail to produce explosions. In the following, we take models of series A 
as a reference and look 
more in detail the reason why the stochastic nature of the GWs is broken (or weakened) 
 due to the inclusion of rotation.
 
\clearpage

\begin{figure}[hbt]
\epsscale{.8}
\begin{center}
\plottwo{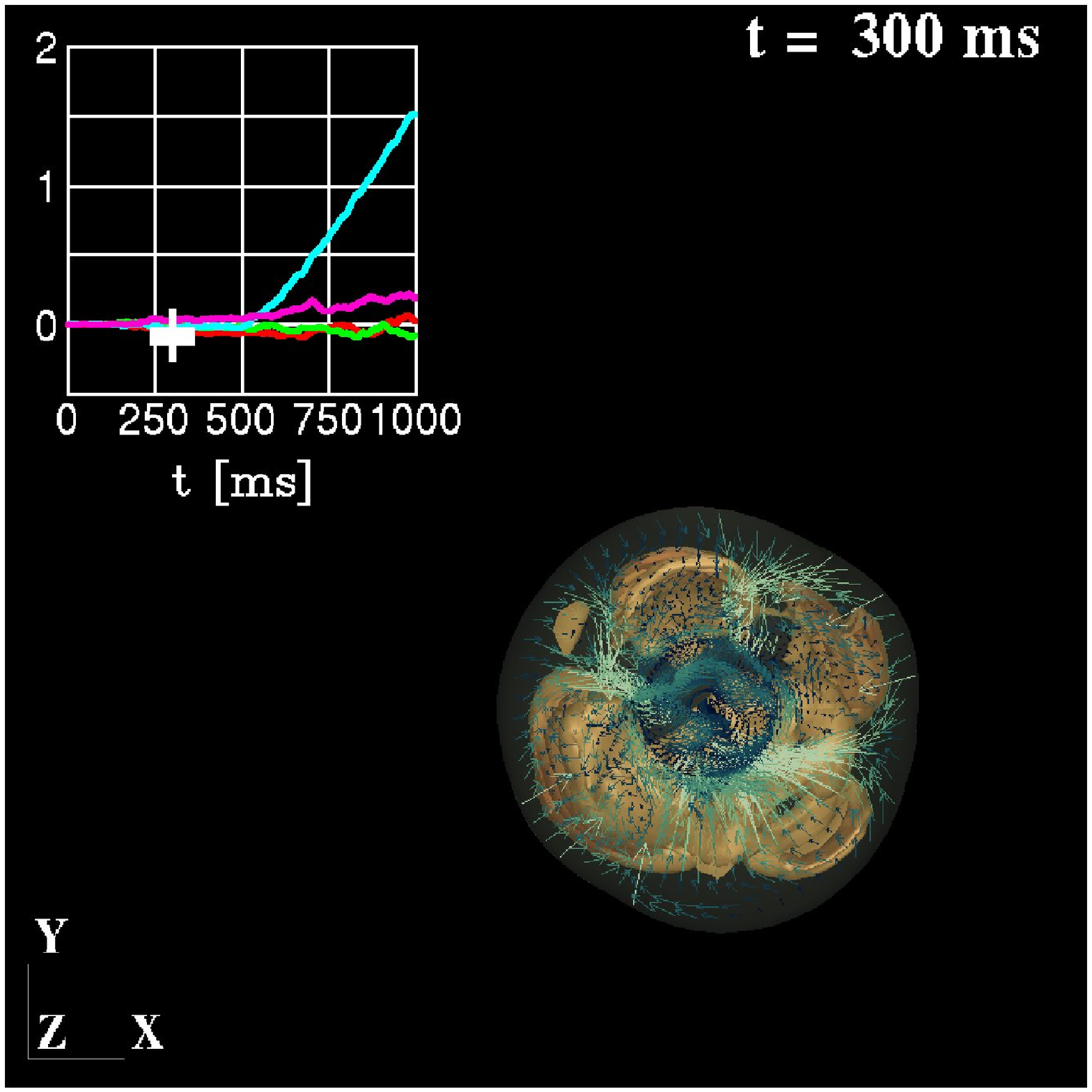}{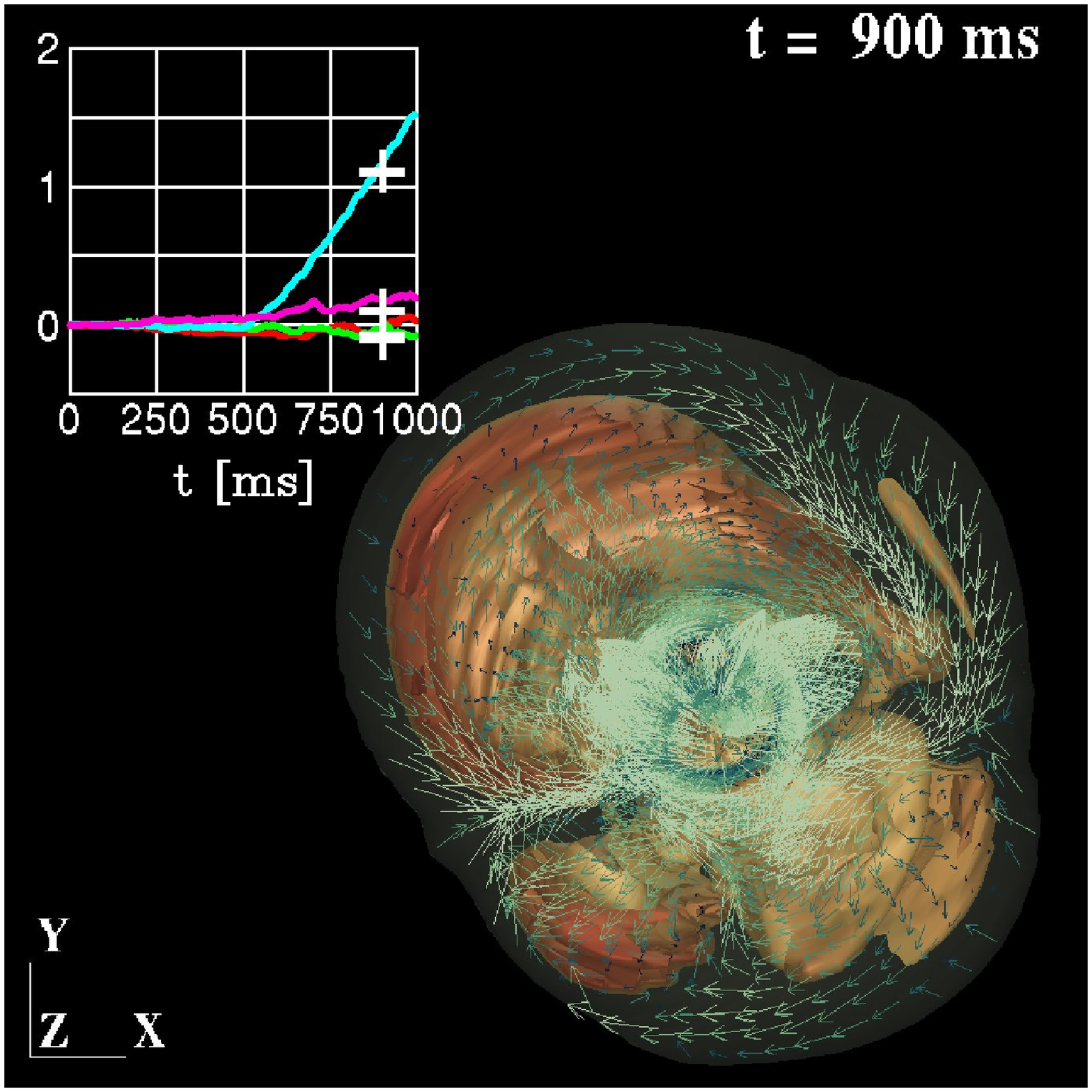}
\plottwo{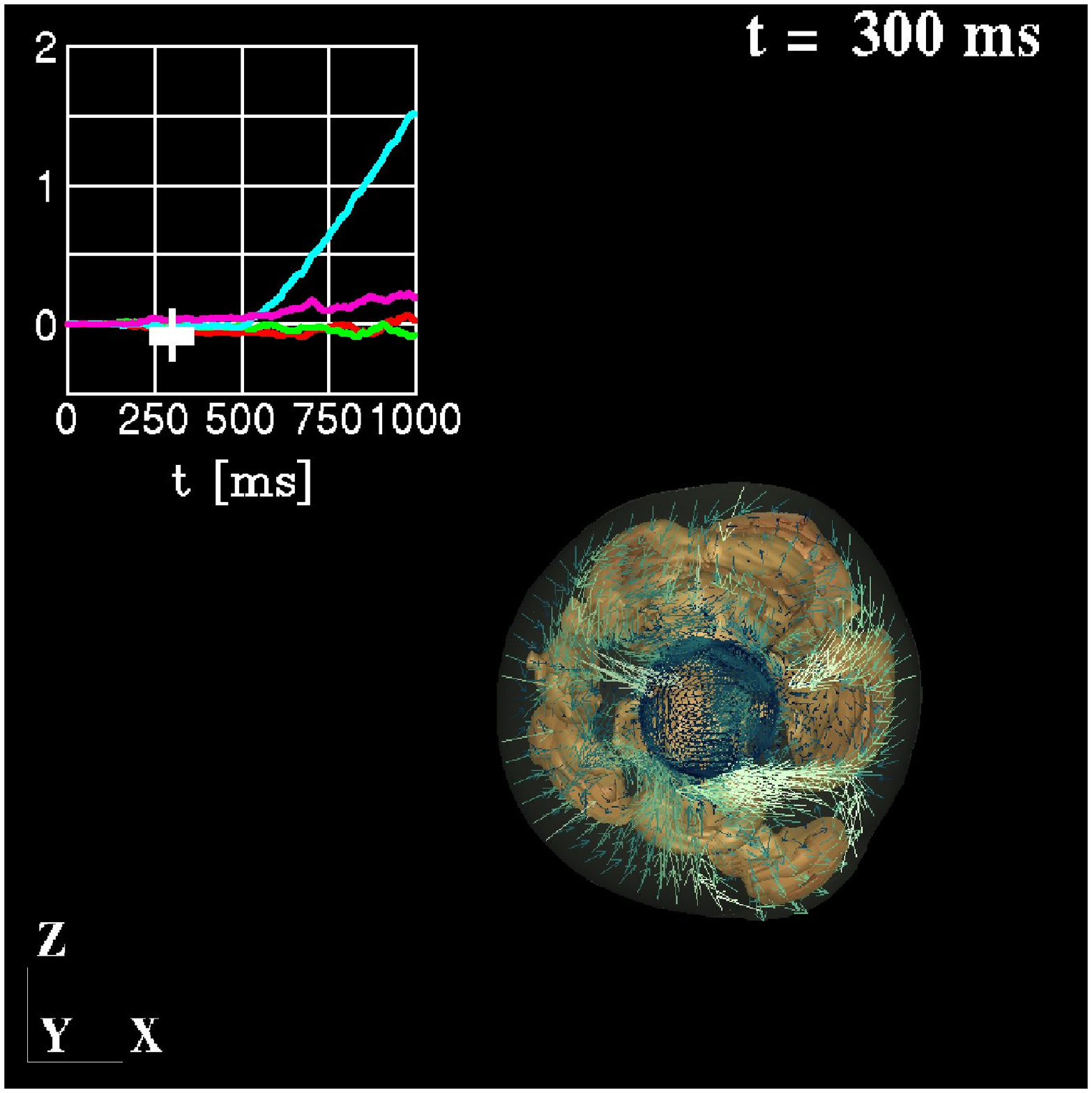}{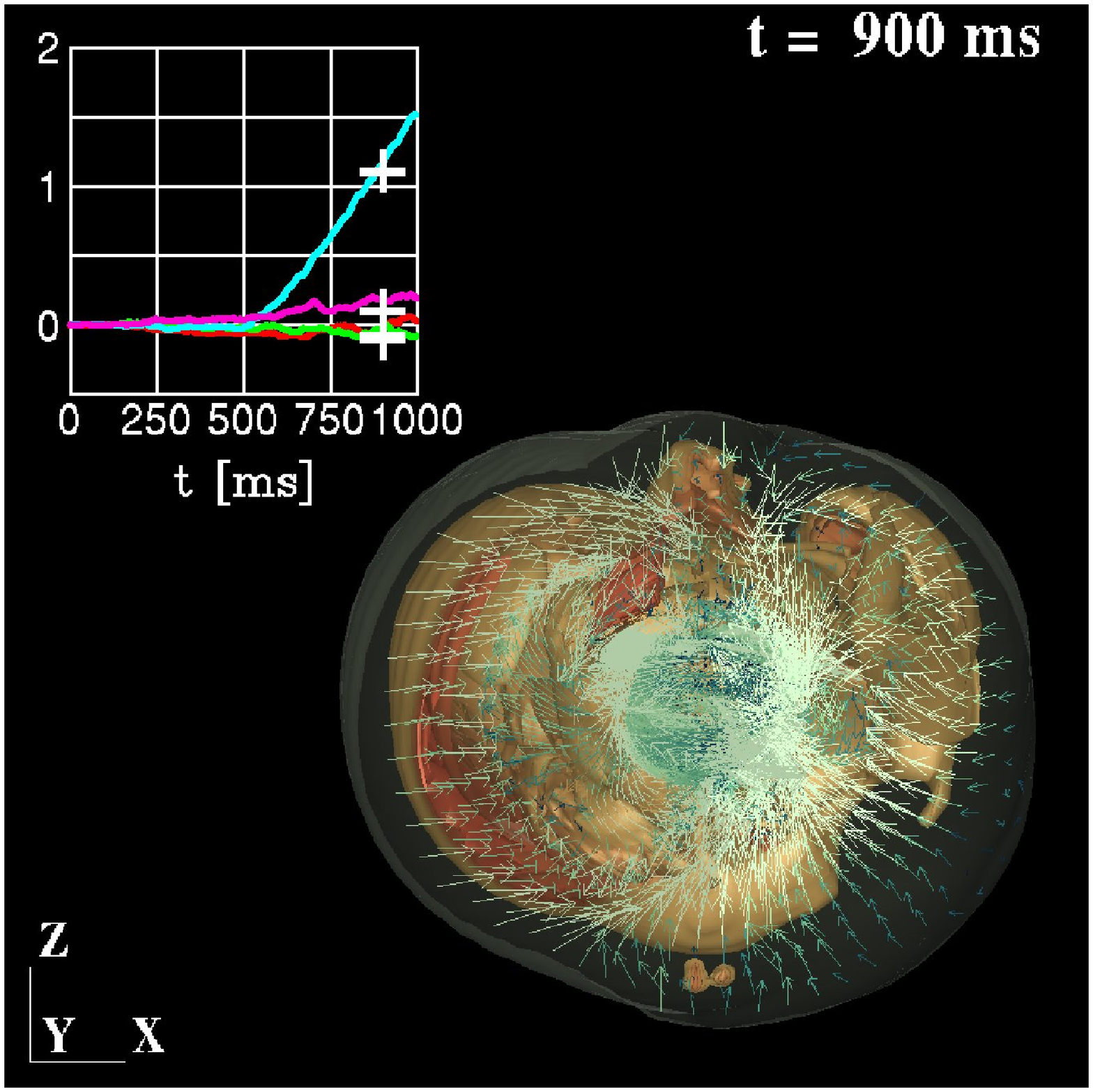}
\end{center}
\caption{Partial cutaway of the entropy isosurfaces and the velocity vectors on the 
 cutting plane for model A2. Left and right panels are at $t=300$ ms and
 $t=900$ ms corresponding to the epoch before and after the rotational flow approaches 
to the PNS surface, respectively. Top and bottom panels are for the polar and equatorial 
 observer, respectively. The insets show the gravitational waveforms with '$+$' on each curves 
representing the time of the snapshot. Note that the colors of the curves are
taken to be the same as the top panel of Figure \ref{fig3}.}
\label{fig8}
\end{figure}


\begin{figure}[hbt]
\epsscale{}
\plotone{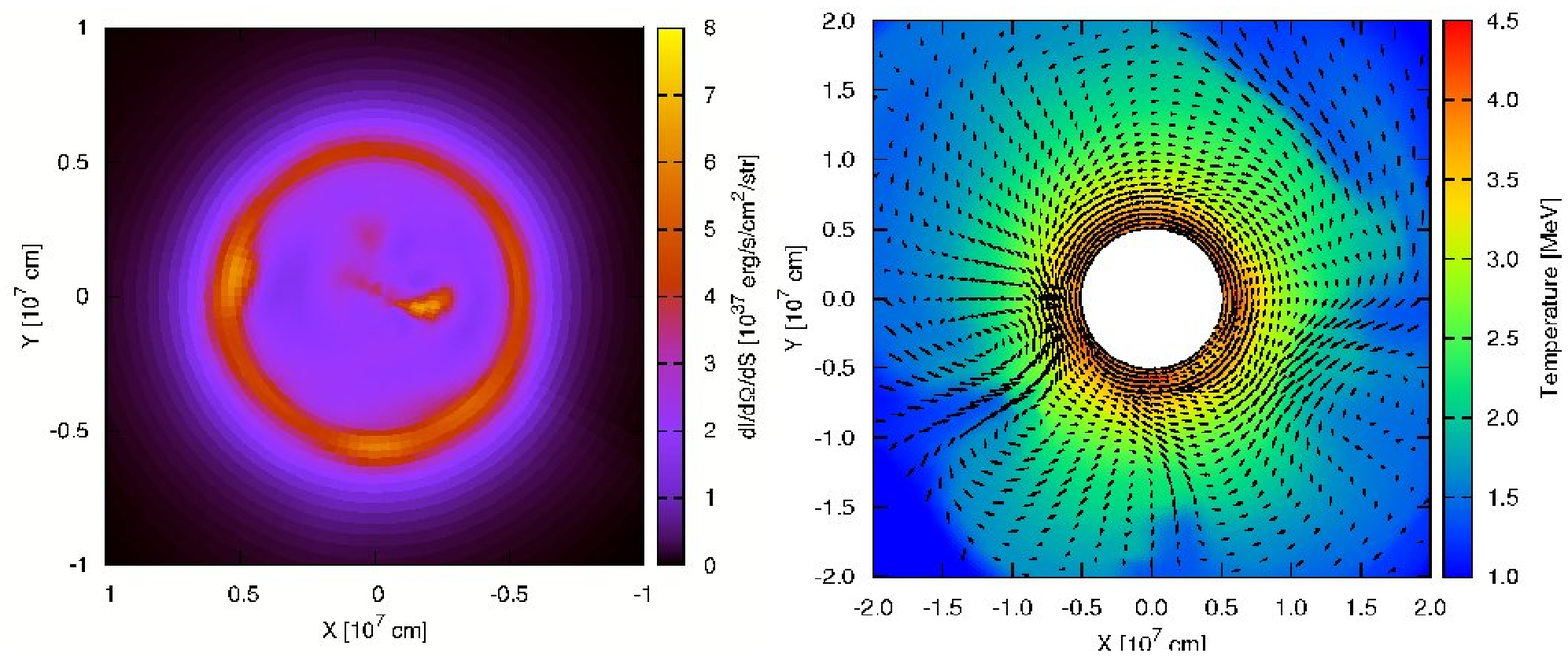}
\caption{Left and right panels show the neutrino energy fluxes of
$dl_{\nu}/({d\Omega dS})$ (equation(\ref{flux})) 
seen from the northern hemisphere and the temperature distributions in the 
equatorial plane for model A2 at $t=900$ ms.}
\label{fig9}
\end{figure}

\subsection{Breaking of  Stochasticity Due to Rotation  \label{sec3.2}}

Figure \ref{fig8} illustrates two typical snapshots of the flow fields for model A2 
 before and after the rotational flow approaches to the PNS surface (left and right 
panels), seen from the pole (top) or from the equator (bottom), respectively.
 The left panels indicate that the deformation of the shock surface is 
only mild for this model, although the SASI has already entered the non-linear phase 
at around $t = 100$ ms. From the bottom right panel, one may guess the presence of
 the sloshing modes that happen to develop along the rotational axis ($z$-axis) at this 
 epoch. It should be emphasized that the dominance of $h^{\rm equ}_{\nu,+}$ 
 observed in the current 3D simulations have nothing to do with the one
 found in our previous 2D studies \citep{kotake_ray}. 
Free from the 2D axis effects, the major axis of the SASI changes stochastically
 with time, and the flow patters behind the standing shock simultaneously change 
in every direction.
 As a result, the sloshing modes can make only a small contribution to the GW emission
 (e.g., the inset in the left panels of Figure \ref{fig8}). The remaining 
possibility is that the spiral flows clearly seen in the top right panel should be a 
key importance to understand the GW feature that we pointed out in the last section.
   
 The left panel of Figure \ref{fig9} shows the local neutrino energy fluxes 
(:$dl_{\nu}/({d\Omega dS})$, equation (\ref{flux})) seen from the northern hemisphere.
Seen as bright arcs in the left panel, the higher values 
of the neutrino energy fluxes are shown to 
coincide with the high temperature regions in the right panel. 
This is because the compression of matter is more enhanced in the vicinity of 
the equatorial plane due to the presence of 
 the spiral flows (see velocity vectors in the right panel).

\begin{figure}[hbt]
\epsscale{.8}
\begin{center}
\plottwo{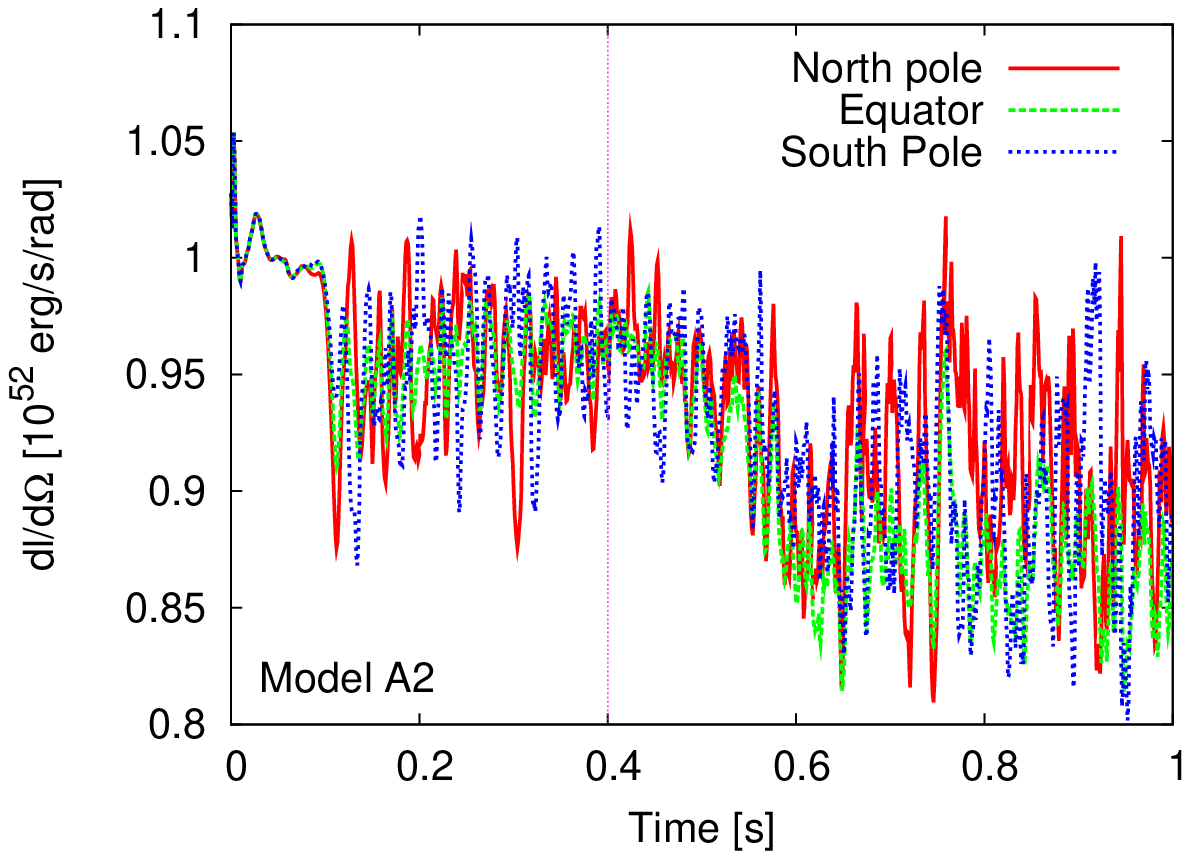}{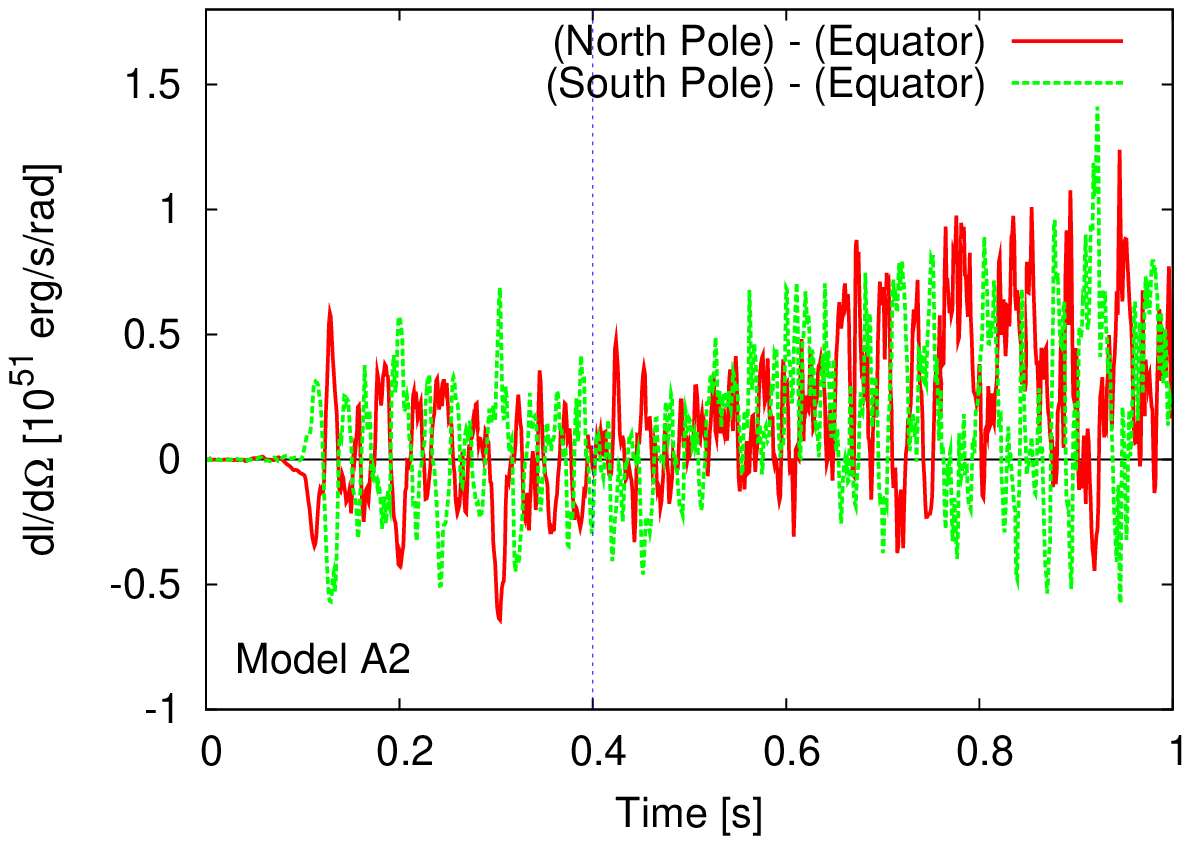}
\plottwo{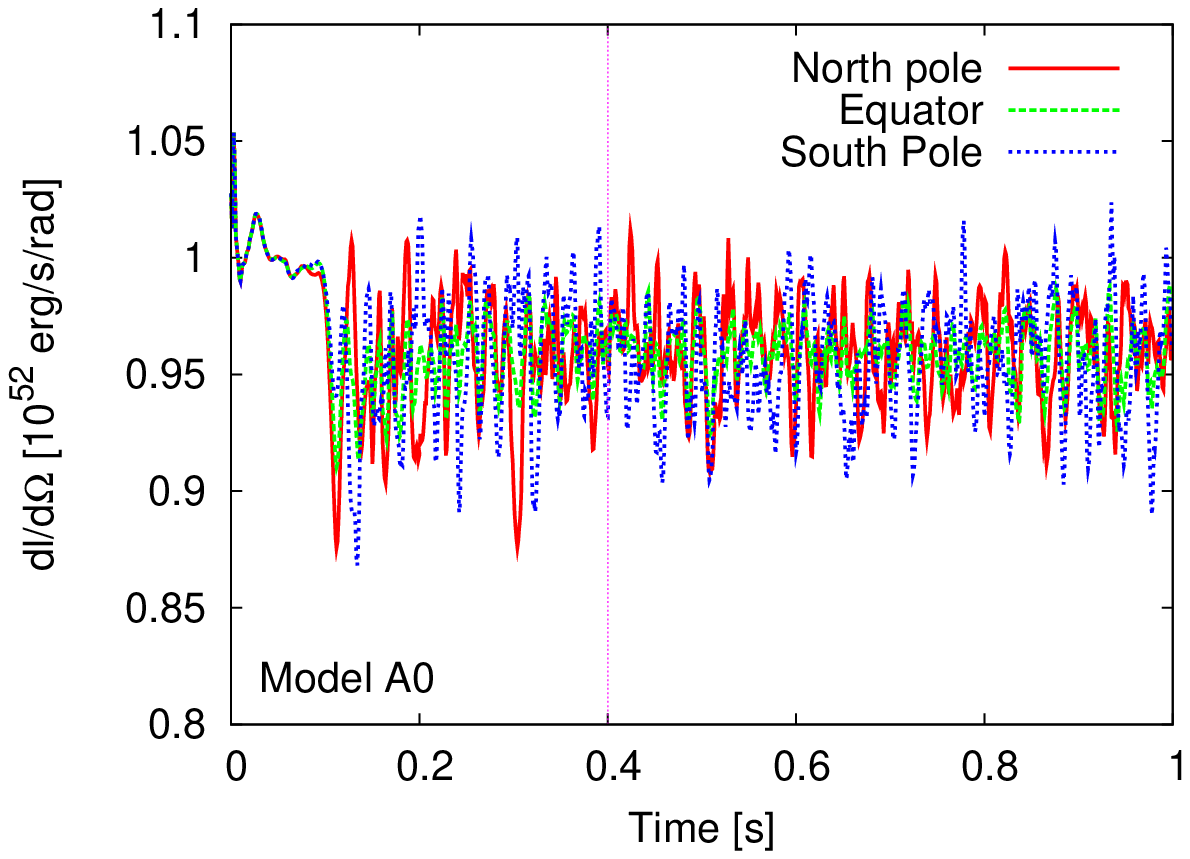}{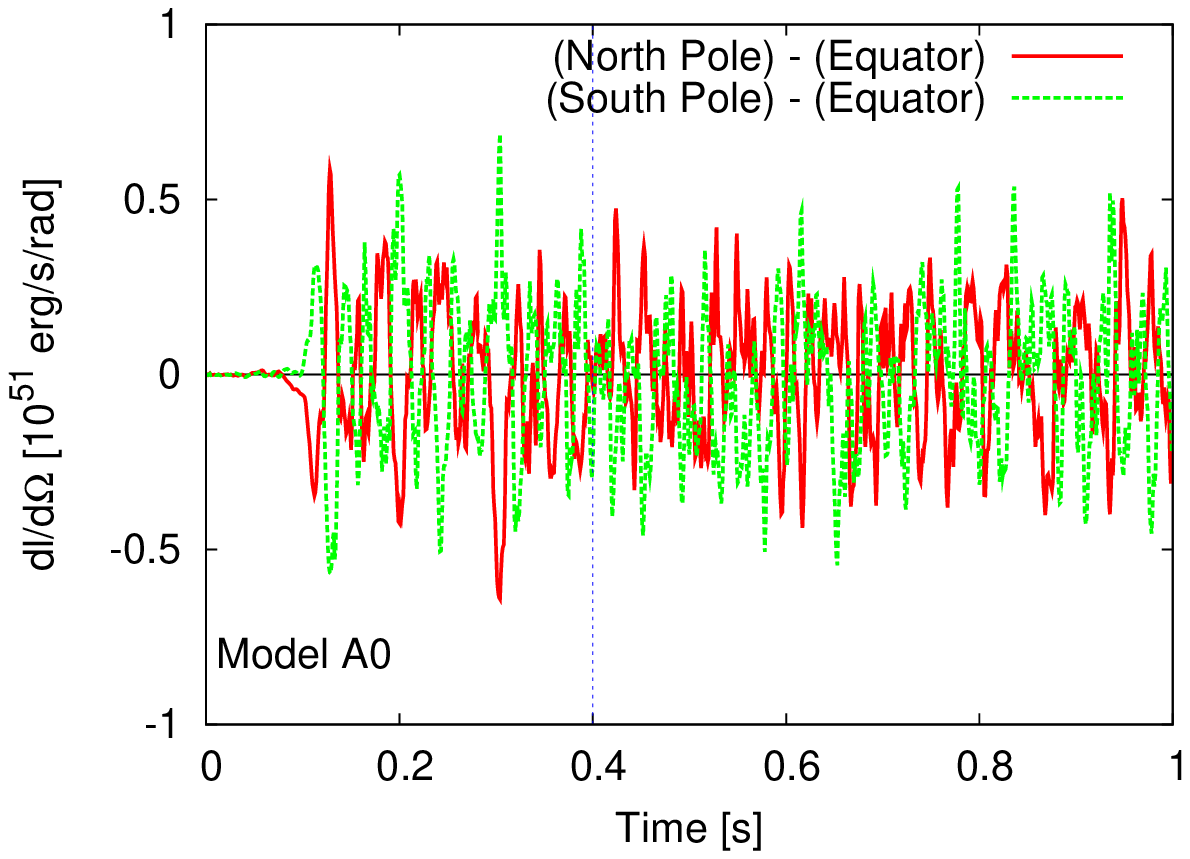}
\end{center}
\caption{Time evolution of the directional dependent neutrino luminosity: $dl_{\nu}/d\Omega$ 
(left panels) in the vicinity of the north pole, the equator, and the south pole, 
and their differences from the equator 
(right panels) for models A2 (top panels) and A0 (bottom panels).
 Vertical lines in the right panels represent the epoch of $t$ = 400 ms when the 
 rotational flow advects to the PNS surface.}
\label{fig10}
\end{figure}

 Figure \ref{fig10} shows the time evolution of 
$dl_{\nu}/d\Omega$ (equation (\ref{final})) in the vicinity of the north pole ($\theta=0$), the equator 
($\theta=\pi/2$), and the south pole ($\theta=\pi$) (left panels), 
and their differences from the equator (right panels) for models A2 (top panels) and 
A0 (bottom panels), respectively. A common feature seen from the right panels is that 
 the dominance of the neutrino emission in the north (red line) and south poles 
(green line) is occasionally
  anti-correlated. This is a consequence of the low-mode nature of SASI, here 
of $\ell=1$. The correlation shown here as well
 as the resulting GWs is much weaker 
than the ones in our 2D study (\citet{kotake_ray}), because 
 the major axis of the SASI changes much more randomly. 

\begin{figure}[hbt]
\epsscale{}
\plottwo{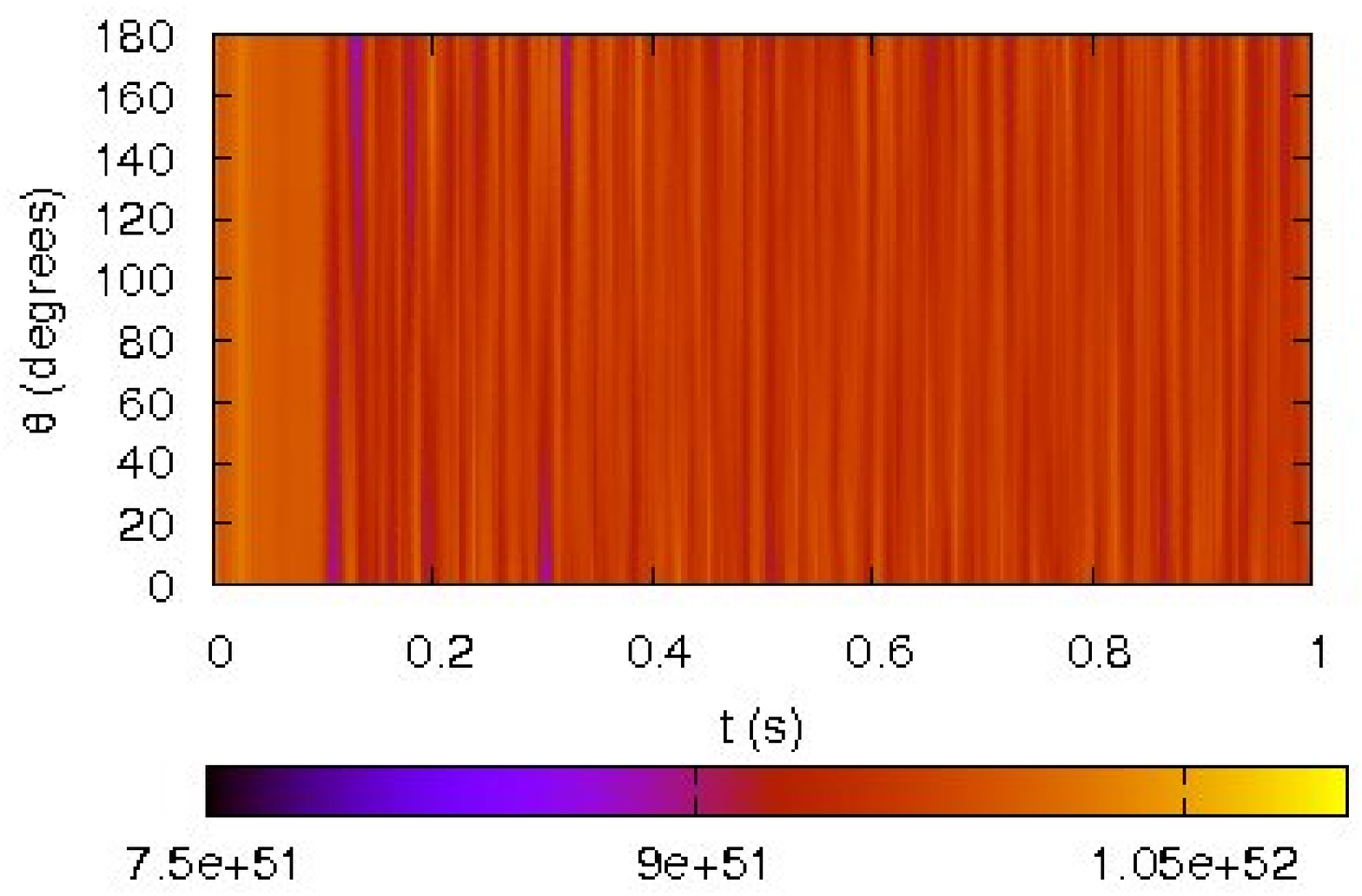}{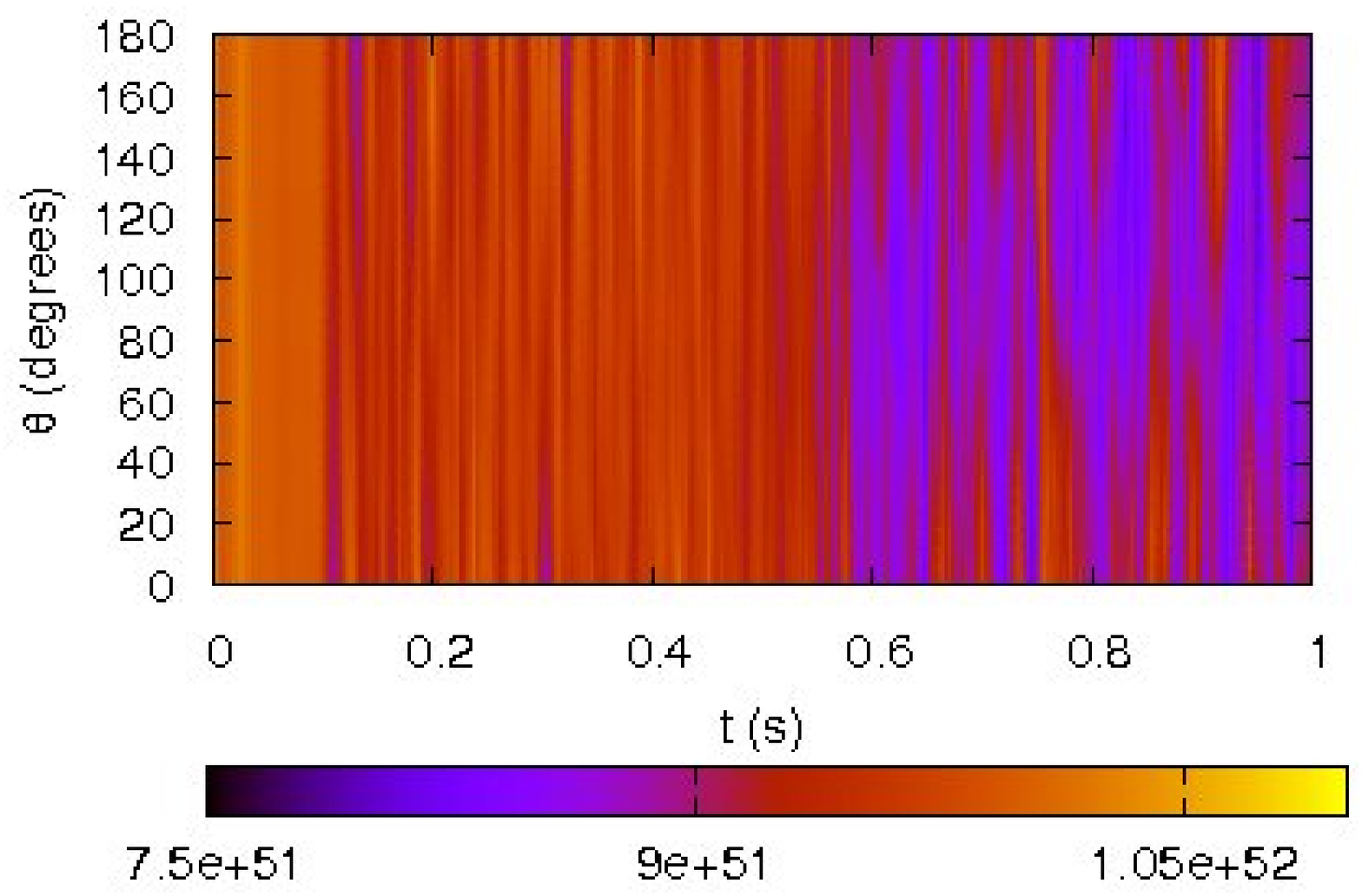}
\caption{Contour of $d l_{\nu}/d\Omega$ ((azimuthal)angle-averaged) as a function 
of time (horizontal axis) and the polar angle (vertical axis) for models A0 (left
 panel) and A2 (right panel).}
\label{fig11}
\end{figure}

The most important message in Figure \ref{fig10} is the gradual deviation of 
$d l_{\nu}/d\Omega$ for model A2 seen in the 
 top right panel. This feature becomes remarkable only after $t \sim 500-600$ ms,
  because it takes $\sim 100-200$ ms for the spiral SASI modes to develop
 behind the standing shock after the spiral flows approached to the PNS 
surface (:$t= 400$ ms denoted by the vertical line).
As already mentioned, the dominance of the neutrino luminosity seen from the 
 polar directions is due to the 
spiral flows that develop near in the vicinity of the equatorial plane. 
 Comparing the two panels in Figure \ref{fig11}, 
a contrast of the neutrino luminosity in
 the polar regions ($\theta \sim 0^\circ$ or $\theta \sim 180^\circ$) to other regions 
 becomes prominent only after $t \sim 600$ ms for model A2 (right panel), 
which is visualized as stripes bridging between the north and south pole. 
Without rotation, the anisotropy of neutrino radiation remains to be much smaller (left 
 panel in Figure \ref{fig11}), leading to much smaller GW amplitudes.
 Remembering again that $\Phi(\theta^{'})$ 
in equation (\ref{tt}) is positive near 
the north and south polar caps, the dominance of the polar neutrino luminosities 
leads to make the positively growing feature in $h^{\rm equ}_{\nu,+}$, so far depicted
in models that include rotation (e.g., Figures \ref{fig3}-\ref{fig7}). It is noted 
 here that the lateral-angle ($\theta$) dependence of equation (1) can be approximated 
by equation (3) because
 the neutrino anisotropy in the azimuthal direction is much weaker than the one 
 in the lateral direction.

\begin{figure}[hbt]
\epsscale{}
\begin{center}
\plottwo{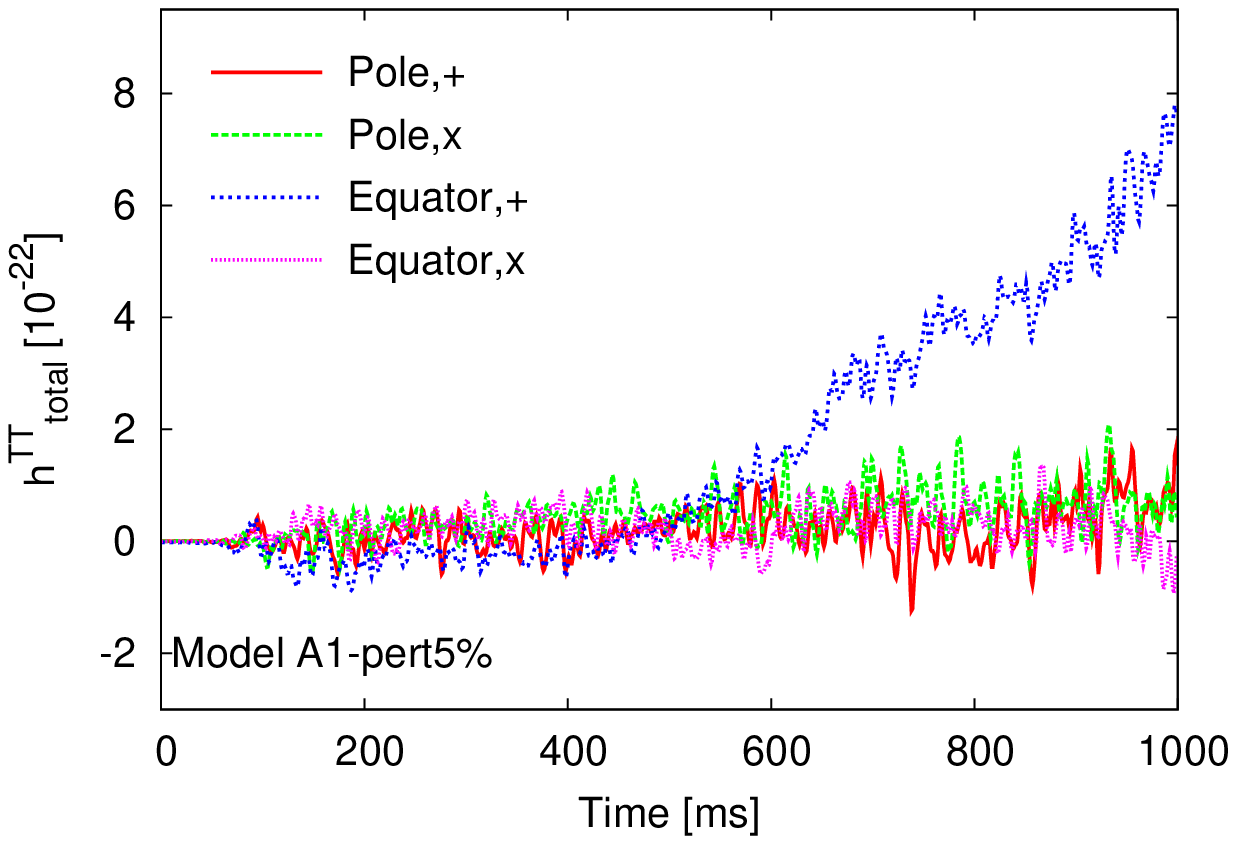}{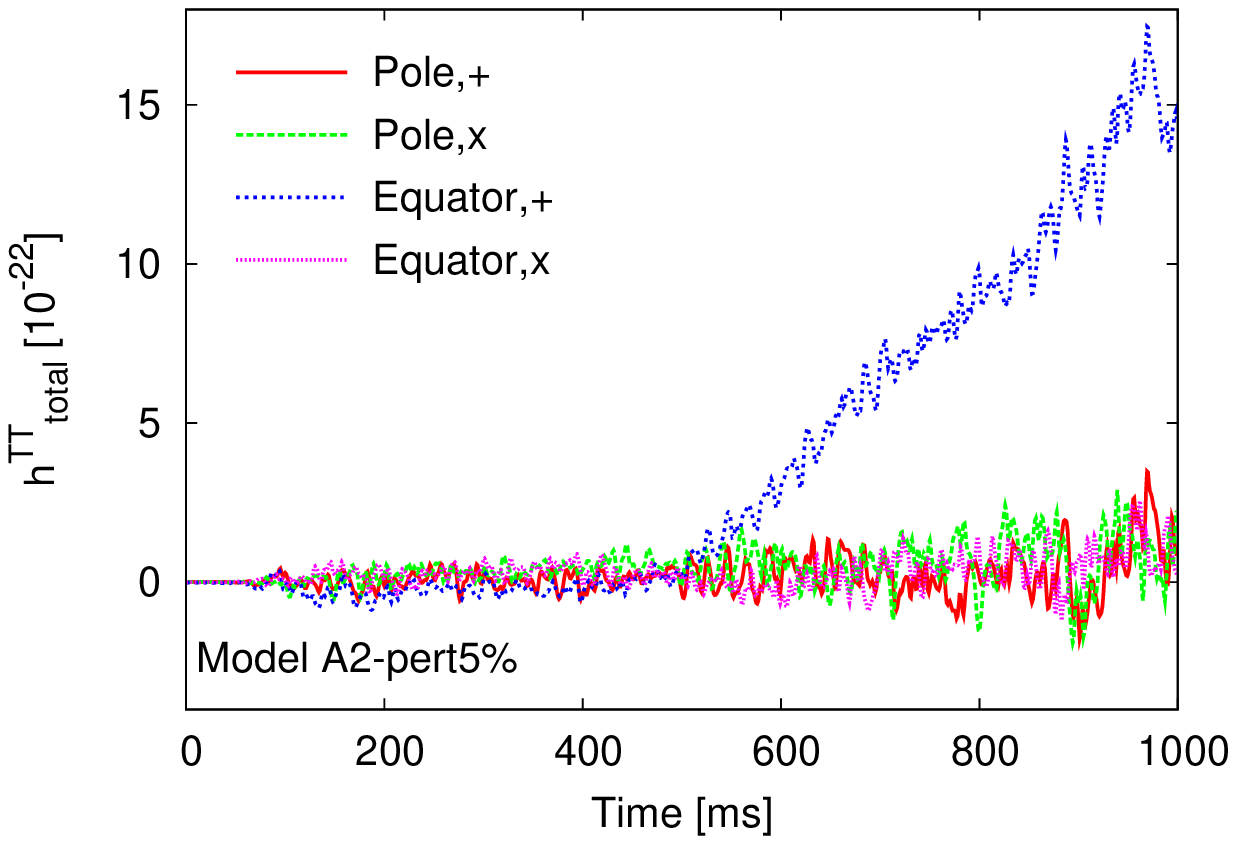}
\plottwo{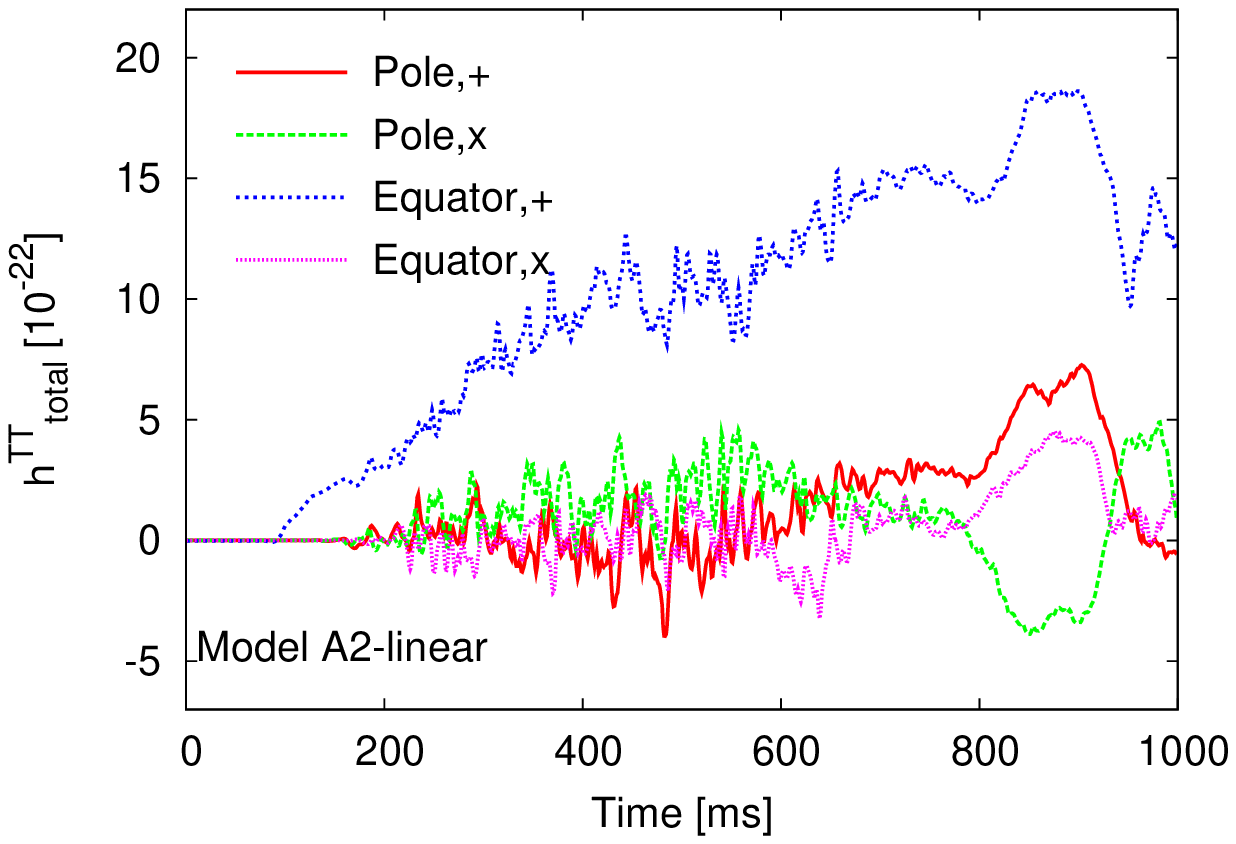}{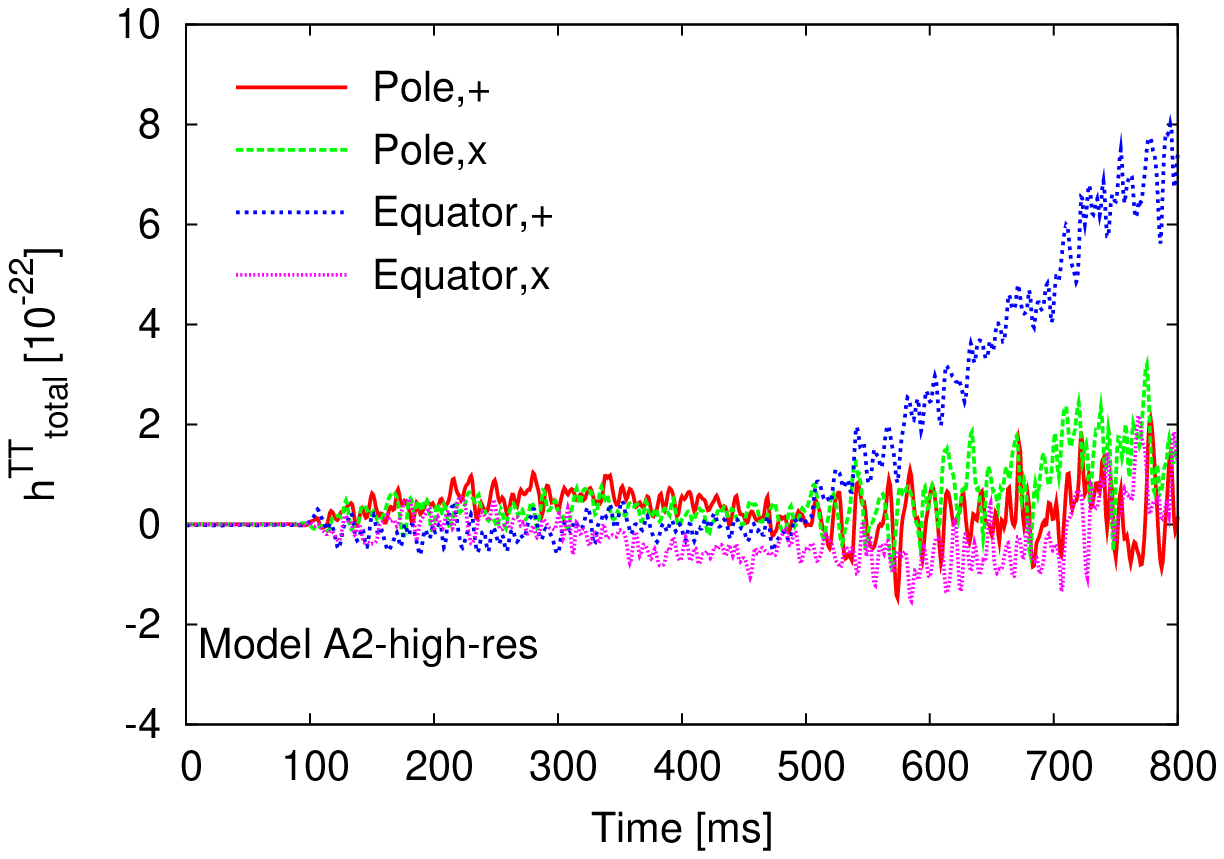}
\end{center}
\caption{Variations of model series of A1 and A2.
 For top panels, larger initial (velocity-)
 perturbations ($5\%$) are imposed (for model A1 (left) and A2 (right) indicated 
 by pert5\%) in contrast to the fiducial value of $1\%$ (e.g., section 2.2).
 For the bottom left panel, the initial rotational flow is tuned to advect to the PNS
 surface in the linear SASI phase for model A2 (indicated by A2-linear). 
In the bottom right panel, the numerical resolution for the azimuthal 
 direction is doubled compared to the fiducial value of 60 mesh points (indicated 
 by ``high-res'' for model A2).} 
\label{fig11_rev}
\end{figure}

 Figure \ref{fig11_rev} shows how the above GW features could or could not change if
 we change the strength of the initial velocity perturbation (top panels), the epoch
 when the initial rotational flow advects to the center (bottom left panel), and
 the numerical resolution (bottom right panel), respectively. In the top two panels, 
 larger initial (velocity-) perturbations ($5\%$) are imposed (for model A1 (left) 
and A2 (right) as indicated by pert5\%) in contrast to the fiducial value of 
$1\%$ (e.g., section 2.2). Due to the large perturbation, the timescale when the 
 non-linear phase sets in, becomes typically $30-40$ ms earlier
compared to the fiducial models (compare the middle and bottom panels of Figure 3 to 
 the top two panels of Figure \ref{fig11_rev}). Regardless of the difference, 
the increasing trend of the GWs (blue lines) is shown to be unaffected. This is also 
 the case when the rotational flow is adjusted to advect to the PNS in the 
 linear SASI phase (bottom left panel), and when a finer numerical resolution is taken
 (bottom right panel).  In the bottom right panel, the numerical resolution for the 
azimuthal direction is doubled compared to the fiducial value of 60 mesh points 
(indicated by ``high-res'' for model A2). 
For the bottom left panel, the initial perturbation 
is taken to be as small as $0.1\%$ to make the linear phase longer. For this model,
 the rotational flow advects to the PNS about $t \sim$ 100 ms in the linear 
 SASI phase, and the transition to the non-linear phase takes place at $t \gtrsim$ 
200 ms. As is shown, the increasing trend appears also from the linear SASI phase
 ($t \sim$ 100 ms).

\begin{figure}[hbt]
\epsscale{1}
\begin{center}
\plottwo{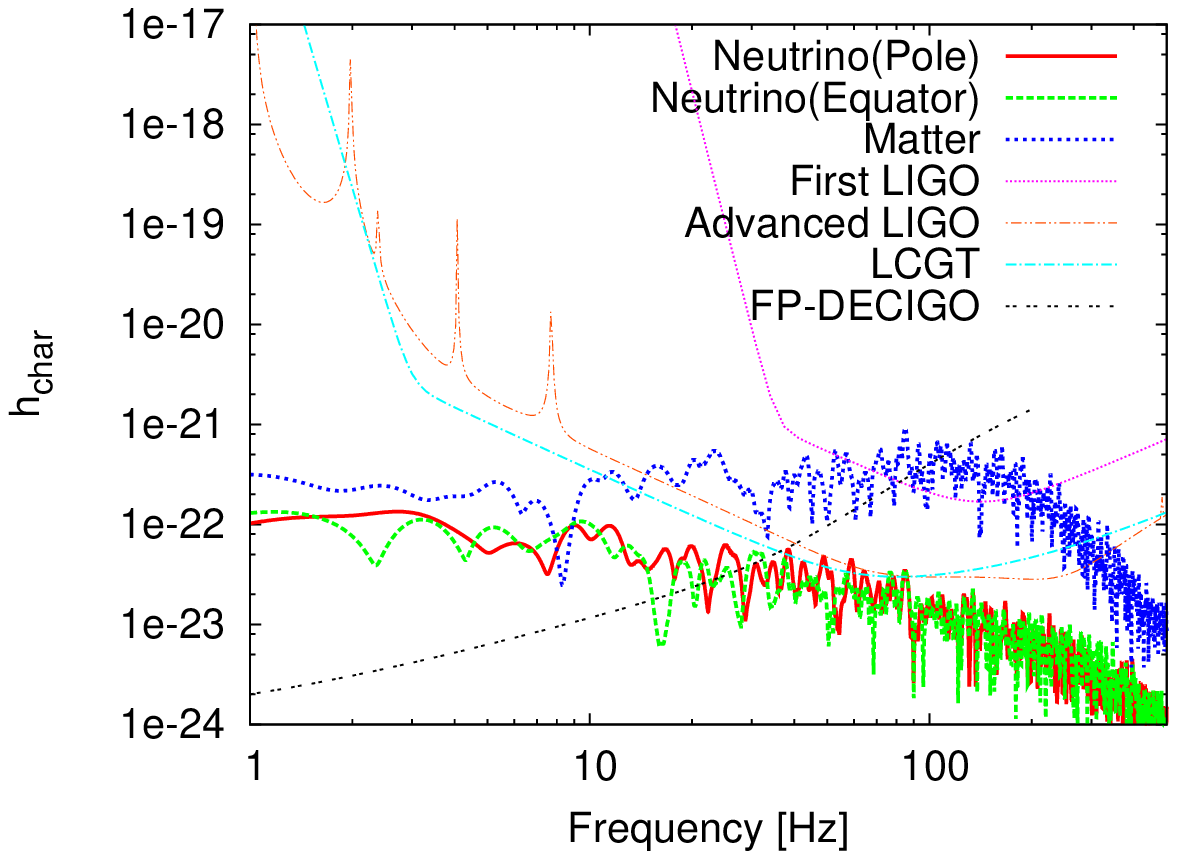}{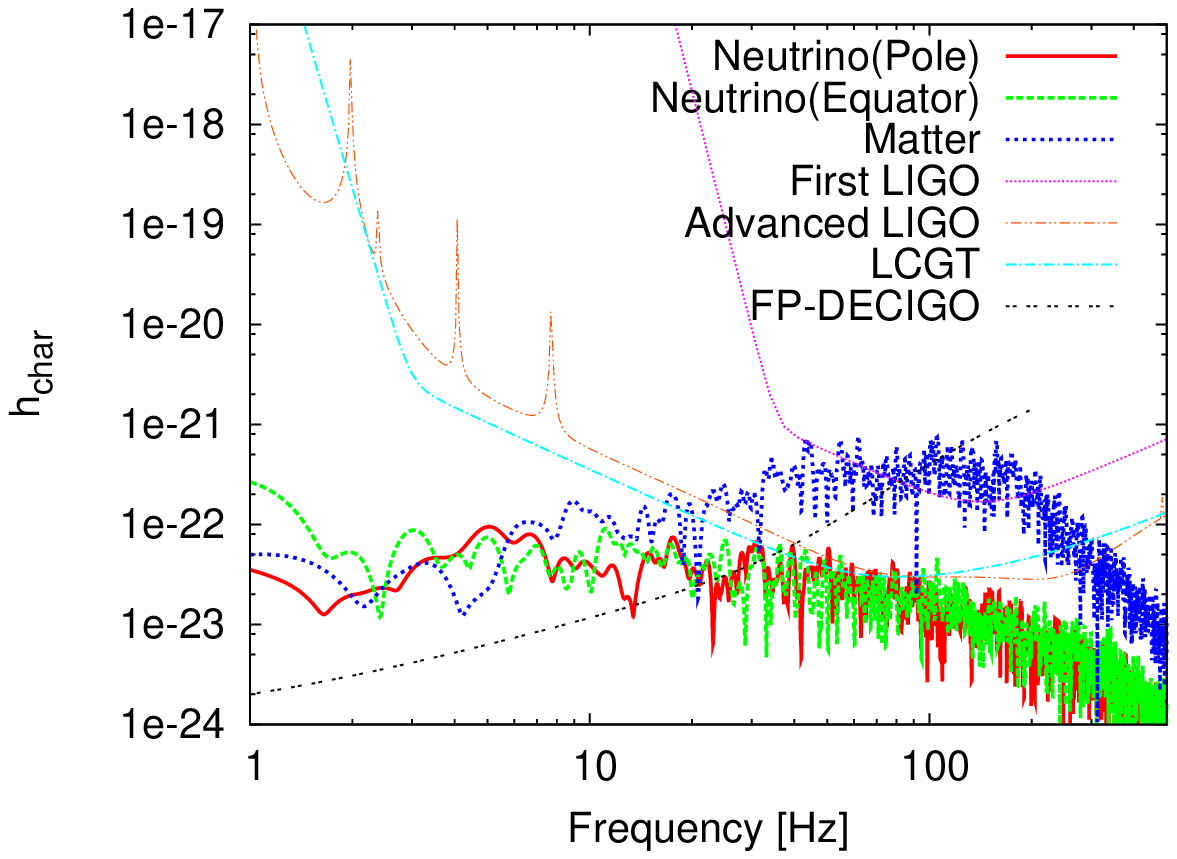}
\end{center}
\caption{Spectral distributions of GWs from matter motions (``Matter'') 
 and neutrino emission (``Neutrino'') seen from the pole or the equator 
for models A0 (left panel) and A2 (right panel) with 
the expected detection limits of TAMA \citep{tamanew}, first LIGO \citep{firstligonew}, 
advanced LIGO \citep{advancedligo}, Large-scale
 Cryogenic Gravitational wave Telescope (LCGT) \citep{lcgt} and Fabry-Perot type 
 DECIGO \citep{fpdecigo,kudoh}. The distance to the supernova is assumed to be 10 kpc.
 Note that for the matter signal, the $+$ mode seen from the polar direction is plotted. }
\label{fig12}
\end{figure}

 Finally Figure \ref{fig12} depicts
 the GW spectra for model A0 (left panel) and A2 (right panel).
 The neutrino GWs seen from the equator (green line, right panel) for model A2
 is larger than the one seen from the pole, and it is slightly larger compared 
to model A0 in the lower frequencies 
below $\sim 10$ Hz. 
It is true that the GW signals from neutrinos are very difficult to detect 
for ground-based detectors whose sensitivity is 
limited mainly by the seismic noises at such lower frequencies 
\citep{tamanew,firstligonew,advancedligo,lcgt}. However these signals may be detectable 
 by the recently proposed future space interferometers like Fabry-Perot type DECIGO 
(\citet{fpdecigo}, black line in Figure \ref{fig12}). The sensitivity curve is 
taken from \citet{kudoh}. Contributed by the neutrino GWs in the lower frequency domains,
 the total GW spectrum tends to become rather flat over a broad frequency range
 below $\sim 100$ Hz. These GW features obtained in the context of the 
SASI-aided neutrino-driven mechanism are different from the ones expected in the 
other candidate supernova mechanism, such as the MHD mechanism (e.g., 
\citet{ober06b,taki_kota}) and
 the acoustic mechanism \citep{ott_new}. Therefore the detection of such signals is 
 expected to provide an important probe into the long-veiled explosion mechanism.

Regarding the GW amplitudes from matter motions, we find no significant change or 
increase due to the spiral flows either seen 
from the polar or equatorial directions. This is mainly because the spiral 
flows can develop only outside the PNS taken to be 50 km in our idealized simulations,
 leading to have small changes in the mass-quadrupole (due to its small masses). 
This situation might be akin to the one 
observed in the early 3D simulations by \citet{rampp}. 
 To produce sizable GW amplitudes, the non-axisymmetric instabilities that 
globally develop in the more central regions to the PNS should be needed 
(e.g., \citet{ott_3D,simon1}). To explore these fascinating phenomena,
 one apparently needs to perform full 3D radiation-hydrodynamic simulations 
covering the entire stellar core and starting from gravitational collapse to 
explosions in a consistent manner.

\clearpage
\section{Summary and Discussion \label{sec4}}


We studied how the spiral modes of the SASI can have impacts on the properties of 
 GWs by performing 3D simulations that mimic the 
 SASI-aided core-collapse supernovae. To see the effects of rotation,
 we imposed a uniform rotation on the flow advecting from the outer boundary 
of the iron core (as in \citet{iwakami2}), whose specific angular momentum is 
assumed to agree with recent stellar evolution models. 
We computed fifteen 3D models in which the initial angular momentum as well as the 
input neutrino luminosities from the PNS
 are changed in a systematic manner. By performing 
a ray-tracing analysis, we accurately estimated 
 the GW amplitudes from anisotropic neutrino emission. With these computations,
 we found that the gravitational waveforms from neutrinos in models that include rotation 
 exhibit a common feature otherwise they 
vary much more stochastically in the absence of rotation. The breaking of the 
 stochasticity stems from the excess of the neutrino 
emission parallel to the spin axis. This is because the compression of matter 
is more enhanced in the vicinity of the equatorial plane due to the growth of the 
spiral SASI modes, leading to the formation of the spiral flows with higher 
temperatures circulating around the spin axis. We pointed out that a recently 
proposed future space interferometers like Fabry-Perot type DECIGO would permit the 
detection of these signals for a Galactic supernova. 

It should be noted that the approximations taken in the simulation, 
such as the excision inside the PNS with its fixed inner boundary and the light bulb
approach with the isentropic luminosity constant with time, are only a very first step 
towards realistic 3D supernova simulations.
As already mentioned, the excision of the central regions inside PNSs 
 hinders the efficient gravitational emission there, such as by the g-mode 
oscillations \citep{ott_new} and the non-axisymmetric instabilities 
 \citep{ott_3D,simon1} of the PNSs, and the enhanced neutrino emissions inside 
the PNSs \citep{marek_gw}. It should also affect the peak frequency of 
 the GW spectra as pointed out by \citet{murphy}. The rotational
 flows advecting into the inner core should spin up the PNS, which is expected to 
conversely affect the dynamics outside \citep{rantsiou}.
 These important feedback should be taken into account 
to unravel the effect of rotation more precisely.  Remembering these caveats in mind,
 one encouraging news to us is that the gravitational waveforms obtained in the 
 2D radiation-hydrodynamic simulations \citep{yukunin}
are similar to the ones obtained in our 2D study \citep{kotake_ray}. 
 It is also noted that magnetic effects should have impacts
 not only on the growth of the SASI \citep{endeve,fogli_B}, but also on the GW 
waveforms, which are remained to be studied in the 3D MHD simulations.
Nowadays, the next 
 generation 3D results are being reported by using new simulation techniques
 such as the Yin-Yang grids \citep{annop} and the adaptive mesh refinement approach
\citep{nordhaus}.
We hope that
 our results may give a momentum to theorists for making the GW prediction
 with a better quantitative precision based on the sophisticated 3D supernova modeling. 

\acknowledgements{K.K. would like express thanks to K. Sato and
S. Yamada for continuing encouragements.
Numerical computations were in part carried on XT4 and 
general common use computer system at the center for Computational Astrophysics, CfCA, the National Astronomical Observatory of Japan.  This
study was supported in part by the Grants-in-Aid for the Scientific Research 
from the Ministry of Education, Science and Culture of Japan (Nos. 19540309 and 20740150).}
\clearpage
\bibliographystyle{apj} 
\bibliography{ms}





\end{document}